\documentclass[traditabstract]{aa} % for the abstract without structuration 
\usepackage{appendix}
\usepackage{graphicx}
\usepackage{txfonts}
\usepackage{longtable}
\usepackage{lscape}
\usepackage{afterpage}
\usepackage{rotating}
\usepackage{cancel}
\usepackage{soul}

\begin{document}%

\title{Ionized gas outflows and global kinematics of low-z luminous star-forming galaxies}

     \author{S.~Arribas\inst{\ref{inst1}}\and L.~Colina\inst{\ref{inst1}}\and E.~Bellocchi\inst{\ref{inst1}}\and R.~Maiolino\inst{\ref{inst2},\ref{inst3}}\and M.~Villar-Mart{\'{\i}}n\inst{\ref{inst1}} }  %\and
          %\fnmsep\thanks{Just to show the usage
          %of the elements in the author field}
   \institute{CSIC - Departamento de Astrof{\'{\i}}sica-Centro de Astrobiolog{\'{\i}}a (CSIC-INTA),Torrej{\'o}n de Ardoz, Madrid, Spain \\ \email{arribas@cab.inta-csic.es}\label{inst1} \and Cavendish Laboratory, University of Cambridge 19 J. J. Thomson Avenue, Cambridge CB3 0HE, UK\label{inst2} \and Kavli Institute for Cosmology, University of Cambridge, Madingley Road, Cambridge CB3 0HA, UK\label{inst3}
}

\date{}

% \abstract{}{}{}{}{} 
% 5 {} token are mandatory

  \abstract
{We study the kinematic properties of the ionised gas outflows and ambient interstellar medium (ISM) in a large and representative sample of local luminous and ultra-luminous infrared galaxies (U/LIRGs) (58 systems, 75 galaxies) at galactic and sub-galactic (i.e., star-forming clumps) scales, thanks to integral field spectroscopy (IFS)-based high signal-to-noise integrated spectra.
   
The velocity dispersion of the ionized ISM in U/LIRGs ($<$$\sigma$$>$ $\sim$ 70 kms$^{-1}$ ) is larger than in lower luminosity local star-forming galaxies ($<$$\sigma$$>$ $\sim$ 25 kms$^{-1}$ ).  While for isolated disc LIRGs star formation appears to sustain turbulence, gravitational energy release associated with interactions and mergers plays an important role in driving $\sigma$ in the U/LIRG range. We find that $\sigma$ has a dependency on the star formation rate density ($\Sigma_{SFR}$), which is weaker than expected if it were driven by the energy released by the starburst.  The relatively small role of star formation (SF) driving the $\sigma$ in U/LIRGs is reinforced by the lack of an increase in $\sigma$ associated with high luminosity SF clumps.  We also find that the impact of an active galactic nucleus (AGN) in ULIRGs is strong, increasing on average $\sigma$ by a factor 1.5. Low-z U/LIRGs cover a range of velocity dispersion ($\sigma$ $\sim$  30 to 100 km s${^{-1}}$) and star formation rate density  ($\Sigma_{SFR}$ $\sim$ 0.1 to 20 M$_{\odot} yr^{-1} kpc^{-2}$) similar to those of  SFGs. Moreover, the observed weak dependency of  $\sigma$ on $\Sigma_{SFR}$ for local U/LIRGs ($\sigma \propto \Sigma_{SFR}^{+0.06}$) is in very good agreement with that measured in some high-z samples.
 
The presence of ionized gas outflows in U/LIRGs seems universal based on the detection of a broad, usually blueshifted, H$\alpha$ line.  The observed dependency of the {\it maximum} velocity of the outflow  (V$_{max}$) on the star formation rate (SFR) is of the type V$_{max}(non-AGN) \propto SFR (L_{IR})^{+0.24}$.  We find that AGNs in U/LIRGs are able to generate faster ($\sim$ $\times$ 2) and more massive ($\sim \times$ 1.4) ionized gas outflows than pure starbursts. The derived ionized mass loading factors ($\eta$) are in general below 1, with only a few AGNs above this limit. The escaping gas fraction is low with only less massive (log(M$_{dyn}/ M_{\odot}) <$ 10.4) U/LIRGs having outflowing terminal velocities higher than their escape velocities, and more massive galaxies retaining the gas, even if they host an AGN. The observed average outflow properties in U/LIRGs are similar to high-z galaxies of comparable SFR. However, while high-z galaxies seem to require $\Sigma_{SFR} >$ 1 $M_{\odot} yr^{-1} kpc^{-2}$ for launching strong outflows, this threshold is not observed in low-z U/LIRGs even after correcting for the differential fraction of the gas content.  

In the bright SF clumps found in LIRGs, ionized gas outflows appear to be very common (detection rate over 80\%). Their observed properties are less extreme than those associated with the entire galaxy. The clumps  in LIRGs follow the general size-L-$\sigma$ scaling relations found for low- and high-z clumps, though they are in general smaller, less luminous, and are characterized by lower $\sigma$ than at high-z.  For a given observed (no internal extinction correction applied) star formation surface density, outflows in LIRG clumps would be about one to two orders of magnitude less energetic than the outflows launched by clumps in high-z SF galaxies. 

}

 \keywords{galaxies -- sizes -- 
               luminous infrared galaxies --
               integral field spectroscopy
               }
 \maketitle
%
%________________________________________________________________

\section {Introduction}

Understanding the physical mechanisms that drive the dynamical status of the interstellar medium in galaxies, and how they relate to processes that are either global (i.e. interactions/mergers or steady intergalactic gas accretion) or local (i.e. star formation and/or active galactic nucleus, AGN), is key to advancing our understanding of how galaxies form and evolve.  Numerical simulations indicate that both mergers and gas accretion not only trigger an increase of star formation, but also a change of the interstellar medium towards a dynamically hotter system  (Bournaud et al. 2011; Dekel, Sari $\&$ Ceverino 2009; Ceverino, Dekel $\&$ Bournaud 2010).  
The predicted massive star formation and/or  feeding of central black holes would expel the surrounding interstellar medium as a consequence of the mechanical and radiative energy liberated, generating strong outflows (di Matteo et al. 2005;Hopkins et al. 2012; Murray et al. 2011; Ostriker $\&$ Shetty 2011). The role of these outflows governing the subsequent galaxy evolution is believed to be crucial, as they can regulate and quench both star formation and black hole activity, being  also the primary mechanism by which dust and metals are redistributed over large scales within the galaxy, or even expelled into the intergalactic medium (IGM, see Veilleux, Cecil, \& Bland-Hawthorn 2005 for a review).  Hence most of the models of galaxy formation require energetic outflows to reproduce the observed properties of massive galaxies (e.g., Silk and Rees 1998;  Di Matteo et al. 2005; Hopkins et al. 2006; 2008; 2012).  Although this general theoretical framework is reasonably well established, the relative role of the different mechanisms involved is not fully understood. Also, it is not well determined how the dynamical status of the ISM and the properties of the outflows depend on the mass of the host galaxy, large-scale tidal forces, star formation rate, and/or type and luminosity of the AGN.

In this context, the study of global gas kinematics in low-z luminous and ultra-luminous infrared galaxies (i.e., LIRGs: $L_{IR} \equiv L[8-1000 \mu m]= 10^{11-12}L_{\odot}$; ULIRGs: $L_{IR} > 10^{12}L_{\odot}$), which represent the most extreme star-forming galaxies in the local universe, is particularly relevant.  These objects have a variety of morphologies and dynamical phases, from regular isolated spirals for low luminosity LIRGs to mergers for most ULIRGs (Borne et al. 2000; Farrah et al. 2001; Bushouse et al. 2002; Veilleux et al. 2002; Arribas et al. 2004), covering a wide range of dynamical mass (Bellocchi et al. 2013). Although massive star formation is believed to be their main energy source, a fraction of LIRGs and ULIRGs host luminous AGNs. This AGN activity has a more important relative contribution for the most luminous objects (i.e., ULIRGs; Veilleux et al. 1995, 1999; Narnini et al. 2010; Iwasawa et al. 2011; Pereira-Santaella et al. 2011; Alonso-Herrero et al. 2012). Although gas outflows and large-scale galactic winds are associated with different types of AGNs and star-forming galaxies (e.g., Veilleux et al.  2005), LIRGs and ULIRGs exhibit the most conspicuous cases for ionized (Heckman, Armus, \& Miley 1990; Colina, Arribas \& Borne 1999; Arribas, Colina \& Clements 2001; Bedregal et al. 2009; Colina et al. 2012; Westmoquette et al. 2012; Rupke \& Veilleux 2013a; Rodr{\'{\i}}guez-Zaur{\'{\i}}n et al. 2013), neutral (Rupke, Veilleux \& Sanders 2005a,b; Rupke \& Veilleux 2011; Rupke \& Veilleux 2013a), and molecular (Feruglio et al. 2010; Fisher et al. 2010; Sturm et al. 2011; Spoon et al. 2013; Veilleux et al. 2013; Rupke \& Veilleux 2013b; Cicone et al. 2014) gas outflows in the local universe. Therefore, LIRGs and ULIRGs are an important population of galaxies to study in detail how their interstellar medium and outflows relate to the SF activity, the presence of AGNs, and the different dynamical phases along the merging process.

In recent years, the study of ISM properties and outflows at high-z has also gained much interest. Thanks to the advent of the integral field spectroscopy (IFS), an important progress has been made in characterizing the internal kinematic structure of the ionized gas in high-z star-forming galaxies (SFGs, Forster-Schreiber et al. 2006, 2009; Law et al. 2009; Gnerucci et al. 2011; Wright et al. 2008, Epinat et al. 2012; Swinbank et al. 2012).  One important discovery has been that high-z SFGs are dynamically hot systems with very large velocity dispersions ($\sigma$ $\sim$ 60-90 kms$^{-1}$; e.g., Forster-Schreiber et al. 2009) as compared with the general population of disc galaxies in the low-z universe ($\sigma$ $\sim$ 20-30 kms$^{-1}$; e.g., Epinat et al. 2009).  The physical mechanism sustaining the gas velocity dispersion in high-z galaxies has not been fully established though (e.g., Genzel et al. 2011).  Another remarkable characteristic of high-z SFGs is that outflows seem to be very prominent and ubiquitous (Shapley et al. 2003; Alexander et al. 2010; Maiolino et al. 2012;  Cano-Diaz et al. 2012). During the peak of star formation (z$\sim$ 2-3) outflows are believed to play a pivotal role in shaping galaxies, as they regulate both star formation and black hole growth.  Recently, IFS studies are starting to characterize ionized gas outflows at high-z (Shapiro et al. 2008; Newman et al. 2012; Harrison et al. 2012; Swinbank et al. 2012; F{\"o}rster Schreiber et al. 2014), providing intriguing relations between their properties and the characteristics of their galaxy hosts.

Most galaxies at z$\sim$ 1.5-2.5 are actively building up their stellar mass, forming stars at rates of 20-200 M$_{\odot}$ yr$^{-1}$ and specific SFR, sSFR= SFR/M$\star, \sim$1-2 Gyr$^{-1}$ (i.e., the {\it main sequence (MS) of star-forming galaxies}; Rodighiero et al. 2011), with a fraction of galaxies (above MS) forming stars up to 10$\times$ faster (starbursts like SMGs). In the low-z universe, LIRGs and ULIRGs form stars at rates $\sim$ 10$-$1000 M$_{\odot}$ yr$^{-1}$, thus covering the SFR range of the z$\sim$ 2 MS (and above). Low-z LIRGs and ULIRGs differentiate each other not only on their amount of star formation, but also on their overall kinematics. There is a clear dominance of rotating  discs in LIRGs and of more turbulent, dispersion dominated mergers in ULIRGs (Bellochi et al. 2013). Studies of z$\sim$2 star-forming galaxies demonstrate that while most MS galaxies are large, rotating  discs (Forster-Schreiber et al. 2009), above the MS starbursts appear more turbulent and associated with mergers (Alaghband-Zadeh et al. 2012; Menendez-Delmestre et al. 2013). Therefore, our understanding of the formation and evolution of high-z SFGs will largely be benefitted from a detailed study of low-z analogs not suffering from the limited linear resolution (even with adaptive optics, AO, techniques) and sensitivity affecting the observations of high-z galaxies.

In this paper, we study the main integrated kinematic properties of the ionized interstellar medium in the largest sample of local U/LIRGs observed so far via IFS (Arribas et al. 2008). Specifically, we focus the paper on the study of i) the velocity dispersion of the ISM, and ii) on the properties of their ionized gas outflows, and how they relate to the characteristics of the system like the global star formation rate, star formation density,  morphological merging phase, and the presence of AGNs. For the present study, we follow a methodology often used at high-z, which consists in integrating the IFS spectra for deriving the (flux weighted) average properties of the system.  The sample includes a wide range of properties, including a good representation of the less  studied luminosity range of LIRGs. This luminosity range may be of special relevance for comparison with high-z sources, as several authors have suggested that some high-z populations are scale-up versions of local LIRGs (e.g., Takagi et al. 2010; Muzzin et al. 2010).  In addition to the global galaxy, we also analyse properties of a number of bright clumps of star formation, which are compared with those found at high-z. 

The paper is structured as follows. In Section 2, we briefly describe the sample and the IFS data used for the analysis. In Section 3, we provide the details of the methodology and data analysis.  The results are presented and discussed in Section 4, which has two clearly distinct parts: the dynamical status of the ISM in U/LIRGs  and the ionized gas outflows, which are treated respectively in 4.1 and 4.2.  The AGN effects are discussed in sections 4.1.4 and 4.2.2, while 4.1.5 and 4.2.7 are specifically focussed on the high-z comparison.   The clumps related properties are discussed in  4.1.6, 4.1.7, 4.2.8., and 4.2.9. Finally the main conclusions are summarized in Section 5.  Throughout the paper, we consider H$_{0}$ = 70 kms$^{-1}$Mpc$^{-1}$, $\Omega_{\rm \Lambda}$ = 0.7, $\Omega_{\rm M}$ = 0.3.

\section {The sample  and observations} 

The sample consists of 58 local LIRG and ULIRG systems for which we have obtained optical IFS data with Visible Multi-Object Spectrograph (VIMOS, LeF$\grave{e}$vre et al. 2003) and INTEGRAL (Arribas et al. 1998) instruments.  The mean (median) distance for the whole sample is 226 (137) Mpc, ranging from 40.4 Mpc (z= 0.0093) to 898 Mpc (z= 0.185). The infrared luminosity, $L_{IR}[8-1000\mu m]$, spreads over the range $10^{10.8} - 10^{12.6}L_{\odot}$. 
The sample also includes all types of nuclear activity and interaction phases, and therefore it is representative of the general properties of local U/LIRGs. About 30 percent of the objects have evidence of hosting an AGN as indicated by either the emission line ratios (see compilations of Rodr{\'{\i}}guez-Zaur{\'{\i}}n et al. 2011, hereafter RZ11, and Garc{\'{\i}}a-Mar{\'{\i}}n et al. 2009a, hereafter GM09a, and references there in) or the presence of (hints of) a very broad $H\alpha$ line (Table A1).  The AGN effects are discussed along the paper (mainly in Sections 4.1.4 and 4.2.2). 

The VIMOS sub-sample (38 U/LIRG systems) has been presented elsewhere (e.g.,  Arribas et al. 2008; Monreal-Ibero et al. 2010; RZ11) and is drawn from the IRAS Revised Bright Galaxy Sample (Sanders et al. 2003), the IRAS 1 Jy sample of ULIRGs (Kim et al. 1998), and the Hubble Space Telescope/ Wide-Field Planetary Camera 2 snapshot sample of bright ULIRGs (ID 6346 PI: K.Borne). These are mainly southern objects. We used the high resolution mode with the HR-orange grating, which provides a spectral resolution of about 1.8 \AA. After combining four dither pointings, the total field of view (FoV) is about 30 x 30 arcsec$^2$, with a spaxel scale of 0.67 arcsec (square). Details about the observations, data reduction, and calibration can be found in Monreal-Ibero et al. (2010), RZ11, and Bellocchi et al. (2013). 

The INTEGRAL sub-sample (20 systems) is that presented in GM09a, and it mainly covers the ULIRG luminosity range. It consists of northern objects selected  from the IR-bright samples of  Sanders et al. (1988), Melnick \& Mirabel (1990), Leech et al. (1994), Kim et al. (1995), Lawrence et al. (1999), and Clements et al. (1996). For the present analysis we exclude Infrared Astronomical Satellite (IRAS) sources F09427+1929 and F13469+5833, for which the H$\alpha$ data have low S/N. The data were taken with a 600 lines mm$^{-1}$ grating, which provides an effective spectral resolution (FWHM) of 6.0 \AA. The INTEGRAL observations were carried out with bundle SB2 (i.e., optical fibers of 0.9 arcsec in diameter and FoV of 12.3 x 16 arcsec$^2$), except for a few cases. Further details about these data can be found in GM09a. 

The sources in the whole U/LIRG  sample have been morphologically classified into three main types, following the simplified version of the scheme proposed by Veilleux et al. (2002). Briefly, the three morphological classes are defined as follows (see RZ11 for further details):
\begin{itemize}
\item Class 0: objects that appear to be single isolated objects, with relatively symmetric disc morphologies and without evidence for strong past or ongoing interaction (hereafter, we call these objects $disc$).
\item Class 1: objects in a pre-coalescence phase with two well differentiated nuclei separated by a projected distance of at least 1.5 kpc (hereafter, $interacting$). For these objects, it is still possible to identify (in some cases using HST imaging) the individual merging galaxies and their corresponding tidal structures due to the interaction.  We distinguish when the galaxies  can be individually analyzed with the IFS data (Class 1.1 ; generally in distant pairs), and when the whole system is globally studied (Class 1.0). In a few cases, the individual galaxies as well as the whole system are studied (F06035$-$7102, F23128$-$5919, F08572+3515, Mrk463, F18580+6527).    
\item Class 2: objects with two nuclei separated by a projected distance $\leq$ 1.5 kpc or a single nucleus with a relatively asymmetric morphology, suggesting a post-coalescence merging phase (hereafter, $merger$).
\end{itemize}
The sample contains 13 objects of class 0, 32 individuals and 14 systems of class 1, and 21 objects of class 2.

In terms of star formation rate (SFR), the sample covers the 5 to 372 M$_{\odot}$yr$^{-1}$ range, considering the Kennicutt (1998) relations for the infrared luminosity and a Chabrier (2003) IMF.  The star formation rates inferred from H$\alpha$, SFR(H$\alpha$), are obtained from the works by RZ11 and Garcia-Marin et al. (2009b; hereafter GM09b), which make use of the 2D H$\alpha$ flux distribution provided by the IFS data. In addition, GM09b correct the observed fluxes using the 2D reddening maps obtained from H$\alpha$/H$\beta$.  For those objects without  H$\alpha$/H$\beta$ values we correct  by reddening considering the median values derived by Veilleux et al. (1999).  We consider A$_{H\alpha}$ = 0.82 $\times$ A$_V$ according to Cardelli, Clayton, and Mathis (1989).  The SFR(H$\alpha$) for the sample expands from 0.5 to 180 M$_\odot$yr$^{-1}$, which indicates that the A$_v$-corrected H$\alpha$ SFRs do not recover the SFRs derived from the IR-luminosities (e.g., GM09b, RZ11, Piqueras-Lopez et al. 2013).   The main properties of the sample are summarized in Table 1.

We have also selected 26 bright extranuclear star forming clumps  in ten galaxies (Fig. 1; Table B1). The main selection criteria were to be clearly defined in the H$\alpha$ emission maps against the local background (RZ11), and to have a high surface brightness emission. The selection was not intended to be complete, as the main goal was to define a sample representative of the bright clumps in these sources. We have restricted the study of the clumps to VIMOS data only because the relatively low spectral (and spatial) resolution of the INTEGRAL observations make the analysis more  uncertain. From the ten galaxies hosting the selected clumps eight are LIRGs and two are ULIRGs. They cover the luminosity range $10^{10.9}L_{\odot} < L_{IR}[8-1000\mu m] < 10^{12.3}L_{\odot}$. Most of these systems are classified as interacting (6), while a lower fraction are isolated (3) or mergers (1).  When we apply reddenning corrections to the clumps, we assume a value of A$_V$=1.9 according to the median values obtained by GM09b.   \ 

\begin{table*}[]
\scriptsize
\caption {Properties of the sample\label{table:tabla1}}
%\centering
%\begin{tiny}
\vskip -0.3cm
\begin{tabular}{l c c c c c c c c c}
\hline \hline
Sample / ID   & z &  D (Mpc) & class & log(L$_{IR}$/L$\odot$) & r$_{1/2} ( kpc )$ & SFR(H$\alpha$) ( M$\odot$yr$^{-1}$)    & $\Sigma$(H$\alpha$)(M$\odot$yr$^{-1}$kpc$^{-2}$)       & Mdyn(10$^{10}$M$\odot$) & AGN \\               
        (1)           & (2) & (3) & (4) &      (5)                      &    (6)         &            (7)                   &               (8)                       & (9) & (10) \\               
\hline
\multicolumn{10}{l}{ VIMOS sample}\\
\hline

F01159$-$4443N   &0.02290 &  100.& 1.1 & 11.48$\pm$0.20& 0.3 $\pm$0.1&  17.7 $\pm$  4.8&  23.0 $\pm$ 13.3& 5.0 $\pm$ 4.4& y\\
F01341$-$3735N   &0.01731 &   75.& 1.1 & 10.99$\pm$0.20& 1.4 $\pm$0.4&   2.9 $\pm$  0.5&   0.3 $\pm$  0.1& 3.5 $\pm$ 2.3 & y\\
F01341$-$3735S   &0.01731 &   75.& 1.1 & 10.72$\pm$0.20& 0.2 $\pm$0.1&   0.9 $\pm$  0.2&   2.8 $\pm$  1.8& 6.1 $\pm$ 5.5 & \\
F04315$-$0840    &0.01594 &   69.& 2 & 11.69$\pm$0.06& 0.5 $\pm$0.1&  45.3 $\pm$ 12.4&  27.7 $\pm$ 15.7& 1.7 $\pm$ 0.7 & \\ 
F05189$-$2524    &0.04256 &  188.& 2 & 12.19$\pm$0.06& (0.3) &  10.2 $\pm$  5.1&  (20.7 $\pm$ 10.3) & ...  & y\\
F06035$-$7102    &0.07946 &  361.& 1.0 & 12.26$\pm$0.06& 3.7 $\pm$1.1&  50.2 $\pm$ 12.1&   0.6 $\pm$  0.3& 9.5 $\pm$ 8.1 & \\
F06035$-$7102N   &0.07946 &  361.& 1.1 &       ...     &      ...    &  16.4 $\pm$  3.9&        ...      & ... &\\
F06035$-$7102S   &0.07946 &  361.& 1.1 &       ...     &      ...    &  33.8 $\pm$  8.2&        ...      & ... & y\\
F06076$-$2139N   &0.03745 &  165.& 1.1 & 11.67$\pm$0.20& (0.3) &   1.4 $\pm$  0.7&   (3.5 $\pm$  1.7) & 3.3 $\pm$ 1.2 &\\
F06076$-$2139S   &0.03745 &  165.& 1.1 &       ...     & 0.5 $\pm$0.2&   1.6 $\pm$  0.8&   1.1 $\pm$  0.9& 3.5 $\pm$ 1.5 &\\
F06206$-$6315    &0.09244 &  423.& 1.0 & 12.27$\pm$0.06& 2.5 $\pm$0.8&  54.0 $\pm$ 15.1&   1.3 $\pm$  0.8& 4.8 $\pm$ 1.5 & y \\
F06259$-$4708N   &0.03879 &  171.& 1.1 & 11.91$\pm$0.20& 0.7 $\pm$0.2&   4.7 $\pm$  1.0&   1.5 $\pm$  0.8& 2.6 $\pm$ 1.7 &\\
F06259$-$4708C   &0.03879 &  171.& 1.1 &       ...     & 1.2 $\pm$0.3&  23.4 $\pm$  6.1&   2.8 $\pm$  1.6& 1.1 $\pm$ 0.5 &\\
F06259$-$4708S   &0.03879 &  171.& 1.1 &       ...     & 1.4 $\pm$0.4&   3.8 $\pm$  1.9&   0.3 $\pm$  0.2& 7.9 $\pm$ 2.9 &\\
F06295$-$1735    &0.02130 &   93.& 0 & 11.27$\pm$0.08& 1.9 $\pm$0.6&   2.5 $\pm$  0.4&   0.1 $\pm$  0.1& 1.9 $\pm$ 0.6 &\\
F06592$-$6313    &0.02296 &  100.& 0 & 11.22$\pm$0.06& 0.3 $\pm$0.2&   3.6 $\pm$  1.0&   4.7 $\pm$  3.1& 2.0 $\pm$ 0.8 &\\
F07027$-$6011N   &0.03132 &  137.& 0 & 11.04$\pm$0.20& 0.5 $\pm$0.2&  14.0 $\pm$  3.6&  10.1 $\pm$  6.8& ... & y\\
F07027$-$6011S   &0.03132 &  137.& 0 & 11.51$\pm$0.20& 0.6 $\pm$0.2&  19.3 $\pm$  9.6&   7.7 $\pm$  5.4& 2.4 $\pm$ 1.1 &\\
F07160$-$6215    &0.01081 &   47.& 0 & 11.16$\pm$0.06& 0.5 $\pm$0.2&   4.6 $\pm$  2.3&   2.5 $\pm$  1.8& 8.0 $\pm$ 1.4 &\\
08355$-$4944     &0.02590 &  113.& 2 & 11.60$\pm$0.11& 0.6 $\pm$0.2& 116.9 $\pm$ 58.5&  53.4 $\pm$ 37.8&  0.5 $\pm$ 0.4 &\\
08424$-$3130NE  &0.01616 &   70.& 1.1 &       ...     &      ...    &   1.2 $\pm$  0.6&        ...      & 4.6 $\pm$ 1.6 &\\
08424$-$3130SW  &0.01616 &   70.& 1.1 & 11.04$\pm$0.20&      ...    &   2.4 $\pm$  1.2&        ...      & 2.8 $\pm$ 2.3 &\\
F08520$-$6850    &0.04632 &  205.& 1.0 & 11.83$\pm$0.06& 1.0 $\pm$0.3&  30.0 $\pm$ 15.0&   4.6 $\pm$  3.2& 13.3 $\pm$ 3.0& \\
09022$-$3615     &0.05964 &  267.& 2 & 12.32$\pm$0.08& 1.3 $\pm$0.4& 141.6 $\pm$ 70.8&  13.3 $\pm$  9.4& 8.1 $\pm$ 2.5 &\\
F09437$+$0317N   &0.02047 &   89.& 1.1 & 10.99$\pm$0.20& 2.2 $\pm$0.7&  12.4 $\pm$  6.2&   0.4 $\pm$  0.3& 12.3 $\pm$ 2.8 & \\
F09437$+$0317S   &0.02047 &   89.& 1.1 & 10.82$\pm$0.20& 1.9 $\pm$0.6&   9.8 $\pm$  4.9&   0.4 $\pm$  0.3& 8.6 $\pm$ 1.7 & \\
F10015$-$0614    &0.01686 &   73.& 0 & 11.31$\pm$0.09& 1.9 $\pm$0.6&   4.1 $\pm$  0.7&   0.2 $\pm$  0.1& 7.7 $\pm$ 1.6 &\\
F10038$-$3338    &0.03410 &  150.& 2 & 11.77$\pm$0.09& 0.5 $\pm$0.3&   4.0 $\pm$  2.0&   2.4 $\pm$  1.9& 1.3 $\pm$ 0.7 &\\
F10257$-$4339    &0.00935 &   40.& 2 & 11.69$\pm$0.07& 1.0 $\pm$0.3&  22.5 $\pm$ 11.2&   3.5 $\pm$  2.5& 3.2 $\pm$ 1.7 &\\
F10409$-$4556    &0.02101 &   91.& 0 & 11.26$\pm$0.09& 1.8 $\pm$0.5&   2.4 $\pm$  0.4&   0.1 $\pm$  0.1& 5.3 $\pm$ 1.2 &\\
F10567$-$4310    &0.01720 &   75.& 0 & 11.07$\pm$0.09& 1.6 $\pm$0.5&   3.0 $\pm$  0.7&   0.2 $\pm$  0.1& 5.1 $\pm$ 1.7 &\\
F11255$-$4120    &0.01635 &   71.& 0 & 11.04$\pm$0.11& 1.3 $\pm$0.4&   1.2 $\pm$  0.2&   0.1 $\pm$  0.1& 2.9 $\pm$ 0.9 &\\
F11506$-$3851    &0.01078 &   47.& 0 & 11.30$\pm$0.09& 0.7 $\pm$0.2&   6.8 $\pm$  1.6&   2.4 $\pm$  1.3& 4.9 $\pm$ 1.2 &\\
F12043$-$3140N   &0.02320 &  101.& 1.1 &       ...     & 0.2 $\pm$0.1&   1.2 $\pm$  0.3&   3.4 $\pm$  2.2&  0.3 $\pm$ 0.2 &\\
F12043$-$3140S   &0.02320 &  101.& 1.1 & 11.37$\pm$0.20& 1.4 $\pm$0.4&   0.9 $\pm$  0.2&   0.1 $\pm$  0.0& 6.3 $\pm$ 1.6 &\\
F12115$-$4656    &0.01849 &   80.& 0 & 11.11$\pm$0.08 & 1.3 $\pm$0.4&   2.4 $\pm$  0.4&   0.2 $\pm$  0.1& 11.5 $\pm$ 3.2 &\\
12116$-$5615     &0.02710 &  118.& 2 & 11.61$\pm$0.06 & 0.3 $\pm$0.1&   7.7 $\pm$  3.8&  15.6 $\pm$ 12.1&  0.8 $\pm$ 0.4 &\\
F12596$-$1529    &0.01592 &   69.& 1.0 & 11.07$\pm$0.08&      ...    &   2.9 $\pm$  0.7&        ...      & 3.5 $\pm$ 1.4 &y \\
F13001$-$2339    &0.02171 &   95.& 2 & 11.48$\pm$0.08 & 0.9 $\pm$0.3&   1.2 $\pm$  0.3&   0.2 $\pm$  0.1& 11.4$\pm$ 8.3 &\\
F13229$-$2934    &0.01369 &   59.& 0 & 11.29$\pm$0.08 & 0.5 $\pm$0.2&   1.5 $\pm$  0.3&   0.9 $\pm$  0.5& 3.5 $\pm$ 1.6 & y \\
F14544$-$4255E   &0.01573 &   68.& 1.1 & 10.80$\pm$0.20& 0.9 $\pm$0.3&   0.5 $\pm$  0.2&   0.1 $\pm$  0.1& 10.4 $\pm$ 3.0 &\\
F14544$-$4255W   &0.01573 &   68.& 1.1 & 10.80$\pm$0.20& 0.6 $\pm$0.2&   1.1 $\pm$  0.2&   0.5 $\pm$  0.3& 2.4 $\pm$ 2.3 & y\\
F17138$-$1017    &0.01733 &   75.& 2 & 11.41$\pm$0.08 & 0.6 $\pm$0.2&   2.1 $\pm$  1.0&   1.0 $\pm$  0.7& 2.6 $\pm$ 0.7 &\\
F18093$-$5744N   &0.01734 &   75.& 1.1 & 11.47$\pm$0.20& 1.1 $\pm$0.3&   7.1 $\pm$  1.2&   0.9 $\pm$  0.5& 3.6 $\pm$ 1.1&\\
F18093$-$5744C   &0.01734 &   75.& 1.1 & 10.87$\pm$0.20& 0.3 $\pm$0.1&   5.8 $\pm$  1.5&   7.5 $\pm$  4.3&  0.4 $\pm$ 0.3 &\\
F18093$-$5744S   &0.01734 &   75.& 1.1 &       ...     & 0.8 $\pm$0.3&   2.7 $\pm$  1.3&   0.6 $\pm$  0.4& 3.4 $\pm$ 1.3 &\\
F21130$-$4446    &0.09255 &  424.& 2 & 12.09$\pm$0.08& 1.6 $\pm$0.5&  15.9 $\pm$  3.3&   1.0 $\pm$  0.5& 20.7 $\pm$ 16.8& \\
F21453$-$3511    &0.01615 &   70.& 2 & 11.41$\pm$0.06& 1.0 $\pm$0.3&   5.1 $\pm$  1.0&   0.8 $\pm$  0.4&  0.9 $\pm$ 0.6 & y\\
F22132$-$3705    &0.01141 &   49.& 0 & 11.22$\pm$0.06& 1.6 $\pm$0.5&   3.1 $\pm$  0.6&   0.2 $\pm$  0.1& 10.5 $\pm$ 1.8 &\\
F22491$-$1808    &0.07776 &  353.& 1.0 & 12.17$\pm$0.11& 1.7 $\pm$0.5&  11.9 $\pm$  3.0&   0.7 $\pm$  0.4& 3.4 $\pm$ 1.7 &\\
F23128$-$5919    &0.04460 &  198.& 1.0 & 12.06$\pm$0.08& 2.0 $\pm$0.6&  22.9 $\pm$  5.8&   0.9 $\pm$  0.5& 2.7 $\pm$ 1.2 &y \\
F23128$-$5919N   &0.04460 &  198.& 1.1 &       ...     &      ...    &  16.6 $\pm$  4.2&        ...      & ... & \\
F23128$-$5919S   &0.04460 &  198.& 1.1 &       ...     &      ...    &   6.3 $\pm$  1.6&        ...      & ... & y \\
\hline
\multicolumn{10}{l}{ INTEGRAL sample}\\
\hline
06268$+$3509     &0.16900 &  813.& 1.0 & 12.51$\pm$0.14& 3.7 $\pm$1.5&  56.2 $\pm$ 28.1&   0.6 $\pm$  0.5& 9.7 $\pm$ 3.6 &\\
06487$+$2208     &0.14400 &  682.& 2 & 12.57$\pm$0.12& 1.0 $\pm$0.5&  54.4 $\pm$ 27.2&   9.0 $\pm$  7.0& 6.2 $\pm$ 1.6 &\\
F08572$+$3515    &0.05800 &  259.& 1.0 & 12.17$\pm$0.11& 1.9 $\pm$0.7&   7.2 $\pm$  3.6&   0.3 $\pm$  0.2& 1.8 $\pm$ 0.5 &\\
F08572$+$3915N   &0.05800 &  259.& 1.1 & 12.17$\pm$0.20&      ...    &   6.5 $\pm$  3.3&        ...      &  0.96 $\pm$ 0.4 & \\
F08572$+$3915S   &0.05800 &  259.& 1.1 &       ...     &      ...    &   0.7 $\pm$  0.3&        ...      & 1.8 $\pm$ 0.7 &\\
F11087$+$5351    &0.14300 &  677.& 2 & 12.13$\pm$0.15& 2.5 $\pm$1.0&   5.1 $\pm$  2.6&   0.1 $\pm$  0.1& 30.4 $\pm$ 6.5 & y \\
Arp299E / IC694  &0.01000 &   43.& 1.1 & 11.58$\pm$0.20& 0.7 $\pm$0.3&   7.4 $\pm$  3.7&   2.6 $\pm$  1.9& 2.3 $\pm$ 1.6 & \\
Arp299W/ NGC3690  &0.01000 &   43.& 1.1 & 11.43$\pm$0.20& 0.5 $\pm$0.2&  13.5 $\pm$  6.8&   7.6 $\pm$  5.7& 1.7 $\pm$ 0.3 & y\\
F12112$+$0305    &0.07300 &  330.& 1.0 & 12.37$\pm$0.11& 2.6 $\pm$1.0&  99.0 $\pm$ 49.5&   2.3 $\pm$  1.7& 4.6 $\pm$ 2.4 &\\
F12490$-$1009    &0.10100 &  465.& 2 & 12.07$\pm$0.16& 1.8 $\pm$0.7&   5.6 $\pm$  2.8&   0.3 $\pm$  0.2& 8.7 $\pm$ 4.4 &\\
Mrk 231          &0.04200 &  186.& 2 & 12.57$\pm$0.06&      ...    &        ...      &        ...      & ... & y\\
F13156$+$0435N   &0.11300 &  525.& 1.1 & 11.73$\pm$0.20& 2.0 $\pm$0.8&   9.8 $\pm$  4.9&   0.4 $\pm$  0.3& 6.9  $\pm$ 3.4  &\\
F13156$+$0435S   &0.11300 &  525.& 1.1 & 11.91$\pm$0.20& 2.7 $\pm$1.1&   4.2 $\pm$  2.1&   0.1 $\pm$  0.1& 14.5  $\pm$ 6.0 & \\
F13342$+$3932    &0.17900 &  866.& 1.0 & 12.51$\pm$0.13&      ...    &        ...      &        ...      & ... & y\\
Mrk273           &0.03800 &  167.& 2 & 12.18$\pm$0.06& 1.3 $\pm$0.5&  16.6 $\pm$  8.3&   1.6 $\pm$  1.2& 19.0 $\pm$ 4.7 & y\\
Mrk463           &0.05000 &  222.& 1.0 & 11.81$\pm$0.12& (0.4) &  25.9 $\pm$ 12.9&  (21.3 $\pm$ 10.6) & 10.5 $\pm$ 2.6 & y \\
Mrk463E          &0.05000 &  222.& 1.1 &       ...     &      ...    &        ...      &        ...      & 7.9 $\pm$ 6.6 & y \\
Mrk463W          &0.05000 &  222.& 1.1 &       ...     &      ...    &        ...      &        ...      & 3.2 $\pm$ 0.3 &\\
F14060$+$2919    &0.11700 &  545.& 2 & 12.18$\pm$0.16& 1.3 $\pm$0.7&  20.9 $\pm$ 10.5&   2.0 $\pm$  1.5& 5.4 $\pm$ 1.9 &\\
F14348$-$1447    &0.08300 &  378.& 1.0 & 12.39$\pm$0.11& 3.8 $\pm$1.5& 177.1 $\pm$ 88.2&   1.9 $\pm$  1.4& 7.4 $\pm$ 1.9 &\\
F15206$+$3342    &0.12400 &  580.& 2 & 12.27$\pm$0.11& (1.0) & 149.4 $\pm$ 74.7&  (23.8 $\pm$ 11.9) & 2.6 $\pm$ 0.6 &\\
F15250$+$3609    &0.05500 &  245.& 2 & 12.09$\pm$0.06& (0.5) &  40.6 $\pm$ 20.6 &  (28.0 $\pm$ 14.2) & 3.7 $\pm$ 1.3 &\\
Arp220           &0.01800 &   78.& 2 & 12.20$\pm$0.06& 0.6 $\pm$0.3&   2.3 $\pm$  1.1&   0.9 $\pm$  0.7& 5.5 $\pm$ 1.1 & y\\
F16007$+$3743    &0.18500 &  808.& 1.0 & 12.11$\pm$0.24& 3.7 $\pm$1.5&  70.6 $\pm$ 35.3&   0.8 $\pm$  0.6& 31.6 $\pm$ 23.7 &\\
F17207$-$0014    &0.04300 &  190.& 2 & 12.43$\pm$0.07& 0.9 $\pm$0.4&  74.7 $\pm$ 37.6&  14.7 $\pm$ 11.0& 2.1 $\pm$ 0.6 &\\
F18580$+$6527    &0.17600 &  850.& 1.0 & 12.26$\pm$0.16& 2.4 $\pm$0.9&  35.9 $\pm$ 17.6&   1.0 $\pm$  0.8& 44.8 $\pm$ 34 & y \\
F18580$+$6527W   &0.17600 &  850.& 1.1 &       ...     &      ...    &        ...      &        ...      & 19.4 $\pm$ 1.5 & y\\
\hline
\end{tabular}
\vskip -0.3cm
\tablefoot{\scriptsize Columns: (1) Sub-sample and identification according to the IRAS code, or other common name. For those systems with enough spatial resolution to separate the individual galaxies they are identified with N, S, E, W, and C (i.e. centre) according to their relative location; (2) redshift according to NED; (3) Luminosity distance; (4) Morphological class, where 0, 1,2, distinguish isolated  discs, interacting, and mergers, respectively. For interacting systems 1.0 and 1.1 distinguish between the total system and individuals, respectively. In five cases (F06035$-$7102, F23128$-$5919, F08572+3515, Mrk463, F18580+6527) the individuals members, as well as the global system are considered; (5) Logarithmic of the infrared luminosity in solar units. For individual galaxies in multiple U/LIRGs, these are estimated dividing the total luminosity taking into account the relative fluxes in the MIPS images. The uncertainty has been estimated as three times the typical error in the IRAS fluxes. For individual objects in multiple systems, an uncertainty of 0.2 dex has been considered; (6) H$\alpha$ half-light radius using the A/2 method (Arribas et al.  2012). Values in brackets correspond to upper limits. Lower limits in the radii values due to limitations in the FoV are not distinguished, as the effect is estimated to be within the quoted uncertainties and the FoV effect is (partially) compensated in the $\Sigma$(H$\alpha$) determination; (7) SFRs obtained from the reddening corrected H$\alpha$ luminosity from RZ11 and GM09, transformed for a Chabrier (2003) IMF. For Arp220 and Mrk273, the SFRs were obtained from the works by Veilleux et al. (1999) and Colina et al. (2004), respectively.; (8) H$\alpha$ star formation rate density; (9) Dynamical masses for the VIMOS sub-sample are from Bellocchi et al. (2013). For the INTEGRAL data they are derived following the same methodology; (10) "y" indicates evidence of an AGN from the optical spectra, as indicated by either the emission line ratios or the presence of a very broad H$\alpha$ (see also notes in table A1). For IRAS F06035-7102, the spectra show evidence of a (weak) AGN in the southern component, but not in the whole system.}

\end{table*}

\section {Data analysis}

\subsection {Integrated and clumps spectra sets }

From our integral field data cubes  we create a spatially integrated spectrum for each source using the {\it pingsoft} tool (Rosales-Ortega, 2011). Before combining the individual spectra they are shifted to the rest-frame wavelength according to the measured H$\alpha$  (narrow) velocity. In this manner the integrated spectra are not affected by the broadening due to the large-scale velocity field, and have increased S/N. We reject individual spectra with low S/N, following the procedure for an optimal extraction of IFS spectra proposed by Rosales-Ortega et al (2012). As a consequence the final spectra have in general very high S/N (i.e.,$>$100),  which  facilitates the line fitting (see Appendix A1). We have also used our IFS data cubes in a similar manner to extract the spectra of the 26 bright extranuclear clumps of SF (see Appendix B1).

\subsection  {Line fitting}

We fit the integrated spectra for the galaxies in three spectral regions (i.e., H$\alpha$-[NII]$\lambda\lambda6549,6583$, [SII]$\lambda\lambda6717,6731$, and [OI]$\lambda6300$),  fixing the relative wavelengths and intensities of the lines according to their atomic parameters, and assuming equal broadening for the lines of the same kinematic component. Although initially we perform the fits independently in these spectral regions, for the [SII] and [OI] lines we fix in some cases the kinematic parameters (i.e., relative velocity shift between components, line width) according to the H$\alpha$-[NII] fits. This is done to better constrain the fit and/or to guarantee that the derived fluxes for the different lines correspond to kinematically consistent gas components, and therefore their flux ratios are physically meaningful. In some cases,  the [SII] and [OI] fits do not converge to a solution coherent with the H$\alpha$-[NII] fit, and they were rejected. 

After detailed experimentation, we find that a 2-Gaussian component model per line led to a remarkably good fit  of the integrated spectra in the H$\alpha$-[NII] region significantly reducing the residuals with respect to 1-Gaussian fits. We classified these two components according to their line width as {\it narrow} (N) and {\it broad} (B).  In some cases the fit  improves (slightly) by adding an extra very broad (i.e., FWHM $>$  25 \AA) H$\alpha$ line, which is likely associated with an AGN. However, despite the very high S/N, in general, the properties of this extra line are not well constrained by the fit. In any case, the two main components (i.e., N, B) are not significantly affected by the inclusion (or not) of this extra very broad line in the fit. Therefore, we performed the fits using a two-component Gaussian model in all the cases. For the INTEGRAL data, because of the lower spectral resolution, the fits were more uncertain. However, with these data, the 2-component model also produces good fits in most of the cases. Global results are presented in Table 2, while the individual fits and results are presented in the Appendix A1.

We note that the 1D (integrated) approach followed here allows us to obtain average properties for a statistically significant sample, and compare them directly with high-z studies that follow a similar approach. Therefore, detailed properties of individual systems are not considered here. This is complementary to other 2D analyses in U/LIRGs (e.g., Westmoquette et al. 2012; Bellocchi et al. 2013). For instance, using a detailed spaxel-based analysis Westmoquette et al. (2012) find more than two components in some spectra, albeit generally weak. By combining the individual spectra within the data cube we are smearing (weak) details present in individual spectra, but we are increasing the S/N of the resulting spectrum that represents the mean global behaviour. For instance, it is interesting to note that while for IRAS 17207-0014 Westmoquette et al. (2012) do not find traces of an outflow, our integrated high S/N spectrum indicates the presence of a relatively weak  broad component with an associated velocity (V$_{max}$) of - 306 $\pm$ 87 kms$^{-1}$, in good agreement with the finding for the molecular gas outflow  (i.e., - 370 kms$^{-1}$; Sturm et al. 2011).  It is also relevant to note that 2-Gaussian fits in a spaxel-by-spaxel approach are limited to the high surface brightness regions with enough S/N, which in general translates into an underestimation in of the role of the secondary component (flux, extension). This is particularly true in studies like the one by Bellocchi et al (2013) for which the second component is added only if clearly required. A more detailed one-by-one multi-component approach is in general unpractical when dealing with a very large number of spectra ($\sim$ 80000 in their case).

The spectra of the clumps were fitted in the H$\alpha$-[NII] spectral region using the same procedure as the one described above for the integrated spectra. The emission line profiles could be well fitted with a 2-Gaussian model per line for most of the selected clumps. Specifically only four out of 26 do not require 2-Gaussians , i.e., a single component per line gives a good fit already, suggesting that if a secondary component exists in these clumps, it is very weak.  The individual fits and results are in Appendix B.

 \begin{figure*}[]
 \centering  
 \includegraphics[width=0.82\textwidth]{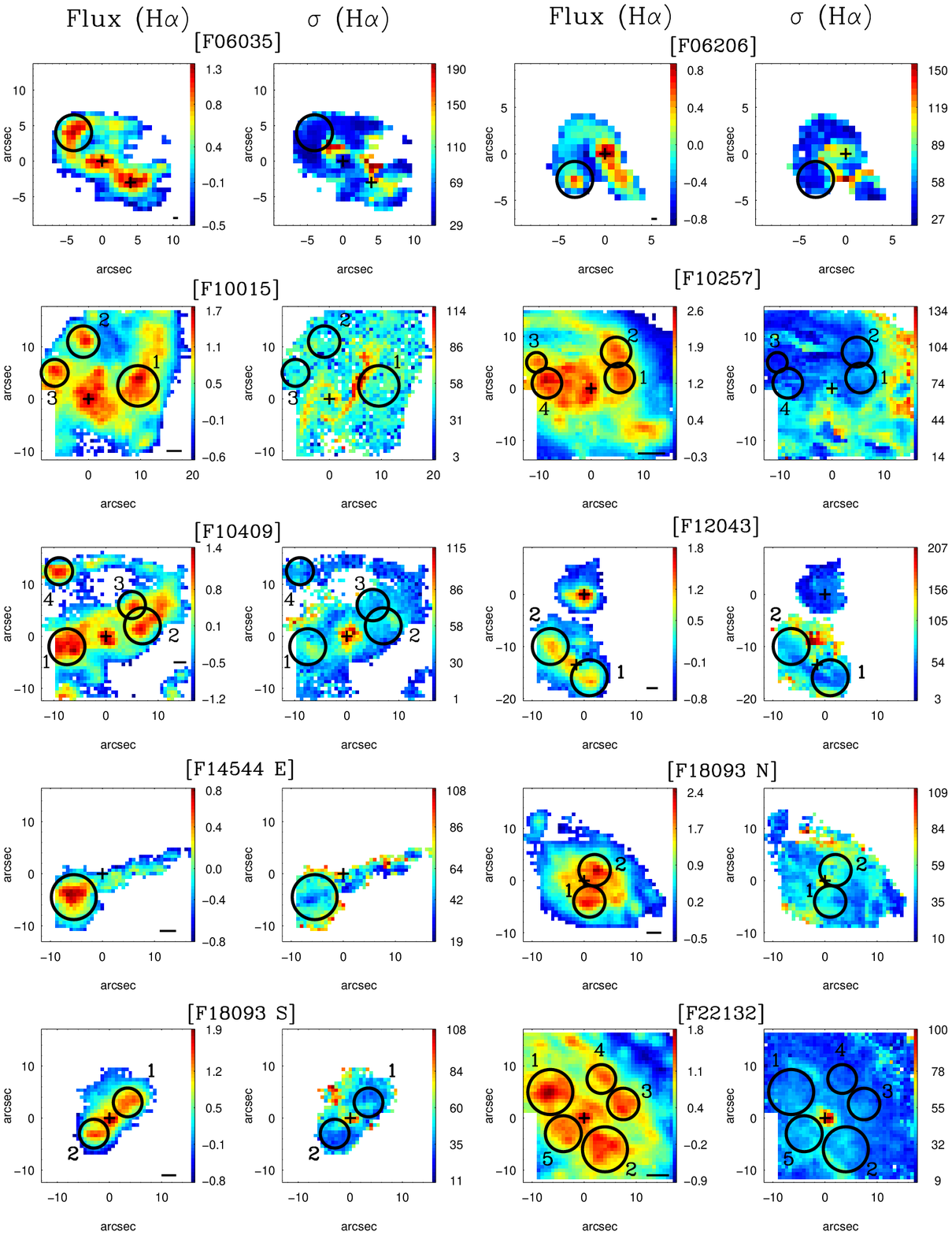}
 \caption{ H$\alpha$ emission and velocity dispersion maps of the narrow component for the sub-sample of galaxies for which extranuclear clumps of star formation are selected (see Bellocchi et al. 2013) . The circles identify the SF clumps in the H$\alpha$ flux and $\sigma$ maps. The cross indicates the position of the optical nucleus. In the case of multiple galaxy systems the secondary nucleus is also indicated. The H$\alpha$ flux emission per spaxel is such that F(ergs$^{-1}$cm$^{-2}$) = 10$^{-16+c}$, where {\it c} is indicated in the coloured bar. The units for the velocity dispersions are  kms$^{-1}$.  North is at the top, and east at the left. The horizontal black bar indicates 1 kpc. }
\
 \label{Fig1}
 \end{figure*}

\subsection{Main kinematic properties from the integrated spectra fits}

The intrinsic (i.e., deconvolved with the instrumental profile) full width half maximum (FWHM )of the narrow component,  as obtained from the  H$\alpha$-[NII] fits, ranges from 79 to 423 kms$^{-1}$, with a median value of 134 kms$^{-1}$ (see Fig. 2 and Table 2, where the line width is presented in terms of $\sigma$). This component traces the ambient ionized gas in the ISM and its main properties will be discussed in Sec. 4.1. The broad component is significantly wider, reaching values of FWHM of up to $\sim$ 1274 kms$^{-1}$, and it is in general blueshifted with respect to the narrow component ($\Delta$V(mean): -66 $\pm$ 11 km$^{-1}$, $\Delta$V(median): -47 km$^{-1}$).  This broad component is commonly interpreted as outflow in a dusty environment (e.g. Heckman et al 1990; Lehnert et al. 1996). In most cases, the broad component is weaker than the narrow (median flux(broad)/flux(narrow)= 0.6), though for some sources it can dominate by factors of up to 3 in flux. Further analysis of this component is done in Sec. 4.2.

 \begin{figure}[h]
 \centering  
 %generated with fwhm_all.sm
 % los ficheros genera con comp_comp.f
 \includegraphics[width=8.5cm]{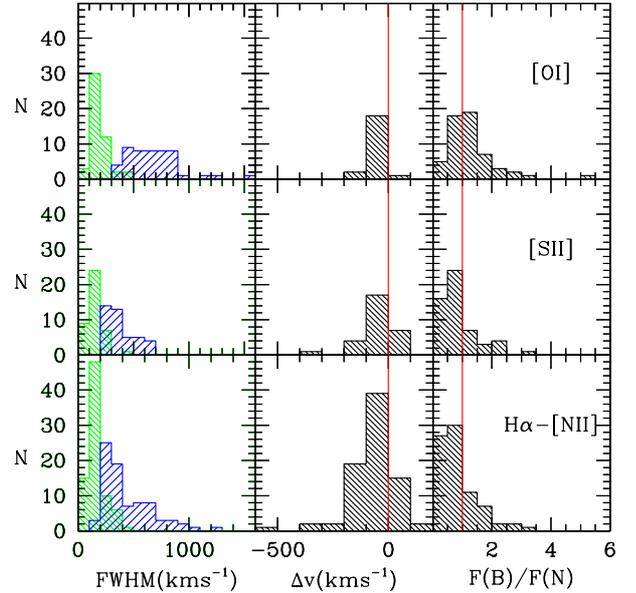}
 \caption{Histograms of the line widths (FWHM), the relative velocity shift ($\Delta$v), and the relative line flux between the broad (blue) and narrow (green) components, as inferred from the H$\alpha$-[NII], [SII], and [OI] fits.  Vertical red lines indicate the zero velocity shift, and the equal intensity level between the two components. The relative fewer values for [SII] and [OI] than for H$\alpha$-[NII] is because of the lower number of acceptable fits, and to the fact that in some cases the corresponding parameter was not independent (i.e., it was fixed according to the H$\alpha$-[NII] fit). }
\
 \label{Fig2}
 \end{figure}

In general, the results for the [SII] fits agree well with those from H$\alpha$-[NII]. For [OI] the fits tend to yield, on average, a wider and stronger broad component than for H$\alpha$-[NII] (Fig.2). Similar results are found by Soto et al. (2012b) who report that the broad component in [OI] is stronger than for H$\beta$. 
Except when explicitly mentioned, throughout the paper we will consider the kinematic properties (i.e., FWHM, $\Delta$v, and F[B]/F[N]) derived from the H$\alpha$-[NII] fits, as they could be obtained, for the whole sample, have higher accuracy than those from the [SII] and the [OI] fits, and allow us to make a direct comparison with other studies.

 \subsection{Kinematic tracers: Integrated versus mean velocity dispersions}
 
In general, the line widths inferred from the integrated spectra are in good agreement with the mean values derived from the two-dimensional maps presented by Bellocchi et al (2013), which were obtained through the fitting of the line profiles on the individual spaxels with adequate S/N. This is shown in Fig. 3 where the comparison is done in terms of the FWHM for the VIMOS sample. The mean ratio FWHM (int) / FWHM (mean) is 0.98 $\pm$ 0.24 without a clear trend with line width.  The systemic (narrow) component for a regular spiral typically has the broadest lines in the nuclear region, while in the disc the lines are relatively narrow and, therefore, one expects FWHM (int) / FWHM (mean) $>$ 1. However, many galaxies in the present sample show an asymmetric velocity dispersion map, with relatively large off-nuclear  values (see Bellocchi et al. 2013). For particular cases, FWHM (int) may be significantly lower than FWHM (mean). For instance, IRAS 09022-3615 shows line profiles considerably narrower in the centre than in the outer extranuclear regions (Bellocchi et al. 2013), with FWHM (int) / FWHM (mean) = 0.64. On the other hand, some objects behave as expected for relaxed regular discs. 
 
 \begin{figure}[h]
 \centering  
 %generated with vint_vmean.sm
 % los ficheros genera xc.f
 \includegraphics[width=8.0cm]{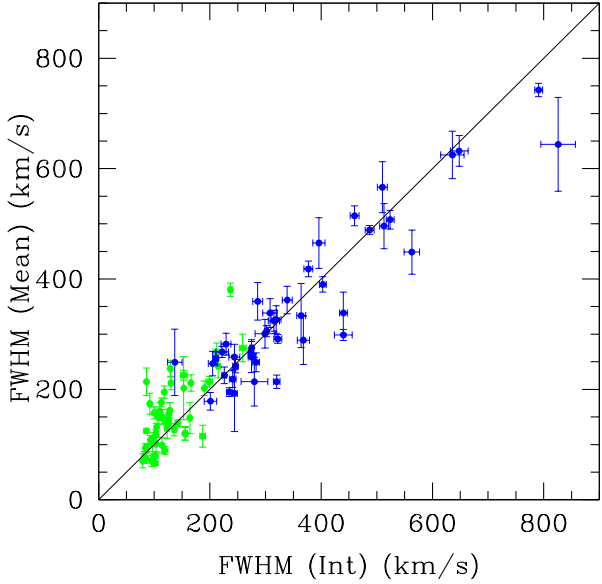}
 \caption{Line width from the integrated spectra versus the mean values obtained from the two-dimensional maps presented by Bellocchi et al. (2013) for the VIMOS sub-sample, for the narrow (green) and broad (blue) components. The line widths are represented in terms of FWHM for the narrow component also, which generally is expressed in $\sigma$ units. The 1:1 line is shown as reference.}
\
 \label{Fig3}
 \end{figure}

\vskip -2cm
\subsection{Excitation and kinematics relationship} 

The [NII]/H$\alpha$,  [SII]]/H$\alpha$, and [OI]/H$\alpha$ line flux ratios indicate that most of the objects are photoionized by stars, though many are about the SB / LINER-AGNs border (Fig. 4, Table 3). These ratios do not show a clear difference between the excitation conditions of the narrow and the broad component although, on average, the latter tends to have higher excitation. 

For the narrow component the data show positive correlations between these ratios and the line width (Fig. 4), indicating a higher excitation for the ionized gas with larger FWHM (i.e., correlation coefficients, r, for the [NII]/H$\alpha$$-$, [SII]/H$\alpha$$-$, and [OI]/H$\alpha$$-$FWHM relations are 0.41, 0.63, and 0.74, respectively). These correlations have been considered as evidence of shocks by several authors (Armus et al. 1989; Dopita \& Sutherland 1995; Monreal-Ibero et al. 2006, 2010). The broad component is affected by errors larger than the narrow, and the trends are less evident. For [SII]/H$\alpha$ there is no correlation (r= -0.01), and it is weak for the other line ratios  (r=0.18 for [OI]/H$\alpha$, and r=0.46 for [NII]/H$\alpha$).  A clear correlation with the broad to narrow component velocity off-set is not found either.

\subsection{Electron density from the narrow and broad components} 

A measurement of the electron density of the ionized gas (Ne) has been obtained using the [SII] $\lambda\lambda$ 6717, 6731\AA~doublet (hereafter S1=[SII] $\lambda$ 6717\AA, and S2=[SII] $\lambda$ 6731\AA). The mean value for the electron density of the narrow component corresponds to 296$\pm$26 cm$^{-3}$ and a somewhat larger value of 459$\pm$66 cm$^{-3}$ for the broad component. These mean density values have been derived after combining a large number ($\sim$ 40) of  high S/N integrated spectra and, therefore, the errors are relatively small.   
    
The individual values of S1/S2 for the narrow component monotonically decrease with FWHM (Fig.4),  indicating that the ionized gas with the larger velocity dispersion is also more dense.  The broad component is consistent with this behaviour, but the much larger uncertainties in the individual measurements do not allow us to further constrain this trend.  
Finally, for the sub-sample of U/LIRGs with identified AGNs and available Ne determinations, the mean densities for the narrow and broad components are 244$\pm$31 and 600$\pm$69 cm$^{-3}$, respectively. Thus, the average electron density of the ionized gas in our sample of U/LIRGs appears to be independent of the level of AGN activity (though see Villar-Martin et al. 2014).

\begin{figure}[h]
 %generated with fw_ion5.sm
 % los ficheros genera con ion.f   
 \includegraphics[width=7.5cm]{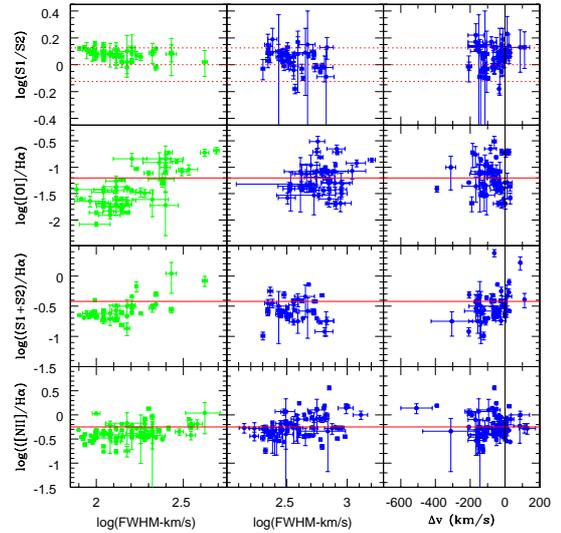}
 \caption{Line ratios as a function of the line width of the narrow  and broad components (left and central panels, respectively), and their relative velocity shift (right panels). S1=[SII] $\lambda$ 6717\AA, and S2=[SII] $\lambda$ 6731\AA. Line widths are derived from the fits of the lines at the numerator. % (i.e. from the H$\alpha$-[NII] fits for [NII]/H$\alpha$; from [SII] fits for (S1+S2)/H$\alpha$ and S1/S2,  and from the [OI] fits for [OI]/H$\alpha$). 
The solid horizontal red lines indicate the values for the border distinguishing between ionization by stars and other mechanisms (i.e., LINER, AGN)  assuming a mean value of [OIII]/H$\beta$=-0.2 (GM09a,  Monreal-Ibero et al. 2010), according to Veilleux and Osterbrock (1987). The dotted red lines in the log(S1/S2) plots correspond to electron densities of 100 (top), 600 (middle), and 1000 (bottom) cm$^{-3}$, assuming T=10$^4$K (Osterbrock, 1989). }
\
 \label{Fig5}
 \end{figure}

%\newpage
%\newpage

\begin{landscape}
\begin{table}
%\begin{sidewaystable}
\scriptsize
\caption{Properties of the narrow and broad components for different galaxy groups$^1$}\label{table:tabla3}
%\centering
%\begin{tiny}
\begin{tabular}{l c c c c c c c c c c c c c c}
\hline \hline
                                    & n$^2$                       & $<log(L_{IR}/L\odot)>$& SFR(H$\alpha$) & $\Sigma$(H$\alpha$) & \multicolumn{2}{c}{$\sigma$(Narrow)} & \multicolumn{2}{c}{FWHM(Broad)} &  \multicolumn{2}{c}{$\Delta(V_B-V_N)$} &  \multicolumn{2}{c}{Flux(B)/Flux(N) }& \multicolumn{2}{c}{V$_{max}$$^3$}  \\ 
                                    
                                     &                                  &                                   &M$\odot$yr$^{-1}$&M$\odot$yr$^{-1}$kpc$^{-2}$ & \multicolumn{2}{c}{kms$^{-1}$} & \multicolumn{2}{c}{kms$^{-1}$} &  \multicolumn{2}{c}{kms$^{-1}$} &  \multicolumn{2}{c}{}& \multicolumn{2}{c}{kms$^{-1}$}  \\                                    
%                            &           &                                 & \multicolumn{3}{c}{(kms$^{-1})$}        & \multicolumn{3}{c}{(kms$^{-1})$}   &  \multicolumn{3}{c}{(kms$^{-1})$}           &  \multicolumn{3}{c}{}                          & \multicolumn{3}{c}{(kms$^{-1})$} \\   
 \hline                                                                        
                            		&               		&         Mean              &    Mean                 & Mean     &Range   & Mean (Median) & Range &   Mean (Median) & Range  & Mean (Median) &  Range  & Mean (Median)& Range      & Mean (Median) \\ 
%  \cline{2-3}\cline{4-6} \cline{8-10}\cline{12-14}\cline{16-18} \cline{20-22}         
All                        	& 80  		& 11.68$\pm$0.06 &  21$\pm$4 &  5$\pm$1	 & 34-180   & 69 $\pm$ 3  (57) & 137- 1274          & 436 $\pm$ 25 (366) & -509,+128   &  -66  $\pm$ 11 (-47)    &  0.13- 3.14      &  0.84$\pm$ 0.07 (0.61)  & 43 - 999       & 285 $\pm$ 19 (260)    \\ 
no-AGN             		& 58 	  	    	& 11.66$\pm$0.07 & 25$\pm$5 &  5$\pm$1 	 & 34-180  & 63 $\pm$ 4  (54) & 137- 979         & 365 $\pm$  22 (310) & -509,+128    &   -63   $\pm$ 13 (-39)  &  0.13- 2.74     & 0.76$\pm$ 0.07  (0.56) & 43 - 526          & 224 $\pm$ 13 (215)  \\
AGN                  		& 22 		  	& 11.74$\pm$0.13 & 15$\pm$3 & 6$\pm$2	 & 42-149  & 86 $\pm$ 7  ( 83)  & 275- 1274     & 666 $\pm$ 53  (636) & -392,+88     &    -74   $\pm$ 21 (-50)  &  0.31- 3.14     &  1.03$\pm$ 0.16 (0.68) & 164 - 999        & 445 $\pm$ 42 (403)  \\
ULIRG             	 	& 26           	& 12.25$\pm$0.03& 47$\pm$10 & 6$\pm$2	  & 39-180   & 87 $\pm$ 8  (78)  & 265- 1274      & 600 $\pm$ 51 (544) & -509,+128    &  -97  $\pm$ 27  (-63) &  0.16- 3.14        &  0.85$\pm$ 0.13  (0.63)  & 115 - 999      & 393 $\pm$ 38 (330)   \\ 
ULIRG (no-AGN)	& 16        		& 12.26$\pm$0.04 & 61$\pm$13 & 7$\pm$2	& 39-148   & 74 $\pm$ 6  (72)  & 265- 749        & 452 $\pm$ 37  (464) & -191,+128     & -67  $\pm$ 20   (-57) & 0.16- 1.53       & 0.72 $\pm$ 0.11 (0.60)  & 116-526          & 293 $\pm$ 25 (295)   \\
ULIRG (AGNs)     	& 10        		& 12.24$\pm$0.05& 19$\pm$6 & 4$\pm$3 	& 43-180   & 109 $\pm$ 14  (98) & 396- 1274      & 814 $\pm$ 74 (766) & -509,+88    & -145 $\pm$ 59 (-90) & 0.31- 3.14         & 1.04$\pm$ 0.27  (0.71)  & 249-999         & 553 $\pm$ 64 (540)    \\ 
LIRG                 		& 41           	& 11.32$\pm$0.05& 11$\pm$3 & 5$\pm$2 	& 34-129   & 57  $\pm$ 3    (52) & 137- 826        & 350 $\pm$ 24 (315) & -208,+112    &  -42  $\pm$ 10   (-31)  &  0.13- 2.74      &  0.84$\pm$ 0.11 (0.58) &  43 - 604          & 219 $\pm$ 18 (190)  \\
LIRG (no-AGN)		& 32            	& 11.34$\pm$0.06& 11$\pm$4 & 5$\pm$2 	& 34-110   & 54 $\pm$ 3   (50) & 137- 513        & 299 $\pm$ 17  (285) & -159,+112    &  -34  $\pm$ 6     (-28)  & 0.13-2.74        & 0.80$\pm$ 0.12 (0.51)  & 43-353           & 186 $\pm$ 14 (166)    \\
LIRG (AGN)       	 & 9             	& 11.19$\pm$0.10&  9$\pm$3 & 8$\pm$3 	& 42- 129  & 66 $\pm$ 9   (60)  & 275- 826        & 530 $\pm$ 60  (510) & -208, -10      & -72 $\pm$ 19     (-50) & 0.53-2.57         & 1.00$\pm$ 0.22 (0.66) & 164-604         & 337 $\pm$ 46  (305)   \\  
Class 0                	& 13           	& 11.21$\pm$0.04& 5$\pm$1 & 2$\pm$1	 	& 34-80     & 47 $\pm$ 3    (43)  & 201- 791       & 348 $\pm$ 45  (315)  & -208,+8        &   -38  $\pm$ 16  (-26)  &  0.24- 2.57       & 0.79$\pm$ 0.20 (0.46) & 94 - 604            & 212 $\pm$ 37 (162)  \\
Class 0 (no-AGN) 	& 11          		& 11.22$\pm$0.04& 5$\pm$1 & 2$\pm$1	 	& 34-80     & 45 $\pm$ 4  (42) & 201- 460       & 289 $\pm$ 23  (246)  & -98,+8            &   -22  $\pm$ 8  (-20)  &  0.24- 2.17       & 0.64$\pm$ 0.17  (0.42) & 94-328            & 166 $\pm$ 19  (148)    \\
Class 0 (AGN) 		& 2                 	& 11.16$\pm$0.09& 8$\pm$4 & 6$\pm$3	 	& 51-60     & 55 $\pm$ 3    (55) & 563-791       & 677 $\pm$ 81  (677)  & -208,-49        &   -128  $\pm$ 56 (-128)  &  0.69- 2.57    & 1.63$\pm$ 0.66 (1.63)& 331-604            & 467 $\pm$ 97 (467)  \\
Class 1 (all)          	& 46		 	& 11.67$\pm$0.10& 21$\pm$5 & 3$\pm$1 	& 36-162   & 68 $\pm$ 4  (64) & 137- 960       & 410 $\pm$ 31  (336)  & -312,+128      &  -62  $\pm$ 13  (-50) &   0.13-2.74       & 0.80$\pm$ 0.09 (0.61)& 43 - 657           & 269 $\pm$ 22  (248)   \\
Class 1 (no-AGN) 	& 32 		 	& 11.67$\pm$0.12& 21$\pm$7 & 2$\pm$0.3 	& 36-90     & 59 $\pm$ 3  (54) & 137- 749       & 326 $\pm$ 26 (286)  & -207,+128      &  -54  $\pm$ 15  (-42) &   0.13-2.74       & 0.77$\pm$ 0.11 (0.57)& 43 - 526           & 220 $\pm$ 19 (206)  \\
Class 1 (AGN)      	& 14 			& 11.70$\pm$0.17& 20$\pm$5 & 7$\pm$3 	& 42-162   & 89 $\pm$ 9  (83) & 275- 960       & 601 $\pm$ 56 (606)  & -312,+11        &  -79  $\pm$ 22    (-51) &   0.31-1.85     & 0.86$\pm$ 0.13 (0.73)& 164 - 657           & 379 $\pm$ 43 (312)  \\
Class 1.1         		& 32 		  	& 11.33$\pm$0.11& 8$\pm$1 & 3$\pm$1	 	& 36-149   & 66 $\pm$ 4   (62) & 137- 960       & 358 $\pm$ 34 (292) &  -207,+112       &  -60 $\pm$ 13  (-47) &  0.13-2.74       & 0.79$\pm$ 0.11 (0.58)& 43 - 614           & 242 $\pm$ 23 (231)   \\
Class 1.1(noAGN)  	& 24 		   	& 11.36$\pm$0.13& 6$\pm$1 & 2$\pm$0.5	& 36-90      & 59 $\pm$ 4  (54) & 137- 618      & 285 $\pm$ 21  (264) &  -207,+112       &  -58 $\pm$ 17  (-48) &  0.13-2.74      & 0.76$\pm$ 0.13  (0.53)& 43 - 439           & 204 $\pm$ 20 (187)  \\
Class 1.1(AGN) 	 & 8            	& 11.35$\pm$0.20& 13$\pm$5 & 8$\pm$5	& 42-149    & 86 $\pm$ 11 (82) & 275- 960       & 577 $\pm$ 82 (551) &  -139,+5            &  -64 $\pm$ 18  (-47) &  0.42-1.85      & 0.88$\pm$ 0.16  (0.64)& 164 - 614           & 353 $\pm$ 54 (306)  \\
Class 1.0       		& 14                    & 12.13$\pm$0.10& 50$\pm$12 & 3$\pm$2 	& 39-162    & 74 $\pm$ 9    (69)  & 265- 826      & 528 $\pm$ 53  (507) & -312,+128      &  -66 $\pm$ 28  (-50) &  0.16-2.22        & 0.82$\pm$ 0.15 (0.81)&  115 - 657          & 330 $\pm$ 43 (283)  \\
Class 1.0(noAGN) 	& 8             	& 12.23$\pm$0.07& 63$\pm$18 & 1.5$\pm$0.5 	& 39-79      & 60 $\pm$ 5   (61) & 265- 749       & 449 $\pm$ 63  (386) & -191,+128      &  -42 $\pm$ 35  (-37) &  0.16-2.22        & 0.81$\pm$ 0.61 (0.77)&  116 - 526          & 267 $\pm$ 45 (261)  \\
Class 1.0(AGN) 	& 5                	& 12.00$\pm$0.19& 28$\pm$8 & 6$\pm$4 	& 46-162    & 94 $\pm$ 17 (88) & 396- 826      & 633 $\pm$ 72  (674) & -312,+11        &  -98 $\pm$ 44  (-79) &  0.31-1.76        & 0.84$\pm$ 0.20 (0.81)&  212 - 657          & 414 $\pm$ 67 (419)  \\
Class 2                  	& 21          		& 12.00$\pm$0.08& 37$\pm$10 & 11$\pm$3 	& 43-180    & 83 $\pm$ 9   (70) & 235- 1274    & 548 $\pm$ 56  (513) & -509,+88         &  -91 $\pm$  28  (-61) &  0.24-3.14       & 0.95$\pm$ 0.14 (0.66)& 128 - 999         & 365 $\pm$ 43 (331)    \\
Class 2 (no-AGN)  	& 15         		& 11.95$\pm$0.09& 47$\pm$13 & 13$\pm$4 	& 43-148    & 73 $\pm$ 7   (66) & 235- 601       & 417 $\pm$ 28 (440) & -170,+9           &  -66 $\pm$  12   (-60) &  0.24-1.53      & 0.83$\pm$ 0.10 (0.79)& 128 - 434         & 275 $\pm$ 21 (283)    \\
Class 2 (AGN)  		& 6                 	& 12.11$\pm$0.14& 8$\pm$2   & 5$\pm$4 	& 43-180    & 108 $\pm$ 21 (114) & 636- 1274    & 875 $\pm$ 95 (842) & -509,+88        &  -154 $\pm$89  (-68) &  0.54-3.14     & 1.23$\pm$ 0.40 (0.61)& 391 - 999         & 592 $\pm$ 86 (515)   \\
                      
\hline
\hline

\end{tabular}

\tablefoot{ $^1$ Objects without infrared luminosity determinations are excluded for the LIRG and ULIRG groups. Objects classified as 1 have been divided in two groups: class 1.1 refers to those galaxies that could be treated individually, while class 1.0 are those treated globally. If a galaxy class 1.1 has $L_{IR} > 10^{10.8}L_{\odot}$ is considered (within 1-sigma estimated uncertainty) to be a LIRG. Those 1.1 class galaxies that form part of a multiple U/LIRGs but have unknown $L_{IR}$ (e.g., without MIPS data) are kept for the analysis but excluded from the LIRG and ULRG specific subsamples. In five cases (F06035$-$7102, F23128$-$5919, F08572+3515, Mrk463, F18580+6527) individuals as well as the global are considered. Class 1 (all) refers to all cases belonging to this group (i.e., individuals and systems). $^2$ Number of objects that enter in the statistics calculation of the results from the fits. Since some systems are multiple, the total number of LIRGs (41) plus ULRGs (26) sources exceed the total number of U/LIRGs systems (58). For the mean values of $<log(L_{IR}/L\odot)>$, SFR(H$\alpha$), and $\Sigma$(H$\alpha$) fewer objects are used in some cases because of the lack of the relevant data (see Table 1) $^3$ The maximum velocity is defined as V$_{max}$=abs(-$\Delta$V+FWHM/2) following, e.g., Rupke et al. (2005b). }
%\clearpage
%\newpage

\end{table}
%\end{sidewaystable}
\end{landscape}
% \clearpage 

\begin{table*}
%\scriptsize
\caption{Line ratios and electron densities for the narrow and broad components}\label{table:tabla2}
%\centering
%\begin{tiny}
\begin{tabular}{l c  c c c c c c  }
\hline \hline
                                                                   & \multicolumn{3}{c}{Narrow} &  & \multicolumn{3}{c}{Broad} \\  
 \hline                                                                        
                                                               &Range           & Mean                 &  Median &&    Range        &   Mean                 &  Median \\ 
\cline{2-4} \cline{6-8}         
log([NII]/$H\alpha$)                                & -0.76,+0.04  & -0.39$\pm$0.16&  -0.41   &&-0.74, +0.59   &-0.22$\pm$0.24  & -0.27    \\
log((S1+S2)/$H\alpha$)                          &-0.87, +0.04  & -0.56$\pm$0.18 &  -0.61   &&-0.99,-0.14    &-0.57$\pm$0.17 & -0.58    \\
log([OI]/$H\alpha$)                                 &-2.08,-0.68    &-1.40$\pm$0.36 & -1.44    &&-1.69,-0.51     &-1.22$\pm$0.28 & -1.29     \\
log([SII]]/[SII])                                          &-0.02,+0.16   & 0.08$\pm$0.05 &   0.08    &&-0.18,+0.19    & 0.04$\pm$0.08  &  0.06     \\
Ne (cm$^{-1}$)                                       & 43- 1025       & 296$\pm$26  &   256   && 16-2777             &  459 $\pm$66    &   315    \\ 

\hline
\hline
\end{tabular}
%\tablefoot{}
\end{table*}

\section {Results and discussion}

\subsection{Dynamical status of the ionized ISM in U/LIRGs}

The mechanism usually invoked to explain merger-induced nuclear starbursts is that the gravitational torques created in the interaction lead the gas to flow inwards, increasing the density in the central regions and, therefore, the SFR (e.g., Barnes $\&$ Hernsquist 1991).  In addition to the nuclear starburst, many interacting and merger galaxies exhibit extranuclear knots of star formation, which are more massive than star clusters in normal spiral galaxies (e.g., Super Star Clusters, SSC; Whitmore et al. 2007; Miralles-Caballero et al. 2012). It has been proposed that this extranuclear star formation is triggered by an increase of the ISM turbulence as a consequence of the tidal interaction (Elmegreen et al. 2000; Struck et al. 2005; Bournaud et al. 2010). A more turbulent ISM has a larger density spread, which favours the creation of dense structures that can further collapse gravitationally forming stars.  Stars, in turn, through winds and supernovae explosions further inject kinetic energy into the ISM contributing to its turbulence (Burkert 2006; Green et al. 2010).  This scenario qualitatively predicts an increase of the velocity dispersion of  the ISM as the merger progresses, but also a dependence on the total SFR.

The narrow kinematic component of the H$\alpha$ line identified in our data primarily traces the ionized gas associated with the dense high surface brightness star forming regions. Therefore, a measure of the dynamical status in these regions is given by the velocity dispersion associated with that component. We emphasize that the measurements of the velocity dispersion used in this paper refer to the width of the line in the spatially integrated spectrum after shifting all individual spectra to the rest frame velocity. Thus, these measurements do not include large-scale velocity gradients (i.e., FoV smearing). 

Our data reveal that the mean velocity dispersion in U/LIRGs  ($\sigma$ = 69$\pm$ 3 kms$^{-1}$; Table 2) is significantly larger than ordinary less active (sub-LIRGs) low-z star-forming galaxies (SFGs, $\sim$ 10 - 30 kms$^{-1}$; e.g. Dib et al. 2006, Epinat et al. 2010), and similar to many SFGs observed at high-z ($\sim$ 30 -90 kms$^{-1}$; e.g. Forster-Schreiber et al. 2009). Since our sample covers a wide range in star formation and morphology, we are in a position adequate to investigate the origin of these high velocity dispersions, analysing how the two proposed mechanisms (tidal forces and stellar winds/SN) affect the overall dynamical status of the ISM in U/LIRGs, as well as the role of the AGNs. Furthermore in this section we will compare with high-z SFGs at both global and local (i.e. clump) scales.

\subsubsection{The LIRG  discs versus the less active local SFGs. The impact of increased star formation in isolated discs} 

In order to isolate the effects of SF on the turbulence of the ISM in U/LIRGs from others like the AGNs or the release of gravitational energy associated to interactions and mergers, we first consider only our sub-sample of U/LIRGs classified as isolated  discs (class 0) and pure starbursts (i.e. with no AGN evidence). This group is formed by 11 objects in the low LIRG luminosity range (hereafter  "LIRG  discs"), with mean SFR(L$_{IR}$) and SFR(H$\alpha$) of 17$\pm$5 and 5$\pm$2 M$_{\odot}$yr$^{-1}$, respectively. The mean velocity dispersion for this group ($\sigma$ = 45$\pm$ 4 kms$^{-1}$), though lower than for the whole sample, is still much larger than the one measured ($\sigma$ = 24$\pm$5 kms$^{-1}$) in less active SFGs as represented by the Gassendi H$\alpha$ survey of SPirals (GHASP, Epinat et al. 2010). 

The GHASP sample consists mainly of isolated spirals and irregulars with a similar range in (dynamical) masses (10$^9$ $-$ 5$\times$10$^{11}$ M$\odot$) as the LIRG discs  (Bellocchi et al. 2013), but with significantly smaller (by a factor of 5) SFRs ( i.e., SFR(H$\alpha$) $\sim$ 1M$\odot$yr$^{-1}$; James et al. 2004, after transforming for a Chabrier 2003 IMF). This suggests that the factor of $\sim$ 2 difference in velocity dispersion between the LIRG- discs and ordinary spirals is likely due to their different SFR regime.  This difference is particularly remarkable because, as reported by Epinat et al. (2010), the velocity dispersion in their sample does not show any dependence on mass or morphological type. Therefore, the main distinct property that could explain their different velocity dispersions is the global SFR, suggesting that the transition from cold (thin) sub-LIRG  discs to hot (thick) LIRG  discs is mainly driven by star formation.  We also note that  IFS  studies of local low luminosity (sub-LIRGs) star-forming galaxies studied at much better linear resolution than the present sample (e.g., Westmoquette et al. 2009, 2011; Rosenberg et al. 2013; Piqueras-Lopez et al. 2012) suggest an important role of SF in driving the dynamics of the ISM at this luminosity range.

\subsubsection{The LIRGs \& ULIRGs.  Saturation of star formation effects on ISM velocity dispersion} 

Over the U/LIRGs luminosity range, the effects of star formation on the dynamical status of the ISM seem to saturate. In fact, the difference in velocity dispersion between non-AGN LIRGs and ULIRGs (54 $\pm$3 and 74 $\pm$6 kms$^{-1}$, respectively) reveals only a slight increase, considering the change by a factor 8 in their corresponding average SFR(L$_{IR}$). This trend is confirmed with an unweighted linear regression to all the non-AGN U/LIRGs in the log($\sigma$) - log [SFR(L$_{IR}$)] plane (see Figure 5, left panels), which shows a rather flat slope (i.e., $\alpha$ = 0.12 $\pm$0.03). This implies a change of only a factor 1.7 in $\sigma$ for a change of $\sim$ 100 in SFR, indicating a weak dependence of the velocity dispersion of the ionized gas on the overall star formation in starburst-dominated LIRGs and ULIRGs.

This dependence is even weaker when the data are analysed in the log($\sigma$) - log [SFR(H$\alpha$)] plane (Fig.5 right panels), where the slope ($\alpha$= 0.04$\pm$0.03) is compatible with no dependence on the SFR.  The fact that the velocity dispersion shows a weaker dependency on SFR (H$\alpha$) than with SFR(L$_{IR}$) could be partially explained by the uncertainties associated with the reddening corrections, which tend to blur a possible, though necessarily weak (according to the results in the log [SFR(L$_{IR)}$] - log($\sigma$) plane), correlation. At any event, the H$\alpha$-based results strengthen previous conclusions of the weak dependence of the velocity dispersion of the ionized ISM on the global SFR in U/LIRGs and, therefore, indicate that the heating of the ionized ISM due to stellar winds and supernova explosions saturates at high star forming rates, i.e., SFR(L$_{IR}$) above 10 M$_{\odot}$ yr$^{-1}$. 

The present data also allow us to investigate whether or not this saturation appears when considering not only the overall star formation in a given U/LIRG but the star formation surface density, i.e., the compactness of the starburst. Some authors (Lehnert et al. 2009) have suggested that if the local/global velocity dispersion of the ISM gas is due to the mechanical energy released by the starburst, a trend of the form $\sigma$ $\propto$ $\epsilon$ $\Sigma_{SFR}^{0.5}$ should be expected, where $\epsilon$ takes the efficiency of coupling between the energy injected and the ISM into account. 

This simple model is not able to explain the relation measured in our sample of  U/LIRGs, which covers almost three orders of magnitude in $\Sigma_{SFR}$(H$\alpha$) from 0.08 to 50 M$_{\odot}$ yr$^{-1}$ kpc$^{-2}$ (see Fig. 6). On the one hand, the slope defined by the bulk of the data (0.06 for either the whole sample or the non-AGNs sub-sample) is much smaller than the one expected by the model (0.5). On the other hand, even for relatively low efficiencies (right edge of shaded region in Figure 6, see also Lehnert et al. 2009), the predicted $\sigma$ values for high star formation surface densities are in general much larger (by factors larger than 2-3 in many cases) than the measured values. This result suggests again that for U/LIRGs, the star formation does not seem to be the main driver regulating the local velocity dispersions observed in the ionized gas, i.e., its dynamical status. Further support to this conclusion comes from the properties of the luminous star-forming clumps in U/LIRGs (crosses in Figure 6) that have a similar trend as the galaxies covering on average relatively low $\sigma$ values. In fact, these high-surface brightness extranuclear clumps of star formation do not have associated increased values on the velocity dispersion maps (see Figure 1 and Bellocchi et al. 2013), clearly indicating that the local star formation is not the dominant source driving the velocity dispersion of the ionized interstellar medium in U/LIRGs, i.e., for SFR(L$_{IR}$) $>$ 10 M$_{\odot}$ yr$^{-1}$ (see 4.1.6 for a more detailed discussion about the clumps).

In summary, while the transition between less active (sub-LIRG) cold ($\sigma$ = 24$\pm$5 kms$^{-1}$) isolated discs and hot ($\sigma$ = 45$\pm$4 kms$^{-1}$) LIRG discs is likely driven by the increased star formation, the heating effects of star formation in the ISM of U/LIRGs seem to flatten, with an average value of 74 $\pm$6 km/s for the most luminous ULIRGs.  In fact, we find a weak dependence of $\sigma$ on both global SFR and $\Sigma_{SFR}$.  This is true not only when considering the integrated light for U/LIRGs, but also for the extranuclear luminous star-forming clumps.  These results suggest that for U/LIRGs, SF has a lower role heating the ISM than for sub-LIRGs systems, and that other mechanisms could also contribute to producing the relatively large $\sigma$ observed.   

\begin{figure*}[]
%generated with fwhmN_lcl.sm  (sin leyenda y con clumps --> fwhmN_l.sm
% los ficheros genera con comp_comp.f
\includegraphics[width=0.85\textwidth]{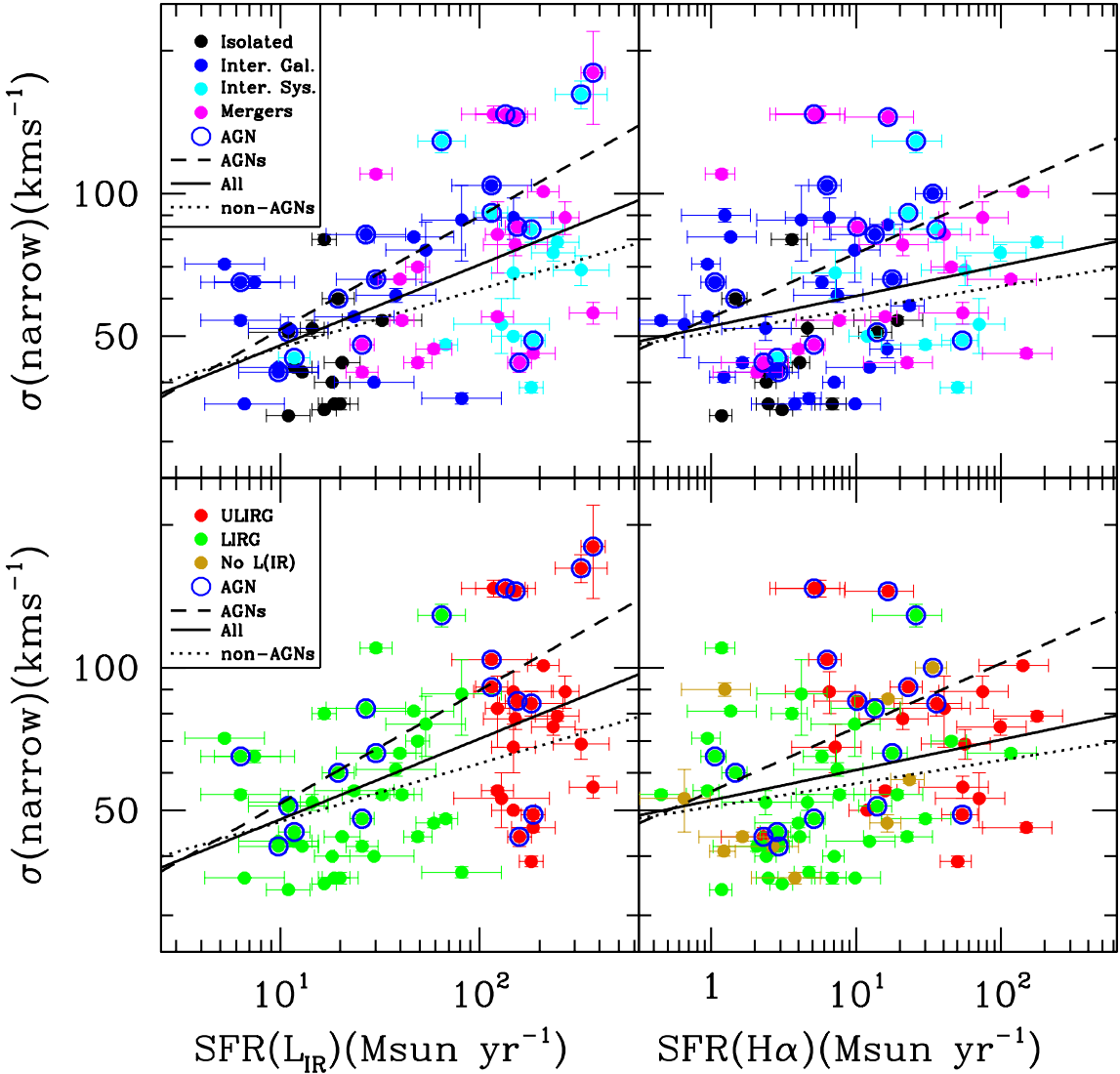}

\caption{Velocity dispersion of the ISM (narrow component) as a function of the SFR as derived from the infrared (left panels) and the (reddening corrected) H$\alpha$ (right panels) luminosities following Kennicutt (1998) with a Chabrier (2003) IMF (see text). Bottom panels: Colours identify LIRGs (green), and ULIRGs (red). Objects without $L_{IR}$ data, are coded in orange. Upper panels: Colours distinguish among the different dynamical classes according to the following code: black are isolated discs (class 0); blue full dots are individual objects in interacting systems (class 1.1); cyan: whole interacting systems (class 1.0); magenta: fully merged (class 2). Blue circles indicate objects with evidence of being affected by an AGN in the optical lines (see text).  For the AGNs Mrk231 and IRAS 13342+3932 (with highest $\sigma$ values in the left panels) are not included in the right panels as they do not have reliable H$\alpha$ absolute fluxes.  Lines correspond to unweighted fits to all the data (continuous), only objects without AGN evidence (dotted), and only objects with AGN evidence (dashed). The error bars in the SFR(H$\alpha$) do not take into account the uncertainties associated with the reddening corrections.}

\
\label{Fig.5}
\end{figure*}

\begin{figure*}
\vskip -5cm
 %generated with sfrd_sigmaHa.sm (fichero previo --> sfrd_sigmaNT.sm)
 % los ficheros genera con comp_comp.f y lehnert.f 
 % para los datos de los clumps generados con comp_comp_knots.f y copiados los ficheros a INTEGRADOS
\includegraphics[width=0.9\textwidth] {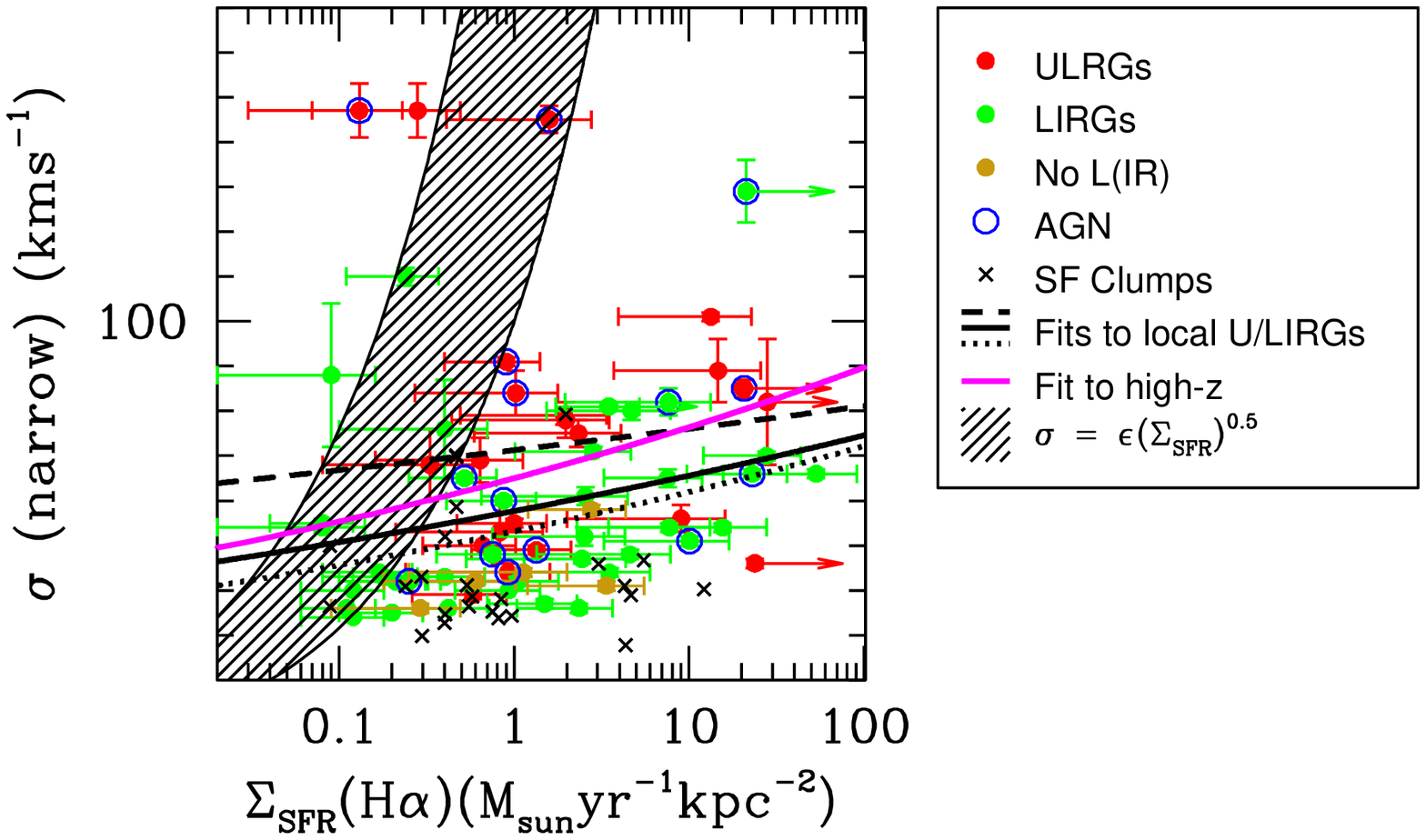}

\caption{Velocity dispersion of the ISM (narrow component) as a function of the star formation density, $\Sigma_{SFR}$, derived considering the extinction corrected H$\alpha$ fluxes (see text), and H$\alpha$ half-light radii (Arribas et al. 2012). The symbol and line codes are indicated in the right panel.The black continuous line corresponds to an unweighted regression to all the  U/LIRGs ($\sigma$  $\propto$ $\Sigma_{SFR}^{n}, n=0.06\pm0.03$, r=0.3), while the dotted and dashed lines are for non-AGNs (n=0.06, r=0.4) and AGNs (n=0.03, r=0.1), respectively. The shaded region corresponds to a behaviour of the type $\sigma$ (kms$^{-1}$)=$\epsilon$$\Sigma_{SFR}^{0.5}$ for efficiencies ($\epsilon$) between 100 and  240, which is expected for SF driven motions (Lehnert et al. 2009). The magenta  line corresponds to the unweighted) fit, found by Genzel et al. (2011), to a large compilation of high-z galaxies and clumps (see text).}

\label{Fig.6}
\end{figure*}

\subsubsection{The LIRG $\&$ ULIRGs. Heating the systems with large-scale tidal forces}

Our data reveal that the velocity dispersion of the ionized gas exhibits a clear trend with the dynamical phase of the system (Fig. 5, top panels). In particular, non-AGN isolated discs, separated pairs, close pairs, and mergers follow a sequence of increasing $\sigma$ (i.e., 45$\pm$4,  59$\pm$4,  60$\pm$5, and 73$\pm$7 km/s, respectively), showing an evolution towards dynamically hotter systems (Table 2).

Since the data show a clear correlation between dynamical phase and SFR (Table 2), one could wonder if the trend observed when comparing the morphological classes is only because of their relative change in SFRs, or, on the contrary, if there is a differential change that cannot be explained by their different SFR regime. To investigate this issue,  we compare, for non-AGN U/LIRGs, the observed change with the predicted change, according to the $\sigma$ - SFR relations found in Sect. 4.2.1 (Fig. 7).  The change in $\sigma$ due to the change in SFR, calculated either from the infrared or H$\alpha$ luminosities, is in general significantly smaller than observed. This suggests that the increase in the velocity dispersion along the merging sequence has an important contribution of the gravitational energy released along the dynamical evolution of the merger, rather than to the change in SF only.   

Therefore, while the comparison of the ISM dynamical properties of normal spirals (i.e., sub-LIRG) and isolated LIRG  discs suggests that SF transforms cold thin rotating  discs (normal spirals) into hot thick  discs (LIRGs discs), gravitational energy associated with interactions and mergers seems to be an important mechanism for generating even hotter systems. During these phases changes in the global SFR and $\Sigma_{SFR}$ have only a modest effect in modifying the dynamical status of the ionized ambient gas. 

Hydrodynamical simulations of the dynamical evolution of the interstellar gas in interacting systems and mergers demonstrate that the velocity dispersion in cold ($\sigma$$\sim$10 kms$^{-1}$) non-interacting spirals increases substantially (factors 3 to 4) in major mergers. A similar relative increase appears in  discs that are already hotter before the merger (Bournaud et al. 2011). Thus, while the observed and simulated values do not necessarily agree in absolute terms, the predicted relative change clearly indicates  that the tidal forces involved in major interactions and mergers are capable of heating the ISM efficiently, producing very hot dynamical systems.

 \begin{figure}[h]
 \centering  
 %generated with sigma_class_SFR.sm 
 \includegraphics[width=8.5cm]{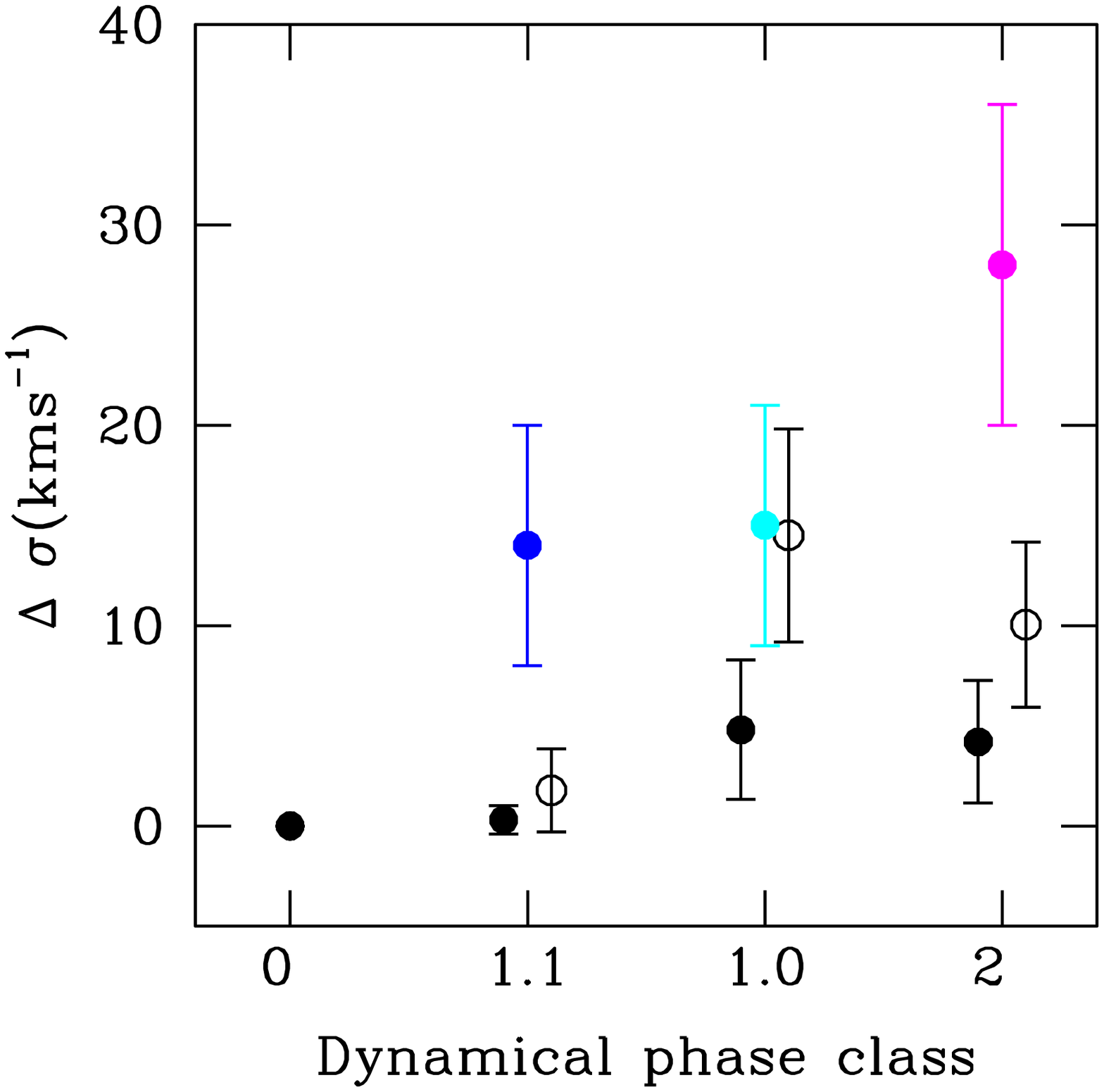}
 \caption{Mean change of the velocity dispersion of the ionized gas (i.e. narrow component) as a function of the dynamical phase for non-AGN U/LIRGs, using the isolated discs (i.e., class 0) as reference. Coloured dots correspond to the observed mean values for individual galaxies in interacting systems (1.1, blue), global interacting systems (1.0, cyan), and mergers (2, magenta).  Black dots correspond to the predicted change due to the mean SFR derived from the infrared luminosity (open dots) and H$\alpha$ (solid dots) for the different classes, according to the relations found in 4.1.2.  The plot shows that only SFR cannot explain in general the observed change in velocity dispersion among the different classes.  The points have been slightly shifted horizontally to avoid overlap.}
\
 \label{Fig3}
 \end{figure}

\subsubsection{The role of AGNs}

A non-negligible fraction (28\%) of the U/LIRGs in our sample have been classified as AGNs\footnote{ For the present study an object is classified as an AGN if its optical spectrum shows evidence of AGN-like ionization or a very broad H$\alpha$ line (see Table 1). Some objects have observational evidence for an AGN according to tracers at other wavelengths (e.g. Mrk463W, Bianchi et al.  2008)}, with a higher fraction (38\%) when considering only ULIRGs. The role of AGNs is clearly visible in the different sigma - star formation relations (see Figures 5 and 6) always yielding steeper and more significant dependences of the velocity dispersion on the derived SF parameters based on the IR and H$\alpha$ luminosities. In LIRGs, AGNs appear to have a small impact in the velocity dispersion of the ionized gas with values a bit higher (65$\pm$8 kms$^{-1}$) than those measured for non-AGNs (54$\pm$3 kms$^{-1}$). However, the net effect of the AGN is  more evident when considering only ULIRGs. While non-AGNs have a mean velocity dispersion of 73$\pm$6 kms$^{-1}$,  AGNs show a dispersion of 109$\pm$14 kms$^{-1}$. These results can be interpreted as due to the presence of more powerful AGNs in ULIRGs than in LIRGs, contributing to a higher fraction of their bolometric luminosity.  This is consistent with previous mid-IR studies that indicate that the AGN contribution to the total luminosity in LIRGs is small, e.g., only 8\%  have an AGN contributing more than 25\% to their 24 microns luminosity (Alonso-Herrero et al. 2012) and about 12\% of the total luminosity emitted by LIRGs (Petric et al. 2011). The AGN contribution increases, however, for ULIRGs with a higher fraction ($\sim$ 40\%) having signatures of an AGN (Armus et al. 2007, Desai et al. 2007, Farrah et al. 2007), with about 20\% - 40\% dominated by the AGN (at least at mid-IR wavelengths). Other studies increase the fraction of ULIRGs with AGN signatures to 70\% and contributing 20\%-30\% of the bolometric luminosity (Nardini et al. 2009, 2010).  Recent X-ray studies (Iwasawa et al. 2011) suggest that 37\% of the galaxies with logL$_{IR} >$ 11.7 L$_{\odot}$ show evidence of an AGN with a minor (10\%) contribution to the total luminosity of the host galaxy. So, it appears as if even with an overall minor energetic contribution, the AGN in a typical ULIRG is able to produce a measurable, i.e., strong, impact on the dynamics of the ionized ambient gas, increasing the velocity dispersion by a factor 1.5 (from 74 km s$^{-1}$ for non-AGN ULIRGs to 109 km s$^{-1}$ for ULIRGs with AGNs). Alternatively, the presence of a strong AGN may be a consequence of a pre-existing turbulent gas.  Indeed, there are models that expect the BH to be fed more efficiently in  discs with higher turbulence, since the latter helps to remove angular momentum (e.g., Bournaud et al. 2011).

\subsubsection{Comparison with the ISM velocity dispersion at  high-z}

Recent IFS observations of luminous star-forming galaxies at high-z have revealed emission lines with relatively large widths \footnote{Note that when characterizing the emission lines for high-z galaxies most of the studies consider one-Gaussian fit model (though see Section 4.2.7). Following this approach, the derived FWHM should be similar to that of the narrow line in a two-Gaussian model as the one considered here, since the broad component usually has a lower peak and intensity affecting mainly to the wings of the profile.}, implying significant random gas motions in the ISM (e.g., Law et al. 2009; Wright et al. 2009; Forster-Schreiber et al. 2009).  For instance, for the SINS sample, Forster-Schreiber et al. (2009) find that many of the massive z $\sim$ 2 star-forming galaxies typically have velocity dispersions ranging from 30 to 90 kms$^{-1}$, similar to those found in the present sample of (U)LIRGs. This, together with the fact that U/LIRGs have a similar range of sizes, star formation rates, and surface densities, as some high-z SFG samples (Arribas et al. 2012), motivates a detailed comparison.

The origin of the high velocity dispersion at high-z is a subject of active discussion (e.g., Lehner et al. 2009, 2013; Green et al. 2010; Genzel et al. 2011; Wisnioski et al. 2012; Swinbank et al. 2012).  On the one hand, it has been suggested that the high velocity dispersions observed are a natural consequence of the intense star formation that is taking place in these objects. Lehner et al. (2009, 2013), on the basis of spatially resolved H$\alpha$ observations of SF galaxies at z$\sim$ 1-3, find that the velocity dispersion of the line emitting gas scales with the star formation gas density as $\sigma$ $\propto$  $\Sigma$$^{n}$$_{SFR}$ with n=0.5, and therefore they conclude that the dispersions are primarily driven by star formation. Similar trends have been proposed by Wisnioski et al. (2012) (i.e. n=0.6) and Swinbank et al. (2012b) (n=0.7) for other high-z samples based on H$\alpha$ IFS observations. 
  
On the other hand, Genzel et al. (2011) have also studied the dependence of the velocity dispersion and the star formation density by combining a large amount of  data for SFGs at z $\sim$ 1.5-3 obtained by several authors (Forster-Schreiber et al. 2006, 2009; Genzel et al. 2006, 2008; Cresci et al. 2009; Wright et al. 2007; Starkenburg et al 2009; Epinat et al. 2009; Lemoine-Busserolle $\&$ Lamareille 2010; Law et al. 2009; Jones et al. 2010),  sampling a wide range of dynamical masses, sizes, and star formation densities.  They find only a modest increase of the velocity dispersion with the star formation density. In particular, an unweighted fit to all their selected data gives a slope of n= 0.07 $\pm$ 0.025, which appears to be inconsistent with the hypothesis that SF is the primary cause for the high velocity dispersion. Although the reason for the disagreement between the different high-z works is unclear, we note the significant impact that different methodological approaches (size definition, reddening corrections, angular resolution, etc) may have on the results, in particular, in these high-z samples showing clumpy surface brightness distributions and complex kinematic structures.     

The results for our sample of low-z non-AGN U/LIRGs are in very close agreement with Genzel et al. (2011) findings. Both low and high-z samples cover similar velocity dispersions and star formation surface densities with most galaxies in the 30 to 100 km s$^{-1}$ and 0.1 to 20 M$_{\odot}$ yr$^{-1}$kpc$^{-2}$ ranges, respectively. Regarding the slope of the $\sigma$ to $\Sigma_{SFR}$ ratio, both agree within uncertainties (i.e., 0.06 $\pm$ 0.03 and 0.07$\pm$0.025 for the low-z U/LIRG and the high-z samples, respectively; see Fig. 6).  As suggested by Genzel et al. 2011 for their high-z sample, we also concluded (Sec. 4.1.1) that for low-z U/LIRGs, the feedback associated with the clumpy star formation does not appear to be the main driver of the velocity dispersion of the ionized gas. Thus, this weak dependence of the dynamical status of the ionized gas on the intensity and compactness of the star formation appears to be universal in luminous star-forming galaxies both locally and at cosmological distances.  
 
Our previous results (Sec. 4.1.1 to 4.1.3) on the present sample of U/LIRGs suggest a greater role than previously thought of the global scale tidal forces in governing the dynamical status of the ionized ISM in the most active star-forming galaxies in the low-z universe. The close agreement between the low- and high-z samples could therefore be interpreted as a similar general behaviour is present at high-z, i.e., global large-scale processes dominate the dynamical status of the ambient ISM.

\subsubsection{Is the velocity dispersion of the clumps in U/LIRGs driven by SF ?}

The luminous star-forming clumps identified in our sub-sample of U/LIRGs (Fig. 1) show a wide range of physical properties (see Table B.1 for specific values) with a median size (half-light radius), velocity dispersion, and (observed H$\alpha$-based) star formation surface density of 0.43 kpc, 39 km s$^{-1}$, and  0.13 M$_{\odot}$ yr$^{-1}$ kpc$^{-2}$, respectively. 
Their sizes are similar to those of giant HII regions (0.1-1 kpc diameter; Kennicutt 1998; Mayya 1994),  and the external H$\alpha$ clumps located mostly in the tidal tails of U/LIRGs (Monreal-Ibero et al. 2007; Miralles-Caballero et al. 2012). The largest clumps, with (light) diameters of 2 kpc and above (see Table B.1) cover the range of sizes measured only in high-z clumps (Wisnioski et al. 2012, Figure 8). The sizes of these H$\alpha$ clumps are  much larger than that of the typical complexes of stellar clusters found in local U/LIRGs observed with HST (Miralles- Caballero et al 2011). Therefore, at least in low-z U/LIRGs, the H$\alpha$ clumps are likely to be aggregates of smaller, unresolved young star clusters and/or star-forming regions (see further discussion in $\S$4.2.8).  

The average velocity dispersion ($\sigma$=43$\pm$3 kms$^{-1}$) of the ambient ionized gas in the luminous clumps, is very similar to the velocity dispersion found from the integrated spectra of their host galaxies (46$\pm$6 kms$^{-1}$)\footnote{Note that the sub-sample of galaxies where the clumps have been identified  has, on average, a lower average $\sigma$ than the entire sample of U/LIRGs. See Table 2.}.  This indicates that the velocity dispersion in the clumps is in general not larger than in the extended low surface brightness regions. This result is also directly confirmed on the two-dimensional maps, where the regions associated with the high surface brightness H$\alpha$ clumps do not appear as regions with increased values in their velocity dispersion (Fig. 1; see also Bellocchi et al. 2013). This result presents direct evidence that in the regions of high star formation (as traced by high H$\alpha$ surface brightness regions), SF is not the main driver of the velocity dispersion of the ionized ISM (see also 4.1.2 and 4.1.3), as it has been also suggested for high-z galaxies (Genzel et al. 2011).

\subsubsection{Luminosity, size, and velocity dispersion scaling relations of the clumps in U/LIRGs}

Early studies of giant HII regions by Terlevich and Melnick (1981) already found that line widths are well correlated with the size and luminosity  (L $ \propto \sigma^{\sim 4}$; r $\propto  \sigma^{\sim 2}$), concluding that these regions are gravitationally bound with $\sigma$ reflecting virial motions. These results have been confirmed by further work at low-z (e.g., Melnick et al. 1988;  Bordalo and Telles, 2011), as well as at high-z  (e.g.,  Wisnioski et al. 2012, and references therein). 

The sizes, luminosities, and velocity dispersions of the clumps in our sample of U/LIRGs obtained from the observed H$\alpha$ line are in general consistent, although with a large scatter, with the mean scaling relations derived by Wisnioski et al. (2012), which are represented by dashed lines in Fig. 8 (i.e.,  $\sigma \propto r^{+0.42\pm0.03}$; L$_{H\alpha} \propto r^{2.72\pm0.04}$; L$_{H\alpha} \propto \sigma^{4.18\pm0.21}$).  These relations cover both local and high-z clumps, and are statistically dominated by the samples of giant HII regions in low-z spirals. 

Neither U/LIRGs nor high-z galaxies alone define the trends, which became apparent when combining both sets of data.  The  L$_{H\alpha }-  \sigma$ relation implies $\sigma \propto$ SFR$^{+0.24}$, i.e., somewhat higher index than that found from the galaxy integrated spectra (Sec. 4.1.2), but still smaller than the one expected if SF were driving the velocity dispersion (i.e., $\sigma \propto$ SFR$^{+0.5}$, Lehnert et al. 2009).

\begin{figure*}
\centering
 %generated with  clumps.sm 
 % los ficheros generados con comp_comp_knots.f . Los de comparacion a mano (algunos con correcciones -- ver pie).   
\includegraphics[width=0.85\textwidth]{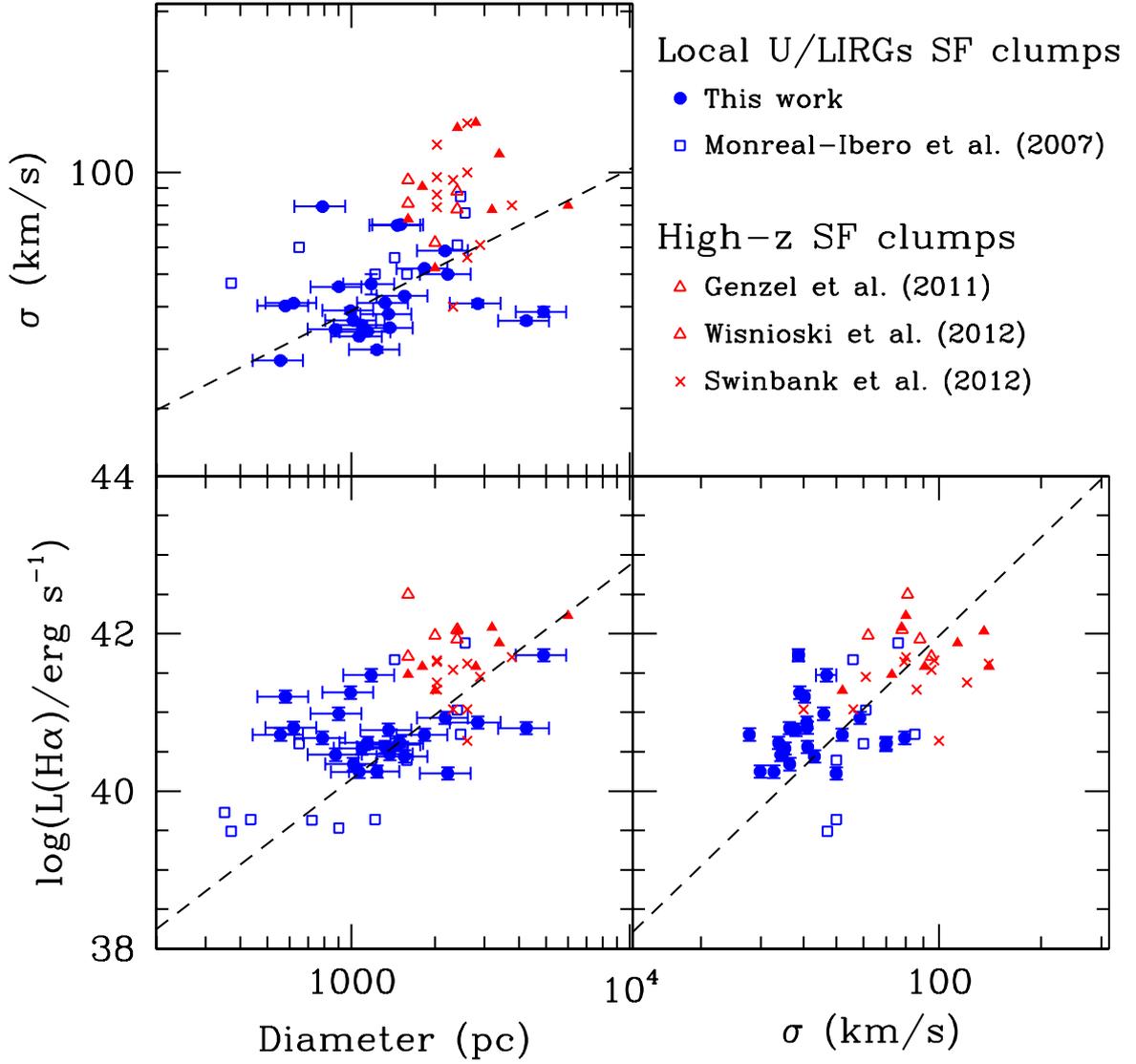}

\caption{Relationship between the size, H$\alpha$ luminosity (no reddening corrected), and velocity dispersion of the clumps of star formation. This figure is similar to that presented by Wisnioski et al. 2012, but here we focus on the comparison between the U/LIRGs clumps (blue) and those for high-z samples (red).  The present values correspond to those of the narrow component in a two-Gaussian fit (see text). Note that we have homogenized luminosity and size measurements among the different studies. For the works by Genzel et al. (2011) and Wisnioski et al. (2012), the H$\alpha$ luminosity correspond to the total value after removing the contribution of the broad component with a mean global value of 40 percent (Wisnioski et al.), and the individual values reported by Genzel et al.. We have only considered individual clumps from Genzel et al (2011). For Monreal-Ibero et al (2007) and Genzel et al. (2001) the observed (i.e., non-reddening corrected) luminosities were calculated from the observed fluxes. In order to homogenize the different methods for inferring sizes, in this plot we have transformed the present half-light measurements given in table B.1 (and those by Monreal-Ibero et al. 2007) into core measurements as those considered by Wisnioski et al. and Genzel et al., following the results by Forster-Schreiber et al. (2009), according to which  0.5 $\times$ FWHM / r$_{hl}$ $\sim$ 1.45.  The clump in IRAS 06206 is not included, as only an upper limit on its size was derived.  The dash lines correspond to the mean behaviour (i.e., least-squares best fit) reported by Wisnioski et al. using a large compilation of local and high-z data, and statistically dominated by the local HII regions and giant HII regions. }
\
\label{Fig7}
\end{figure*}

To the first order, the star forming clumps are consistent with being virialized (i.e., $\sigma \propto r^{0.5}$), and  photon-bounded (i.e., L(H$\alpha$) $\propto r^3$).  For the L(H$\alpha$)-size relation, a high fraction of the small clumps (i.e., diameters of less than 1 kpc), are compatible with being more luminous (at a given size) than local HII regions. This result can be explained by an ISM in the clumps of U/LIRGs denser than in giant HII regions in spirals. Under photon-bounded conditions, the luminosity of the H$\alpha$ has a strong dependency on the electron density and size of the region as given by the expression L(H$\alpha) \propto$ N$_e^2 \times r^3$. Thus, for a given size, the H$\alpha$ luminosity will be higher in the densest star-forming regions. Since the average electron density of the ionized medium in U/LIRGs (see Table 3 for specific values) is about 300 cm$^{-3}$, i.e., factors 2 to 20 higher than those measured in classical, high surface brightness HII regions in local spirals (Kennicutt 1984), more luminous H$\alpha$ clumps would be expected in U/LIRGs, as observed.  Another alternative, also proposed to explain the higher H$\alpha$ luminosity of high-z clumps (Swinbank et al. 2012, Livermore et al. 2012) is that, for a given size, clumps in U/LIRGs would have higher molecular gas surface densities, and therefore higher star formation surface densities (according to the KS law). Consequently, we expect higher H$\alpha$ luminosities. 

In order to compare the mean properties of the clumps in local U/LIRGs with star-forming clumps in high-z galaxies, our sample (mainly clumps in LIRGs) is supplemented with that of Monreal-Ibero et al. (2007), who studied the properties of 12 bright H$\alpha$ external clumps in five ULIRGs on the basis of IFS observations. The comparison among different works is in general difficult as it is affected by systematic effects on the selection, analysis, and  the linear resolution on targets provided by the observations. In order to minimize these effects here we restrict the comparison sample at high-z to those unlensed clumps that have been observed in H$\alpha$ via IFS with AO techniques (Genzel et al. 2011; Wisnioski et al. 2012; Swinbank et al. 2012b), although for other approaches see, for instance, Swinbank et al. (2007) and Livermore et al. (2012). 

The clumps in local U/LIRGs and those at high-z reveal clear differences in their mean properties (Fig. 8). While clumps in U/LIRGs are spread over a wide range in size, with the largest sizes similar to those of high-z clumps, on average low-z clumps are smaller, less luminous, and have lower velocity dispersions than clumps at high-z.  However, this result may be partially affected by systematic effects. In particular, the seeing-limited (local U/LIRGs), and the AO-assisted (high-z galaxies) integral field spectroscopic observations are such that generally the physical scales resolved in local U/LIRGs are smaller (i.e., typically FWHM (PSF) $\sim$ 200-1500 pc and $\sim$ 1000-2000 pc, for low-z ULIRGs and high-z AO observations, respectively).
Therefore, clumps that are identified as single in high-z galaxies could be aggregates of few/several smaller clumps, depending not only on their spatial distribution but also on their relative brightness (i.e., combination of differential mass, extinction, and age of stellar populations on kpc scales).  

Beam smearing could indeed play a role in explaining part of the differences between low-z and high-z clumps.  In this sense it is interesting to note that the two local clumps with properties very similar to that observed at z$\sim$2 in all the panels are in IRAS16007+3743 (Monreal-Ibero et al. 2007), which has been observed with an angular resolution equivalent to $\sim$ 3 kpc (FWHM), even lower than the typical resolutions of $\sim$2 kpc for high-z samples.  This also shows that, when observed with similar physical-scale resolution, there are clumps in local U/LIRGs that resemble high-z clumps  in their global properties (i.e., size, luminosity, velocity dispersion). 

In summary, the clumps in U/LIRGs  seem to follow, although with some scatter, the L-$\sigma$ and size-$\sigma$ scaling relations found for low-z giant HII regions. However,  the L-size relation suggests that  at least a fraction of U/LIRG clumps, like those at high-z, are more luminous than giant HII regions. Finally, on average, clumps in U/LIRGs are smaller in size, and have lower luminosities and velocity dispersions than at high-z. Systematic differences in the linear resolution in low-z U/LIRGs and at high-z for regions identified as star-forming clumps could be playing a relevant role. Observations sampling similar physical scales at low and high-z are required before establishing whether the observed differences are real or not.

\subsection {Ionized gas outflows in LIRGs and ULIRGs}

The broad kinematic component of the 2-Gaussian fits is generally blueshifted with respect to the narrow systemic component. For emission lines a blueshift could in principle be interpreted in two ways:  outflow on the near side of the galaxy or inflow on the far side. For U/LIRGs, the presence of large amounts of dust strongly supports outflows for the general case.  Furthermore, the complex geometries and dust distributions usually observed in U/LIRGs (e.g., GM09b, Piqueras-Lopez et al. 2013) allows us to explain the few redshifted cases as outflow too (e.g.,  dust distributions more severely  blocking the approaching gas than the receding one and/or outflows perpendicular to wrapped inner discs). 

\begin{figure*}
 %generated with  vmaxlir2.sm 
 % los ficheros generados con comp_comp.f
 \includegraphics[width=0.85\textwidth]{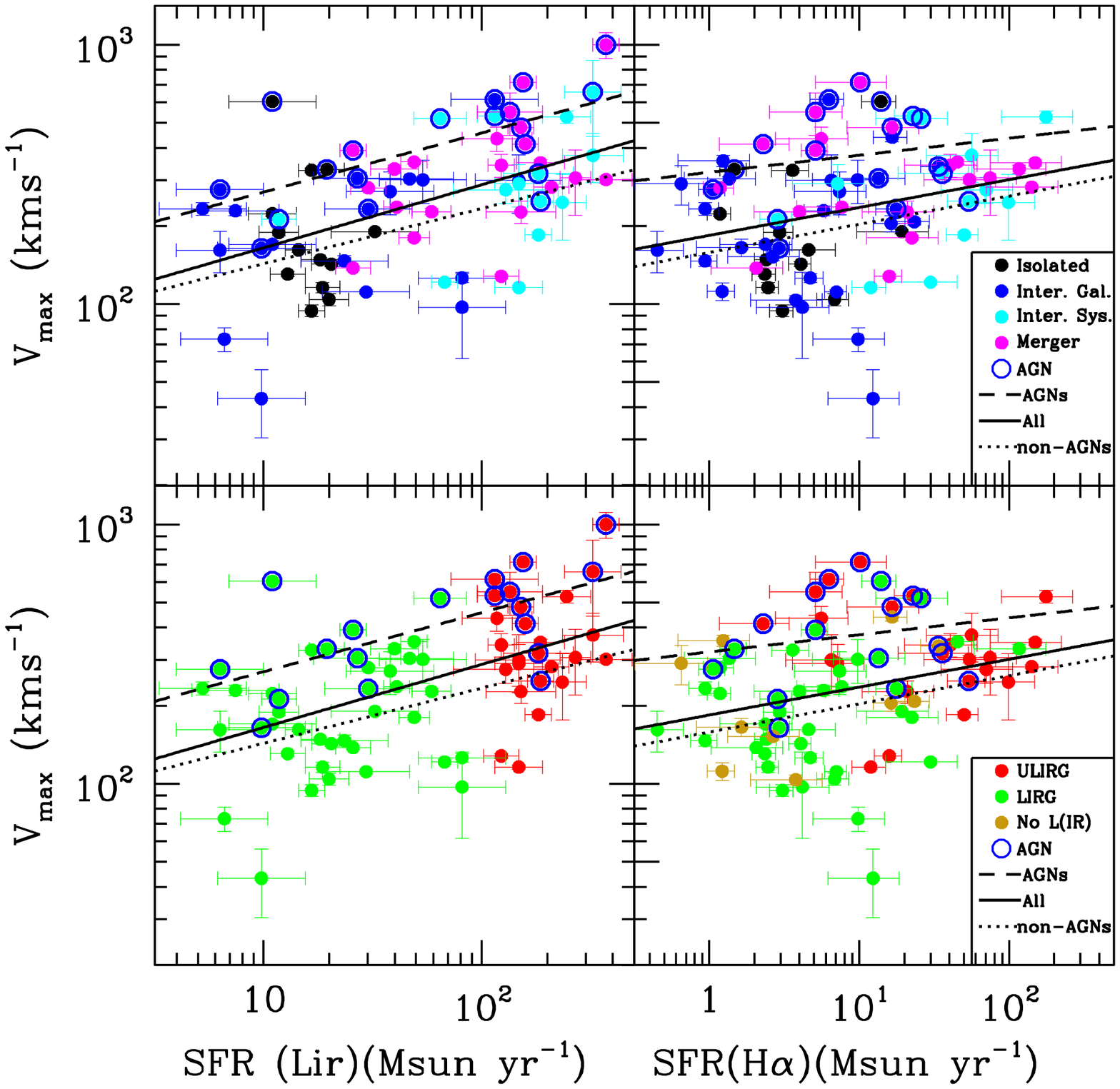}
 \caption{Maximum velocity of the outflow versus the SFR derived from the infrared (left panels) and reddening corrected H$\alpha$ (right panels) luminosities. Lines correspond to unweighted fits to all the data (continuous), only objects without AGN evidence (dotted), and only objects with AGN evidence (dashed). The error bars in the SFR(H$\alpha$) do not take into account the uncertainties associated to the reddening corrections. (see text).}
\
\label{Fig8}
\end{figure*}

In this section, we make use of the properties derived for the broad component to discuss the characteristics of ionized outflows in U/LIRGs.  These will be related to properties of galaxies/systems, and compared  to the predictions by current galaxy models.  We also compare them with high-z ionized outflows, which in some cases have been derived not only using the same tracer, but also with a similar methodology as the one followed here.

\subsubsection{The role of SF launching ionized outflows} 

Although many observations have shown a clear correlation of the outflow properties (velocity, mass, momentum, energy) with SFR, it is still unclear if the effects of star formation on the wind saturate for high levels of star formation found in U/LIRGs  (Rupke et al. 2005b; Martin 2005)\footnote{It is important to note that Martin (2005) and Rupke et al. (2005) works were based on the NaD I interstellar feature, which traces the cold neutral gas clouds entraining the wind. These clouds could be located anywhere  along the line of sight, while the ionized gas probed by H$\alpha$ should reside in dense regions relatively closer to the starburst. Although this may explain some discrepancies between the kinematic properties inferred with these two tracers, the agreement is in general good (e.g., Westmoquette et al. 2012).}.

The present sample includes a relatively large number of LIRGs (41) and ULIRGs (26), which  allows us to compare the basic kinematic properties of the broad outflow component in the luminosity range covered by these galaxies. We consider that the SFRs can be derived from the measured infrared and (reddening corrected) H$\alpha$ luminosities  following the Kennicutt relations (adapted to Chabrier IMF), independent of whether a given galaxy is classified as a pure starburst or as a starburst+AGN. As already discussed (see Sec. 4.1.4), although the fraction of AGNs in U/LIRGs increases with the IR luminosity, their contribution to the total luminosity is generally small (10\%-20\%). Thus, while the derived SFR is an upper limit on the true SFR in galaxies with AGN, the difference will be small in a statistical sense for a large sample like the one considered here. Furthermore, the AGNs in the present sample are distributed fairly homogeneously over the whole luminosity range, and do not show any significantly biased distribution towards LIRGs or ULIRGs. A detailed discussion of the impact of the AGN on the properties of the outflowing gas is deferred to sub-section $\S$4.2.2.

\begin{figure*}
\vskip -5cm
 %generated with sfrd_vmax_v2.sm
 % los ficheros genera con comp_comp.f 
 
\includegraphics[width=0.85\textwidth]{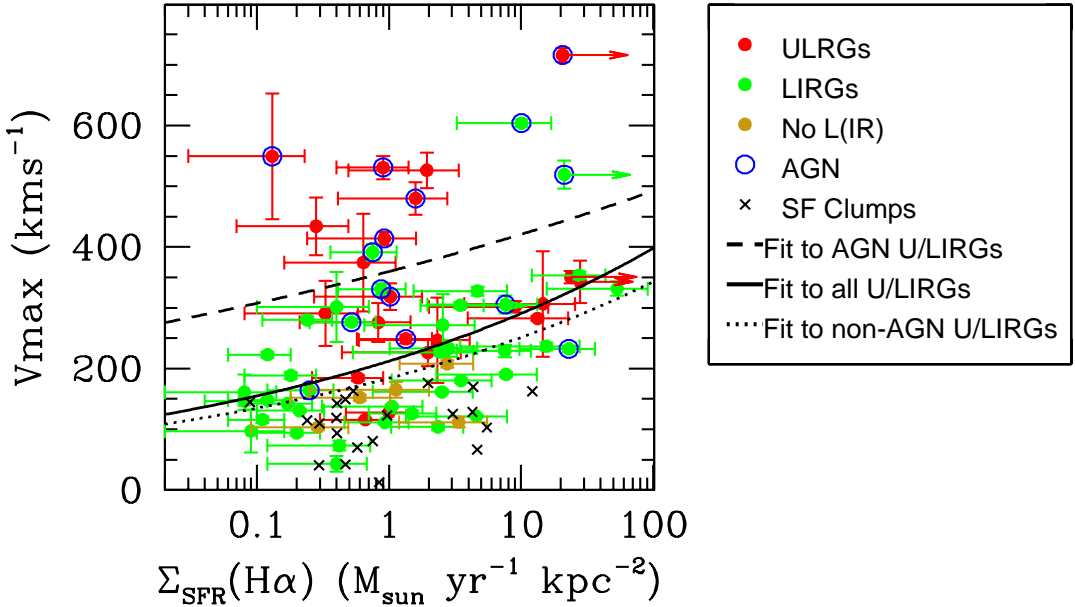} 

\caption{Maximum velocity of the outflow versus the reddening corrected H$\alpha$ star formation rate density. The symbol and line codes are indicated in the right panel.}
\
\label{Fig.9}
\end{figure*}

There are significant differences in the average properties of the outflows as traced by the broad line component in LIRGs and ULIRGs (see Table 2). While the relative flux of the broad to the narrow line component is similar for LIRGs and ULIRGs (F(B)/F(N) $\sim$ 0.85), the velocity shift of the broad component with respect to the narrow line is on average larger for ULIRGs than for LIRGs ($\Delta$V  of -97$\pm$27 and -42$\pm$10  kms$^{-1}$, respectively). Also ULIRGs have significantly broader lines than LIRGs (i.e., FWHM(Broad) of 600$\pm$51 and 350$\pm$24 kms$^{-1}$, respectively), and the projected area where a broad component is detected is generally larger than in LIRGs (Bellocchi et al. 2013). Therefore, outflows in ULIRGs show on average more extreme properties, i.e., line width, velocity offset, and extension, than in LIRGs.  

In order to analyse further the effects of the star formation on the outflow, in the following we focus on its maximum velocity, defined as V$_{max}$ = abs ($\Delta$V-FWHM/2) (e.g., Veilleux et al. 2005; Rupke et al. 2005). This property is relatively well constrained, and has been discussed previously by several authors (Rupke et al. 2005; Martin 2005; Westmoquette et al. 2012). Note that according to the above definition more than 90 percent of the observed gas has a lower velocity than V$_{max}$.

The present U/LIRGs show a large range of V$_{max}$ values (See Table 2) from 43 up to 999 kms$^{-1}$ (for Mrk 231, a well known ULIRG with a Seyfert 1 nucleus), with an average value of 285$\pm$19 kms$^{-1}$,  and a significant difference between LIRGs (219$\pm$18 kms$^{-1}$) and ULIRGs (393$\pm$38 kms$^{-1}$). These results compare relatively well with those obtained by Rupke et al. (2005) for the neutral gas component, who found average maximum velocities of 301 and 408  kms$^{-1}$ for their sample of LIRGs and ULIRGs, respectively. 

Our data also show a clear trend of the maximum velocity of the outflow with SFR. For the infrared luminosity based SFR (Fig. 9, left panels), a linear regression to the data leads to a relation of the type V$_{max}$ $\sim$ SFR$^{0.24\pm0.05}$ (correlation factor, r=0.5). This is in excellent agreement with previous results for the outflowing neutral gas in U/LIRGs by Rupke et al. 2005b  (i.e., slope 0.21 $\pm$ 0.04), also suggesting a flattening with respect to the trend reported by Martin (2005) over the 0.1 $\sim$ 800 $M\odot$ yr$^{-1}$ SFR range (V$_{max}$ $\sim$ SFR$^{0.35}$). This type of relation between the velocity of the outflowing material and the rate of star formation is expected if the mechanical energy is supplied by stellar winds and supernova explosions. In fact,  Heckman et al. (2000) predicted a relationship between the velocity and the luminosity of the starburst of the form V$_{max}$ $\propto$ L$_{bol}$$^{1/4}$, in very good agreement with the relations obtained for the ionized (this work) and neutral gas (Rupke et al. 2005).

When derived from the (extinction-corrected) H$\alpha$ luminosity, SFR(H$\alpha$),  V$_{max}$ shows a lower dependence on the star formation rate (see Fig. 9, right panels). This is understood as a consequence of the relatively large uncertainties associated with the reddening corrections, that tend to destroy a possible (intrinsic) correlation, reducing its slope (0.11$\pm$0.04), and lowering the significance of the correlation (r=0.3). The relation between V$_{max}$ and the star formation density (Fig. 10) has a similar slope (0.13$\pm$0.03) and shows a somewhat better correlation (r=0.5).

\subsubsection {The AGN influence on the strength and velocity of the ionized gas outflow}
 
The U/LIRGs with evidence of hosting an AGN according to their optical spectra show a distinct behaviour in the V$_{max}$-SFR relation with respect to those identified as pure starbursts: they all tend to have the largest maximum velocity for a given SFR(L${_{IR}}$) (Fig. 9).  To quantify the AGN impact more accurately, the V$_{max}$-SFR relation has been derived independently for objects with and without evidence for an AGN. Both groups maintain a similar dependence (slopes of 0.21$\pm$0,05 and 0.23$\pm$0.07, respectively), indicating that the impact of the AGN appears to be independent of the overall luminosity (or SFR-equivalent) of the galaxy. In addition, we find a nearly constant off-set in V$_{max}$ between the pure-starbursts and starbursts+AGN sub-samples, which translates into a maximum velocity of a factor of $\sim$ 2 larger for galaxies with AGNs than without\footnote{The factor 2 increase in V$_{max}$  for AGNs may be due not only to an additional contribution of the active nuclei powering the gas, but also to a change in the geometry of the outflow. If the AGN is more likely to be observed when a relatively high fraction of gas has been removed from the central regions, the outflow could be less collimated.  Consider a simple model in which the flow is collimated and perpendicular to the disc plane. The average inclination is 57 degrees (e.g., Law et al. 2009), and the average measured velocity is (for a given SFR) is 0.5V, where V is the outflowing velocity.  However, for a less collimated outflow the inclination effects are less relevant, and for an isotropic outflow the average velocity is V.}.     

The potential influence of the AGN on the outflow properties have been addressed by several authors, who have reached different conclusions. The NaD based study of Rupke et al (2005c) finds hints (inconclusive evidence) of the effects of the AGN on the neutral gas outflow velocity. However, most of the work  on the ionized gas outflows (e.g., Lipari et al. 2003; Westmoquette et al. 2012; Soto et al. 2012; Rodriguez-Zaurin et al. 2013; Rupke and Veilleux, 2013) conclude that the fastest outflows appear in galaxies hosting a central AGN. Analysis of the neutral outflowing gas in our sample of galaxies is underway (Cazzoli et al. in preparation) to further investigate the effect of the AGN in the neutral gas component, and in particular, in these galaxies showing the extreme values for the V$_{max}$ in the ionized gas. 

The influence of the AGN appears to be not only on the maximum velocity of the outflow but also on its relative flux. The fraction of flux carried out by the broad line component is significantly larger in objects with AGNs than without, with average F(B)/F(N) values of 1.03$\pm$0.16 and 0.76$\pm$0.07, respectively, independent of the IR luminosity (see Table 2). Thus, while AGNs in U/LIRGs are in general minor contributors to the total energy output of the galaxy (see section 4.1.4), they are very efficient at launching outflows of ionized gas. For a given L$_{IR}$, AGN-U/LIRGs are able to generate more massive (x 1.4), and faster (x 2) outflows, therefore releasing about six times more (kinetic) energy than non-AGN U/LIRGs. Cicone et al. (2014) have also shown that the presence of an AGN strongly boosts the outflow rate and kinetic power in molecular outflows.

\subsubsection {Outflow kinematic properties and merger evolution}

Previous works have reached inconclusive results about the variation of the kinematic properties of the outflowing material along the merger process. Martin (2005) does not find any correlation between the outflowing velocity of the neutral gas and the dynamical phase of the system. Rupke et al (2005b,c) tentatively propose an increase in the maximum velocity in the latest merger stages.  These authors, however, warn about the significance of this result because of the low statistics, and the fact that the largest median V$_{max}$ values are not found in the final phases of the merger. 

The first important result from the present new data is that the outflow component in the ionized gas is identified for all sources independent of their dynamical phase, from isolated discs to interacting pairs and mergers, i.e., our data suggest it is a universal phenomenon, at least in the ionized gas, in this type of galaxies. Second, the observed characteristics of the outflow are a clear function of the dynamical phase of the system, with the mergers showing the most extreme properties. Isolated discs (class 0) have on average relatively small line widths  (FWHM(B)=348$\pm$44 kms$^{-1}$), and velocity off-sets ($\Delta(V_B-V_N)$ = $-$38$\pm$56 kms$^{-1}$- see Table 2). Contrary, mergers (class 2) show significantly larger values  (i.e., FWHM(B)=548$\pm$56 kms$^{-1}$,  $\Delta(V_B-V_N)$= $-$91$\pm$28 kms$^{-1}$). Interacting systems (class 1) have intermediate properties, with separated individuals (class 1.1) behaving similar to isolated discs, and  close pairs (class 1.0) like mergers. Regarding the maximum velocity, the different dynamical classes define a sequence of increasing maximum velocity from isolated discs  (212$\pm$37 kms$^{-1}$; median: 162 kms$^{-1}$) to mergers (365$\pm$43 kms$^{-1}$; 331 kms$^{-1}$). Moreover, the spatially resolved velocity maps show that the broad component is more extended in mergers than in isolated discs or interacting pairs (Bellocchi et al. 2013).  Finally, the relative strength of the broad component with respect to the narrow comnponent is quite similar across the different dynamical phases (Table 2), with a marginal increase in the broad-to-narrow flux ratio (F(B)/F(N) from  discs (class 0: 0.79$\pm$0.20) to mergers (class 2: 0.95$\pm$0.14).    

Although the dependence of the mean kinematic properties on the morphological class (i.e., evolutionary phase) is clear, one may wonder if this could be due to the presence of AGNs and/or to the different mean SFRs of the sub-classes, which clearly correlate with the evolutionary phase of the system since more ULIRGs are identified in advanced phases of the merging process. As for the effects of the AGN, when only non-AGN systems are selected, the same trend as the dynamical phase appears in V$_{max}$, FWHM, and $\Delta(V_B-V_N)$, although with lower absolute values (see Table 2 for the values for the different AGN/non-AGN/all sub-samples).  Therefore, the presence of AGNs does not seem to drive the change in the observed properties of the outflowing gas with the evolutionary phase. On the other hand, the observed trend in V$_{max}$ is compatible with the predicted change due to the variation of the SFR across the different phases, within the uncertainties (Fig. 11).  Therefore, our data do not suggest that the release of gravitational energy across the merging phase contributes to powering the outflow, as SF alone can account for the observed changes in V$_{max}$ along the sequence.  

 \begin{figure}[h]
 \centering  
 %generated with plot_class_SFR.sm 
 \includegraphics[width=8.5cm]{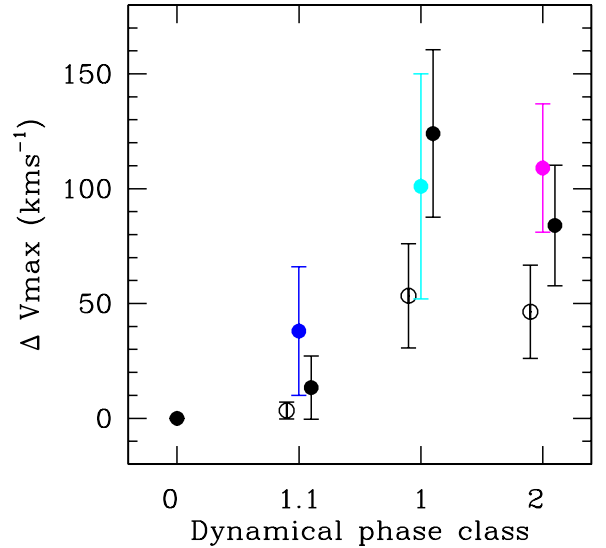}
 \caption{Mean change of the maximum velocity of the outflow as a function of the dynamical phase for non-AGNs, using isolated discs (i.e., class 0) as reference.  Coloured dots correspond to the observed mean values for individual galaxies in interacting systems (1.1, blue), global interacting systems (1.0, cyan), and mergers (2, magenta). Black dots correspond to the predicted change in Vmax due to the mean SFR derived from the infrared luminosity (open dots) and H$\alpha$ (solid dots), according to the relations found in 4.1.2.  The plot shows that the observed change across the different classes is compatible with the expected change due to SFR. The points have been slightly shifted horizontally in order to avoid overlap.}
\
 \label{Fig3}
 \end{figure}

In summary, our data show that ionized gas outflows are universal among U/LIRGs and that the kinematic properties of the outflowing gas evolve with the merging phase. In particular, outflows with the largest velocity dispersions, blueshifts, maximum outflowing velocities, extension, and possibly relative strength occur during the more advanced phases of the merger process. The ultimate cause seems to be the associated change (increase) of the rate of star formation along the merger process.      

These results agree well with the scenario in which merger driven bursts of star formation are expected to be more vigorous during the final dynamical phases of the merger process (e.g., Barnes \& Hernquist 1991; Mihos \& Hernsquist 1996; Genzel et al. 1998; Cox et al. 2006; di Matteo et al. 2007). Recent  hydrodynamical simulations of major mergers of disc galaxies (Teyssier et al. 2010) are able to reproduce the starburst mode with high star formation rates soon after the first pericentre, and reaching its maximum at the final coalescence. As a consequence, stellar winds and supernovae driven outflows should also show a more powerful manifestation during these late phases. 

\subsubsection{Ionized outflowing mass rates and loading factors} 

Momentum- and radiation-driven theoretical models (Murray et al. 2011, Ostriker \& Shetty 2011) predict the generation of large-scale outflows in discs with average star formation surface densities above 0.1 M$_{\odot}$ yr$^{-1}$ kpc$^{-2}$, in agreement with the empirical limit obtained in low-z starbursts (Heckman et al. 2003). The results presented so far show clear evidence that the presence of outflows is universal in LIRGs and ULIRGs. Moreover, ULIRGs tend to produce faster and more extended outflows than LIRGs (Bellocchi et al. 2013). Also, U/LIRGs with AGNs generate faster outflows than non-AGN U/LIRGs. So, the outflowing mass rate and total outflowing material depends on the activity class of the galaxy (i.e., AGN versus non-AGN), and on the overall energy output (i.e., ULIRG versus LIRG). A dependence on the morphology (i.e., interaction phase), is also expected as a consequence of  the increased SF in the later phases of the interaction/merger process. 
We now derive some of the relevant characteristics of this outflowing material, including the total mass, mass outflow rate, and mass loading factors for non-AGN U/LIRGs. 

Following Colina et al. (1991), the total amount of ionized outflowing gas is given by:

\begin{equation}
M_g(M_{\odot}) = 4.44 \times 10^{-33}  \times L_B(H\alpha) \times N_e^{-1}
\end{equation}

\vspace{1mm}
\noindent
where L$_B$(H$\alpha$) is the luminosity of the broad H$\alpha$ line given in erg s$^{-1}$
and N$_e$ is the electron density  of the broad component in electrons per cm$^{-3}$.
Using Kennicutt (1998) expression that relates the total (i.e., broad plus narrow) H$\alpha$
luminosity and the star formation rate (SFR$_{H\alpha}$ in M$_{\odot}$ yr$^{-1}$) for a Chabrier IMF
(note that U/LIRGs with identified AGN are also included here and therefore in these cases, the SFR included in the expression
above is an upper limit on the real value), the expression above can be written as:

\begin{equation}
M_g(M_{\odot}) = \frac{9.65 \times 10^{8} \times SFR_{H\alpha}} {N_e\times (1+(F(B)/F(N)^{-1})} 
\end{equation}

\vspace{1mm}
\noindent
as a function of the  broad to narrow flux ratio (see Table 2 for specific F(B)/F(N) values).
Assuming a dynamical time of t$_{dyn}= d/V_{max}$, the outflow mass rate is given by:
\begin{equation}
\dot{M_g}(M_{\odot} yr^{-1}) = M_g \times V_{max} \times d^{-1}
\end{equation}

\vspace{1mm}
\noindent
or

\begin{equation}
\dot{M_g}(M_{\odot} yr^{-1}) = \frac{9.65 \times 10^{8} \times SFR_{H\alpha} \times V_{max} } {N_e\times (1+(F(B)/F(N)^{-1}) \times d} 
\end{equation}

\vspace{1mm}
\noindent
where V$_{max}$ is the velocity of the outflowing material in km s$^{-1}$ as defined in
section 4.2.1, and $d$ is the radius of the outflowing region in kpc.
Consequently, the corresponding mass loading factor ($\eta$ = $\dot{M_g}$/SFR$_{H\alpha}$) will then be given by:

\begin{equation}
\eta =\frac {0.44 \times V_{max}} {N_e \times d}
\end{equation}

\vspace{1mm}
\noindent
or

\begin{equation}
\eta = \frac { V_{max}}{ N_e \times d \times (1+(F(B)/F(N)^{-1})}
\end{equation}

\vspace{1mm}
\noindent
for the average (0.8) or specific F(B)/F(N) value for a given galaxy, respectively. Note that the derivation of the mass outflowing rate, and consequently the mass loading factor, are subject to assumptions about the unknown structure (shells, filaments, dense clouds) of the outflowing gas and of its velocity (i.e., mass distribution according to velocity, see Hopkins et al. 2012), which may introduce differences by factors of 2-3 (see detailed discussion in Maiolino et al. 2012). Our expressions consider that all the outflowing mass moves at the maximum velocity, therefore setting a lower limit on the dynamical time, and consequently upper limits on the mass outflowing rate and mass loading factors.

As already mentioned (see section 3.6), our data allow us to determine the electron density of the outflowing material with a median (average) value of 315 (459$\pm$66) cm$^{-3}$, based on the ratio of the [SII] doublet. The projected area of the outflowing region is in general small as  derived from our spatially resolved kinematics  (see table 5 in Bellochi et al. 2013). The median area
for the sample corresponds to 1.55 kpc$^2$ (equivalent to a circular area of 0.7 kpc radius),
with only four galaxies (9$\%$) showing very extended regions  ($>$ 10 kpc$^2$). Note that this value should be considered as a lower limit on the real size of the outflowing region, on the one hand, due to projection effects and, on the other hand, to the methodology used to identify the outflows on the two-dimensional maps (i.e., line profile fitting to two components on a spaxel-to-spaxel basis where high signal-to-noise data are required to obtain reliable fits).  

Assuming an electron density of 315 cm$^{-3}$ and a typical radius of 0.7 kpc, the ionized outflowing mass rate and total mass will be 0.1$\leq \dot{M_g}$ $\leq$ 200 M$_{\odot}$ yr$^{-1}$, and 1.4 10$^6$ $\leq$ M$_g$ $\leq$ 2.8 10$^8$ M$_{\odot}$ for an H$\alpha$-derived SFR in the 1 to 200 M$_{\odot} yr^{-1}$ range and velocities (V$_{max}$) in the 50 to 500 km s$^{-1}$ range. The corresponding average mass loading factors for non-AGN LIRGs and ULIRGs are 0.3 and 0.5, respectively. Thus, while only luminous ULIRGs and some AGN-LIRGs will be able to eject large quantities ($\dot{M_g}$ $\geq$ 30 M$_{\odot}$ yr$^{-1}$) of ionized material to the external regions of the galaxy, (see section $\S$4.2.6 for estimates of the expected escape fraction into the intergalactic medium), the amount of ejected ionized material will generally not be larger than the corresponding amount of mass that is converted into stars per year, except for a few AGNs  for which mass loading factors above one are measured.

The ionized mass loading factors derived for our sample of U/LIRGs are similar to the average values obtained for the neutral gas outflows in a sample of low-z LIRGs, but higher than those of ULIRGs  ($\eta$ of 0.33 and 0.19, respectively; Rupke, Veilleux \& Sanders 2005). This result could suggest that while outflows in LIRGs carry a similar amount of mass in ionized and neutral material, the outflowing material in ULIRGs appears to be mostly ionized. The relatively low loading factors found by Martin (2006) for the neutral outflows in ULIRGs  ($\eta$ $\sim$ 0.01-0.1) reinforce this result. Recent findings in a small sample of six ULIRGs (several of them AGN-dominated) indicate that while in some galaxies the outflowing mass is dominated ($> 90\%$) by neutral gas (Mrk 231, IRAS 10565+2448), in others (Mrk 273, IRAS 08572+3915) the ionized phase represents 25\% to 50\% of the total (Rupke \& Veilleux 2013). Further investigation with a larger sub-sample of U/LIRGs for which both NaD and H$\alpha$ IFS data are available (Cazzolli et al. in preparation) is required to extend these studies, and establish fractions of neutral and ionized outflowing material in U/LIRGs.

\subsubsection{Ionized mass loading factor. Observed vs. predicted dependency on the mass and star formation of the host galaxy}

At low-z U/LIRGs represent the extreme star-forming galaxies, covering the upper end of the infrared luminosity function (i.e., largest absolute star formation rates) and the more compact starbursts (i.e., highest star formation surface densities). As already discussed in previous sections, the present study has been able to characterize the outflowing ionized gas in a large sample of U/LIRGs from, for the first time, two-dimensional H$\alpha$ kinematic and flux measurements. These results, covering a wide range in star formation properties and dynamical masses (see also Arribas et a. 2012 and Bellocchi et al. 2013 for detailed derivation of sizes and masses), allow us further to investigate the dependency of the mass loading factors on the overall mass and star formation of the host galaxy in light of recent hydrodynamical simulations (Hopkins, Quataert \& Murray 2012). These models simulate the structure for different discs, including dwarfs, MW-like spirals and LIRG-like galaxies (i.e., increased gas fraction with respect to a MW-like galaxy) including the stellar, gas, and dark matter mass distributions as well as the cooling, star formation, and feedback processes in a multiphase (cold molecular, warm ionized, and hot X-ray) medium (see details for models labelled Sbc in Hopkins et al. 2012). According to the models, the mass in the outflow is dominated by the mass in the warm ionized phase. While this result does not appear to be confirmed empirically in all cases (Rupke \& Veilleux 2013, Cazzoli et al. 2014; Sect. 4.2.4), for the purpose of the following discussion we will be assuming that our estimated H$\alpha$-derived mass loading factors correspond to the total predicted values, as suggested by these models.

An inverse dependency of the mass loading factor on the dynamical mass is obtained for the non-AGN LIRGs  ($\eta \propto M_{dyn}^{-0.43}$), while they are consistent with a slight dependency ($\eta \propto M_{dyn}^{-0.17}$) when both non-AGN LIRGs and ULIRGs are considered (see Fig. 12). For a better comparison with Hopkins et al. (2012) models, we restrict to LIRGs  as a high fraction can be considered discs. The trend (i.e., slope) shown by non-AGN LIRGs  is in excellent agreement with the predicted mass relations where $\eta \propto M_{baryon}^{-(0.3-0.65)}$ or $\propto M_{stars}^{-(0.25-0.5)}$. However, there is a large discrepancy in the absolute value of the predicted versus observed mass loading factor for the mass range ($\geq 10^{10} M_{\odot}$) covered by the U/LIRG sample. While the empirically derived $\eta$ values in U/LIRGs are in general below one (i.e., no effective negative feedback), the theoretical values are in the $\sim$ 1-5 range (i.e., efficient quenching of the star formation) independent of the mass tracer (i.e. either baryon or stellar;  see fig. 7 of Hopkins et al. 2012). For LIRG-like galaxies (i.e., Sbc in Hopkins et al. terminology), these high predicted $\eta$ values are attributable to a strong non-linear coupling between the outflows generated by the radiation pressure, as well as to those directly generated by the momentum deposition due to supernovae and stellar winds. The empirical $\eta$ values (this study; Rupke et al. 2005; Martin 2006) do not support  this strong coupling but rather suggest that outflows could be dominated either by radiation pressure or shock heating from SNe and stellar winds as these mechanisms contribute in a similar manner to the outflow mass (Hopkins et al. 2012, see also Murray et al. 2005). Evidence that both mechanisms (radiation- and shock-driven) coexist in galaxies have been presented in the nearby LIRG NGC 5135. In this galaxy, outflows associated with the nuclear AGN, and strong outflows due to supernovae in an aged circumnuclear star-forming clump (Bedregal et al. 2009, Colina et al. 2012, Colina et al. in preparation) have been identified in different phases of the interstellar medium (coronal, partially-ionized and hot molecular), and on scales of hundreds of parsecs.

It is also worth mentioning that the recent dynamical models of SN feedback by Lagos et al. (2013) predict a somewhat steeper slope in the $\eta$ vs. M$_{stellar}$ plane, but it is still consistent with the slope found by us (Fig. 12). Their predicted $\eta$ are also larger than the observed values for the low mass range, though the agreement improves for largest masses especially if ULIRGs are also include in the comparison.

\begin{figure}[h]
\centering
%generated with mass_eta_log_as.sm con errores asimetricos
%\includegraphics[width=8.5cm]{m_lf_log.ps}
\includegraphics[width=8.5cm]{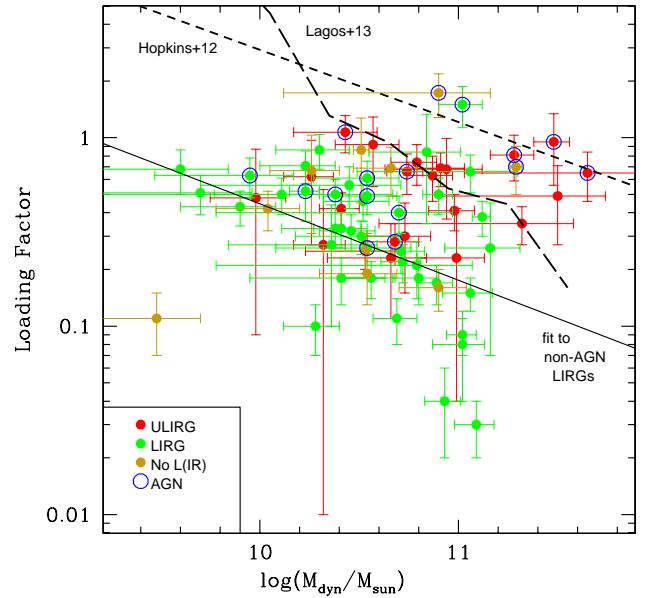}
\caption{Mass loading factor as a function of the dynamical mass. The continuous line represents an unweighted linear fit (in log) to the non-AGN LIRGs. The short- and long-dash discontinuous lines represent the predicted behaviour according to the models by Hopkins et al. (2012) ($\dot{M_g} \propto SFR^{+0.7}$) and Lagos et al. (2013), respectively. For the comparison with the models of Lagos et al.  we consider M$_{stellar}$ $\approx$ M$_{dyn}$. See text.}
\
\label{Fig_Mass_lf}
\end{figure}

From our data, we derive a clear linear correlation of the outflowing mass rate with the H$\alpha$-derived star formation rate ($\dot{M_g} \propto SFR_{H\alpha}^{+1.11}$) for non-AGN LIRGs (see Fig. 13), which is steeper than the predicted sublinear dependency $\dot{M_g} \propto SFR^{+0.7}$ found by Hopkins et al. (2012). When the SFR traced by the IR luminosity is considered, the correlation shows a steeper index ($\dot{M_g} \propto SFR(L_{IR})^{+1.23}$), deviating even more from model predictions.  The observed supralinear correlation is clearly understood as a direct consequence of the linear dependency of the outflowing mass on the SFR and V$_{max}$, with V$_{max}$ $\propto SFR^{+0.24}$ (see $\S$4.2.1).

\begin{figure}[h]
\centering
%generated with sfr_mdot.sm
\includegraphics[width=8.5cm]{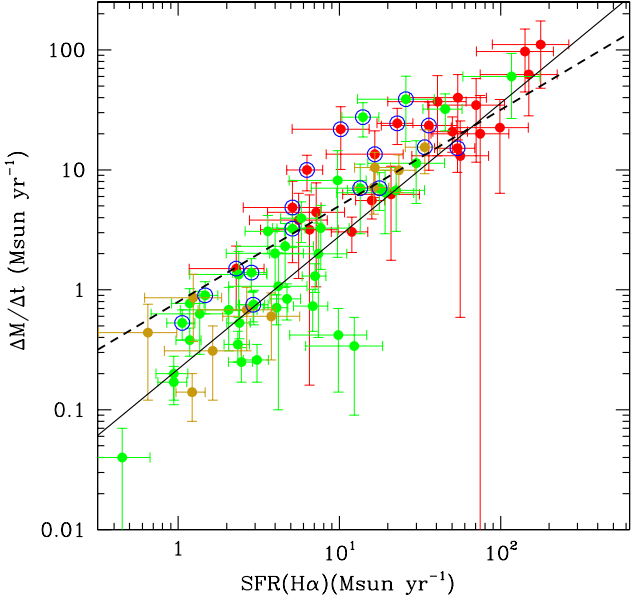}
\caption{Mass outflow rate as a function of the star formation rate derived from the (reddening corrected) H$\alpha$ luminosity. Symbols and colour code as in Fig. 12. The continuous line represents an unweighted linear fit (in log) to the non-AGN LIRGs ($\dot{M_g} \propto SFR_{H\alpha}^{+1.11}$). The discontinuous line is the predicted behaviour according to the models by Hopkins et al. (2012) ($\dot{M_g} \propto SFR^{+0.7}$). See text.}
\
\label{Fig_sfr_omr}
\end{figure}

Finally, we also investigate the dependency of the mass loading factor on the compactness of the star formation as traced by the H$\alpha$-derived star formation surface density (see Fig. 14). The hydrodynamical simulations by Hopkins et al. (2012) predict a scaling relation $\eta \propto \Sigma_{gas}^{-0.5}$, which translate into  $\eta \propto \Sigma_{SFR}^{-0.36}$, if the Kennicutt-Schmidt law is assumed ($\Sigma_{SFR} \propto \Sigma_{gas}^{1.4}$; Kennicutt 1998). However, our data show a positive relation $\eta \propto \Sigma_{SFR}^{+0.17}$ for all non-AGN LIRGs (with an exponent of +0.21 if  non-AGN U/LIRGs are considered), in clear disagreement with the predicted dependency, but consistent with previous claims of positive correlations for the neutral outflowing gas in less luminous low-z galaxies (Chen et al. 2010), and for the ionized phase in high-z star-forming galaxies (Newman et al. 2012).

\begin{figure}
\centering
%generated with sfrd_lf.sm
\includegraphics[width=8.5cm]{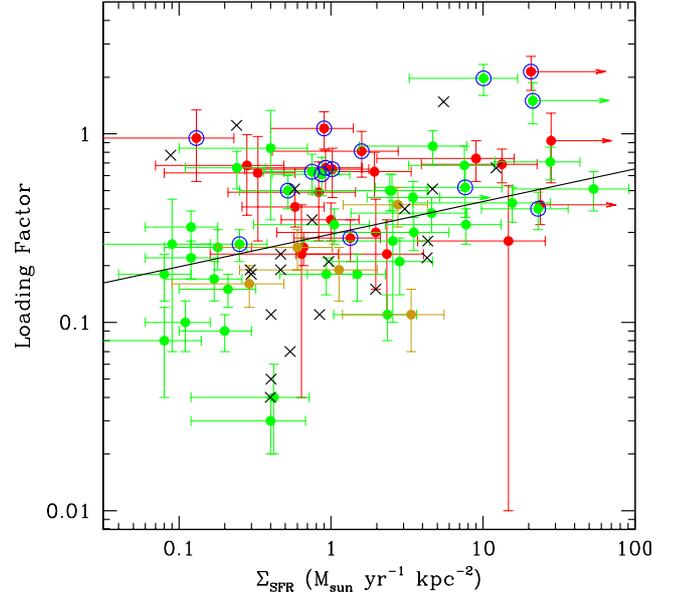}
\caption{Mass loading factor as a function of the star formation density as traced by the (reddening corrected) H$\alpha$ luminosity and half-light radius. Symbols and colour code as in Fig. 12. The continuous line represents an unweighted linear fit (in log) to the non-AGN LIRGs (see text). }
\
\label{Fig_sfrd_loading}
\end{figure}

\subsubsection{Ionized gas escape fraction and IGM metal enrichment}

Massive galaxies with deep potential wells are expected to retain more of their newly synthesized metals than dwarf galaxies (e.g., Larson 1974, Wyse $\&$ Silk 1985, Heckman et al. 2000, Veilleux et al. 2005, and references therein). This galaxy-mass dependent effect has fundamental implications, and it has been invoked in the past to explain both the mass-metallicity relation and radial metallicity gradients in elliptical galaxies and galaxy bulges (e.g., Bender, Burstein $\&$ Faber 1993; Franx $\&$ Illingworth 1990; Heckman et al. 2000; Martin 1999; Garnett 2002; Tremonti et al. 2004). 

Several authors have looked for correlations between the outflow velocities and the galaxy mass as traced by its circular velocity (V$_c$). Studying the neutral gas kinematics, Heckman et al. (2000) do not find a significant correlation between the outflow velocity and the galaxy rotation speed, implying that the outflowing gas is mostly lost by dwarf galaxies, which completely dominate the enrichment of the IGM. However,  Martin (2005) and Rupke et al. (2005) do find that not only are neutral outflows faster in more massive galaxies, but they also appear to increase almost linearly with the galactic rotation speed and, therefore, with the escape velocity. However, this positive correlation is mainly driven by few dwarf starbursts (Schwartz $\&$ Martin, 2004). Moreover, Rupke et al (2005), using only their LIRG and ULIRG data, find instead a (weakly significant) negative correlation between V$_{max}$ and V$_c$. 

Our VIMOS data allow us to investigate the relation between the outflow velocity and the galaxy mass. Dynamical masses (M$_{dyn}$) for the galaxies of the sample observed with VIMOS have recently been obtained (see Bellocchi et al. 2013 for details). For the sake of homogeneity, previous determinations for the INTEGRAL sample (Colina et al. 2005; Garcia-Marin, 2007) have been revisited (Table 1) following the same methodology as in Bellocchi et al. (2013).  

\begin{figure*}
\centering
 %generated with  vmax_mass2as.sm (esta version incluye errores asimetricos)
 % SIMILAR PLOT for morphological classes  vmax_mass3.sm (sin distinguir class 1) y vmax_mass3.sm (distinguiendo class 1). 
 % los ficheros generados con comp_comp.f . Las lineas con vesc.f  
\includegraphics[width=0.75\textwidth]{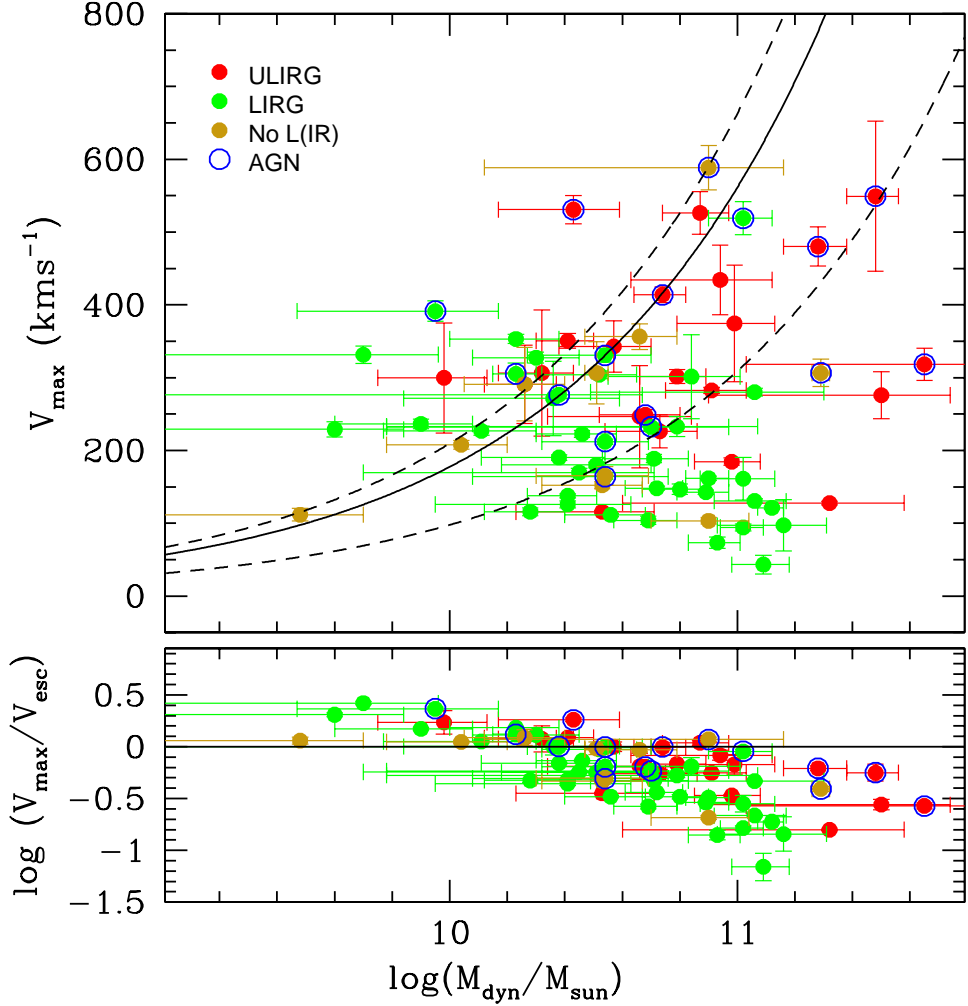}
\caption{Top panel: Maximum outflow velocity as a function of the dynamical mass. Red and green dots distinguish ULIRGs and LIRGs systems. Orange dots are objects without L$_{IR}$ determinations. Blue circles indicate objects with evidence of being affected by an AGN in the optical lines (see text).  The solid line represents the mean escape velocity for r=3kpc, considering a simple model of a truncated isothermal sphere of r$_{max}$/r=10. The dotted lines correspond to values for r$_{max}$/r=1 and 100 (see text). Bottom panel: The maximum outflow velocities are represented in units of the escape velocity for the case r$_{max}$/r=10. }
\
 \label{Fig9}
 \end{figure*}

Within the mass range covered by the entire sample, the distribution of galaxies in the V$_{max}$-M$_{dyn}$ plane is rather scattered, and without any clearly identified trend (Fig. 15). However, there appear to be some differences between LIRGs and ULIRGs.  The maximum velocity of the outflowing gas in LIRGs tends to decrease with the mass of the galaxy from about 200-250 kms$^{-1}$ for the less massive (log(M$_{dyn}/M_{\odot}$)$\leq$10) galaxies to about 100-150 kms$^{-1}$ for the more massive (log(M$_{dyn}/ M_{\odot}$)$\geq$11). This trend is even more clear if galaxies with evidence of hosting an AGN are excluded  (correlation factor, r=-0.54). On the other hand, ULIRGs show a significant scatter with about equal number of galaxies above and below a V$_{max}$ of 300 kms$^{-1}$ independent of their mass. This scatter is partially due to the relatively larger uncertainties in the derived dynamical masses for these objects, which are in a coalescence phase in most of the cases.  A formal unweighted fit to the U/LIRGs (excluding AGNs) is also consistent with a weakly significant negative correlation (r=$-$0.15).   

The fraction of metal-enriched outflowing gas that can escape and pollute the IGM has been estimated comparing the outflow velocity with the local escape velocity. The detailed determination of the escape velocity is difficult as a consequence of the lack of constraints on the halo drag ( c.f. Veilleux et al. 2005).  However, as reference we represent in Fig. 15 (top) the mean escape velocity at r=3kpc for an isothermal sphere for an ample range of truncated values  (i.e., r$_{max}$/r=1, 10, 100), as a function of the dynamical mass. In particular, considering the dependency of Vesc on the circular velocity (e.g.,  Heckman et al. 2000; Veilleux et al. 2005), and the prescriptions used by Bellocchi et al. (2013) for computing Mdyn, we adopt:

\begin{equation}
V_{esc} \approx  (\frac{2  M_{dyn}  G (1 + ln (r_{max}/r)}{3 r})^{1/2} ,
\end{equation} 

where G is the gravitational constant. 

When the outflow velocity is represented in units of the escape velocity (Fig.15 bottom) a clear relation with the dynamical mass is revealed with log(V$_{max}$/V$_{esc}$) $\propto$ $\alpha$  $\times$ log(M$_{dyn}$), where $\alpha$= $-$0.79 and $-$0.62 for non-AGN LIRGs (correlation factor, r = -0.88) and non-AGN U/LIRGs (r = -0.77), respectively.  
This trend indicates that only the less massive galaxies (i.e., M$_{dyn}$ smaller than $\sim$ 10$^{10.4}$ M$_{\odot}$) have outflow velocities higher than the escape velocity, and therefore (part of) their ejected metal-enriched gas can escape the potential well. Massive galaxies  (i.e., M$_{dyn}$$>$ 10$^{10.4}$ M$_{\odot}$) predominantly retain all the gas.  We recall that more than 90 percent of the outflowing gas has velocities lower than the {\it maximun} outflow velocity. 

As expected from our previous results (Sec. 4.2.2), Figure 15 also shows that the presence of an AGN significantly increases the maximum velocity of the outflowing material and, therefore, the escape efficiency of the outflow.  However, the additional contribution of the AGNs to V$_{max}$  in general does not seem to be enough to compensate the increase of the escape velocity as a function of the dynamical mass of the galaxy.

These results are consistent with those of Garnett (2002) who, using a simple closed-box chemical model to interpret the metallicity - rotation velocity relation, suggest that the chemical evolution of galaxies with V$_c$ greater than 125 km s$^{-1}$  (i.e., M $\gtrsim$ 10$^{10.4}$ M$_{\odot}$ is unaffected by GWs, whereas galaxies below this threshold tend to lose a high fraction of their SN ejecta.  Tremonti et al. (2004), using SDSS data from ~53,000 star-forming galaxies at z $\sim$ 0.1, have shown that the gas-phase metallicity of local star-forming galaxies increases steeply with stellar mass from 10$^{8.5}$ to 10$^{10.5}$ M$_{\odot}$. However, they found that the gas-phase metallicity flattens above 10$^{10.5}$ M$_{\odot}$, in good agreement with our finding for the threshold mass above which the outflowing material is retained in the galaxy.  They also interpret these results using a simple chemical model, finding that metal loss is strongly anti-correlated with baryonic mass, and suggesting that GW efficiently remove metals from the galaxy potential for low-mass galaxies.

In summary, our results give strong evidence for a selective effectiveness of outflows polluting the intergalactic medium, even in the strongest starburst galaxies in the low-z universe. Since more than 90\% of the outflowing gas has velocities lower than V$_{max}$, only a low fraction of the ejected ionized material from low-mass LIRGs could reach the intergalactic medium. The outflowing material in massive (log(M$_{dyn} \geq 10.4$ M$_{\odot}$)  LIRGs or ULIRGs, even with AGNs, seems unable to escape the galaxy potential, and therefore to contribute to the metal enrichment of the IGM. These results are in good general agreement with the findings of simple chemical evolutionary models.

\subsubsection {Comparison with outflows in high-z star forming galaxies}

There is ample evidence for powerful galactic outflows in high-z SFGs, as indicated by velocity centroid blueshifts with respect to the systemic in UV metal absorption lines (e.g., Pettini et al. 2000;  Steidel et al. 2010), by broad atomic lines (Maiolino et al. 2012), and broad and shifted H$\alpha$ lines (Shapiro et al. 2009; Genzel et al. 2011; Newman et al. 2012). The H$\alpha$ based studies of high-z populations are particularly relevant in the context of our study because they use the same tracer and follow a similar methodology as the one used here (i.e., IFS - spatially integrated spectra after removing the large-scale velocity field) \footnote{In order to increase the S/N, they also stack the integrated spectra to derive average properties of sub-samples.}. Therefore,  a direct comparison of the overall kinematics of the ionized outflows in local and high-z populations with similar SFRs can be pursued.

It is well established from our previous results that the presence of strong ionized outflows in U/LIRGs is universal.  We find that the observed mean kinematic properties (i.e., FWHM, $\Delta$V) of the ionized outflowing (i.e., broad) component clearly correlate with SFR, while the presence of an AGN contributes to an additional increase in its maximum velocity (by a factor 2 on average) and relative strength (on average a factor 1.4 increase in the broad to narrow emission line flux ratio). These trends and dependencies are also present at high-z, as the speed and strength of the outflow depends on the SFR (Table 4) and on the presence of an AGN (Harrison et al. 2012). Furthermore, a detailed quantitative comparison shows remarkable similarities between the local U/LIRGs and high-z SFGs, especially if this comparison is restricted to those works that have followed a similar methodology as the one used here and deal with objects of comparable SFRs (Shapiro et al. 2009; Newman et al. 2012). Specifically, low- and high-z galaxies with SFRs typical of LIRGs have average values of 350$\pm$24 and 423$\pm$47 kms$^{-1}$, respectively, while galaxies with SFRs typical of ULIRGs have lines $\sim$1.5$\times$ broader (i.e., 510$\pm$12 and 600$\pm$51 kms$^{-1}$, respectively). Regarding the shift of the outflowing gas with respect to the narrow component ($\Delta$V in Table 4), high-z galaxies seem to have slightly smaller values than local U/LIRGs, in particular when galaxies in the ULIRG range are considered ($-$97 $\pm$ 27 km s$^{-1}$ and $\sim$ -23$\pm$4 km s$^{-1}$ for the low- and high-z ULIRGs, respectively; see Table 4). This difference can naturally be explained in terms of internal reddening effects since, on average, local U/LIRGs have larger extinction than the UV/optically selected SINS galaxies (Arribas et al. 2012). However, this difference could also be due to the presence of a higher fraction of  AGNs in the low-z ULIRGs, i.e., outflows with larger blueshifts (see Table 4) than at high-z, which could be mostly pure starbursts. 

Regarding the fraction of outflowing ionized gas, as traced by the flux ratio of the broad to the narrow line components (F(B)/F(N) in Table 4), there is also some evidence that outflows in low-z U/LIRGs could be slightly stronger than the high-z galaxies with similar SFR. Low-z U/LIRGs have a flux in the outflowing component that represents a high fraction (0.84 for the entire sample and 0.76 for non-AGNs) of the flux of the narrow component, showing no dependence with the luminosity (i.e., SFR). On the other hand, outflows in high-z star-forming galaxies appear to represent a lower fraction ($\sim$0.58) of the flux in the narrow component with a hint of an increase as a function of SFR (from 0.50 to 0.65 for galaxies with SFR below and above 100 M$_{\odot}$ yr$^{-1}$ limit (see Table 4 and Newman et al. 2012). 

Differences between the outflows in low- and high-z star-forming galaxies also appear when the strength of the outflow is considered as a function of the compactness of the star formation traced by its extinction-corrected H$\alpha$ surface brightness. There have been recent claims of a strong dependence of the outflows on the star formation surface density in high-z star-forming galaxies (Newman et al. 2012). Based on a sample of 30 z$\sim$ 2 galaxies, Newman and collaborators show that there is a clear threshold around a star formation surface density of $\Sigma_{SFR}$ $\sim$ 1 $M\odot yr^{-1} kpc^{-2}$ such that galaxies with densities above this limit show strong outflows (i.e., F(B)/F(N)= 0.77$\pm$0.027), while galaxies with SF surface densities below that limit show weak outflows (i.e., F(B)/F(N) = 0.16$\pm$0.030).  In our local universe, U/LIRGs cover a wide range in star formation surface density from about 0.1 M$_{\odot}$yr$^{-1}$kpc$^{-2}$ to about 30-50 M$_{\odot}$yr$^{-1}$kpc$^{-2}$. However, the strength of the outflow does not show any evidence of a clear threshold, and not even of a strong dependency with the star formation surface density (Fig. 16). In fact, for a given star formation surface density there is a large scatter in the F(B)/F/N) ratio covering values from about 0.2 to 2.5 (mostly equivalent to $\eta$ values of less than 1 for the adopted electron densities, radius, and range of V$_{max}$, see section $\S$4.2.4). This large scatter could be expected, as the geometry of the outflow and internal extinction also play a role on the observed flux of the broad component (Chen et al 2010) and therefore, on the the F(B)/F(N) ratio. As a consequence of this scatter, it appears to be only a shallow positive dependence, as given by F(B)/F(N) $\propto$  $\Sigma_{SFR}^{+0.12}$ (r= 0.27) for all non-AGN U/LIRGs (Fig. 16).
 
The presence of the $\Sigma_{SFR}$ $\sim$ 1 $M_{\odot} yr^{-1} kpc^{-2}$ threshold in high-z galaxies has been explained as the result of stellar winds and SN driven outflows in turbulence-dominated star-forming discs (Newman et al. 2012). According to models (Ostriker \& Shetty 2011),  the $\Sigma_{SFR}$ threshold for transitioning to strong outflows (i.e., mass loading factors, $\eta \geq$ 1) depends quadratically on the gas surface density in the turbulence dominated regime. Assuming a typical gas fraction and surface density at high-z, the observed threshold agrees with the predicted value (Newman et al. 2012). As the fraction of the gaseous component in galaxies decreases with redshift from about 50\% in z$\sim$ 1-2 star-forming galaxies (Tacconi et al. 2010), to 24\% and 15\% for ULIRGs in the 0.6-1.0 and 0.2-0.4 redshift ranges, respectively (Combes et al. 2013), and 5\% to 7\% (low-z U/LIRGs; Gao \& Solomon 2004; Chung et al. 2009), a lower threshold would be expected at low-z. However, this is not observed (Fig. 14 \& 16).  Assuming similar momentum-driven outflows for low-z U/LIRGs, strong, galactic-scale outflows are expected for star formation surface densities $\Sigma_{SFR} (M_{\odot} yr^{-1} kpc^{-2}) \geq 0.1 \times (\Sigma_g/100 M_{\odot} pc^{-2})^2$ (for typical momentum values of 3000 kms$^{-1}$ per stellar mass; Ostriker \& Shetty 2011). Since the gas surface density in U/LIRGs has typical values of 100 to 1000 $M_{\odot} pc^{-2}$ (Garcia-Burillo et al. 2012), strong outflows would be expected for $\Sigma_{SFR}$ anywhere in the 0.1 to 10 $M_{\odot} yr^{-1} kpc^{-2}$ range and without any clear threshold. This is what is observed in our sample of U/LIRGs (Fig. 14), where the mass loading factor shows a large scatter although with a mild positive dependence with the star formation surface density. Finally, one should note that when the empirical H$\alpha$ broad-to-narrow flux ratio is considered, both low- and high-z star-forming galaxies cover the same range of values (see Fig. 16), suggesting that the threshold identified in high-z galaxies could more likely be part of a continuous variation of the F(B)/F(N) ratio (i.e., of $\eta$) with the $\Sigma_{SFR}$ as observed in low-z U/LIRGs (i.e., $\eta \propto \Sigma_{SFR}^{+0.21}$, section 4.2.5).

 \begin{figure}[h]
 \hskip -1cm
 %generated with  sfrd_ratio.sm 
 % los ficheros generados con comp_comp.f
\includegraphics[width=12cm]{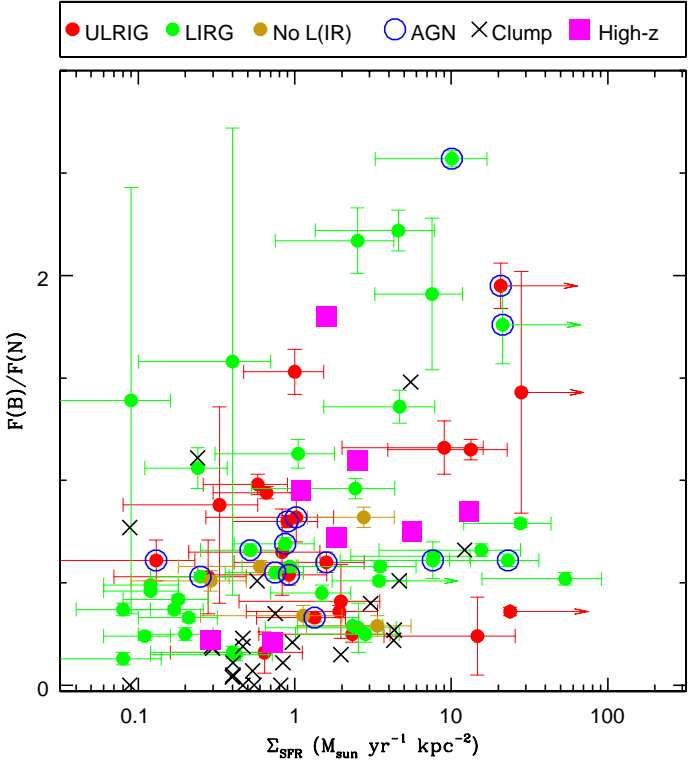}
\caption{Relative strength of the outfow (broad) component, as traced by the ratio of the broad and narrow line fluxes, as a function of the reddening corrected H$\alpha$ star formation rate density.  The magenta squares  represent the high-z (stack) galaxies analysed by Newman et al. (2012) (see text). Symbol code in the upper panel. }
\
 \label{Fig16}
 \end{figure}

\begin{table*}
%\scriptsize
\caption{Comparison of basic kinematic properties for ionized gas (outflow) component for local and high-z (U)LIRGs$^{\star}$.}
\label{table:tabla3}
%\centering
%\begin{tiny}
\begin{tabular}{l l c c c c l }
\hline \hline
%                                                                       & \multicolumn{3}{c}{Narrow} &  & \multicolumn{3}{c}{Broad} \\                                                                          
Sample 		                           			& Ref 				& z				& FWHM(B) 			&  $\Delta$V 		& Flux(B)/Flux(N) 		& Notes    \\ 
                                                                           &                                            &                                   &kms$^{-1}$                       &kms$^{-1}$              &                                             &\\
\hline\hline
LIRGs + ULIRGs                                         	&Present work                    &  local                        & 436$\pm$26                    &  -66$\pm$11           &   0.84$\pm$0.07              & \\
SINS (47  UV/optically selected)	 	&Shapiro et al. 2009	         & $\sim$ 2                   &556$^{+228}_{-87}$        &  +8$^{+14}_{-3}$   &  $\sim$ 0.72$\pm$0.14  & 1 \\
SINS + zC-SINF (30 AO observed)      	&Newman et al. 2012       & 2 - 2.5                        &  $\sim$ 466                      &  -22                           & $\sim$ 0.57                      & 1,2 \\
\hline
LIRGs 		                                         	&Present work                    &  local                        &  350$\pm$24                     &   -42$\pm$10	  &  0.84$\pm$0.11             & \\                                                               
SINS + zC-SINF (SFR $<$100 M$\odot$yr$^{-1}$)&Newman et al. 2012        & 2- 2.5        &  423$\pm$47                    &   -21$\pm$8          &  0.50$\pm$0.041             & 1\\    

\hline
ULIRGs 		                                         	&Present work                    &  local                        &  600$\pm$51                     &   -97$\pm$27            &  0.85$\pm$0.13             & \\                                                               
SINS + zC-SINF (SFR $>$100 M$\odot$yr$^{-1}$)&Newman et al. 2012        & 2 2.5          &   510$\pm$12                   &   -23$\pm$4              &  0.65$\pm$0.074            & 1\\ 
ULIRGs (30 SMGs)                                        &Swinbank et al. 2004      & $\sim$ 2.4               &  1300$\pm$210                &   ...                             &  1.7$\pm$0.3                   & 1, 3 \\

\hline
ULIRGs (no AGN)		                            &Present work                    &  local                        & 452 $\pm$37                      &   -67$\pm$20          &  0.72$\pm$0.11           & \\     
ULIRGs (no-AGN) (SMGs) 			&Swinbank et al. 2004      & $\sim$ 2.4               & 890$\pm$210                     &  ...                              &   0.45$\pm$0.20          & 1, 3\\ 

\hline
ULIRGs (+AGN)                                              &Present work                     &local                           & 814$\pm$74                       &   -145$\pm$59        &  1.04$\pm$0.27            &\\
ULIRGs (+AGN) (4 SMGs)                            &Harrison et al. 2012         &1.4-3.4                       & 1075$\pm$74                    &   -80$\pm$130       &   1.75$\pm$0.50           & 4, 5 \\                       

\hline
\hline
\end{tabular}

\tablefoot{(1): Based on stacked integrated H$\alpha$ spectra, 
            (2): The values  for the table are obtained by averaging of their high- and low- SFR bins
            (3): Swinbank et al. 2004 only consider two components for H$\alpha$, not for the [NII] lines 
            (4): Based on  integrated  [OIII] lines
            (5): For sake of homogeneity we compare with the galaxy integrated data (four sources), though Harrison et al. also provide IFU results.The smearing of the velocity field is not removed. The mean values have been obtained from Table 3 in Harrison et al.,  excluding a third redshifted component by +1350 km/s detected in one of their sources.
$^{\star}$For the SINS galaxies studied in Shapiro et al. 2009 and  Newman et al. 2012, the SFRs (i.e., $\sim$ 100 $M\odot yr^{-1}$) are somewhat larger than our sample ($\sim$ 90 and 21 $M\odot yr^{-1}$, as derived from the infrared  and H$\alpha$ luminosities, respectively; though see effects due to reddening corrections in Sec. 2).  
The Harrison et al. (2012) sample is formed by ULIRGs with SFRs in the range $\sim$ 300-1400 $M\odot yr^{-1}$ and therefore, significantly higher than local U/LIRGs. Similarly for Swinbank et al. (2004), who study a sample with average $L_{IR}$ of $\sim$ 10$^{13} L\odot$, implying $\sim$ 1000 $M\odot yr^{-1}$. Therefore, for these two works  the comparison with local samples is less meaningful. }

\end{table*}

\subsubsection{Star-forming clumps in LIRGs. Outflowing gas properties}

The presence of outflowing ionized gas has been investigated in the sub-sample of 24 H$\alpha$-bright star-forming clumps (see Fig. 1) identified in non-AGN LIRGs characterized by kinematic properties consistent with that of a rotating disc\footnote{For sake of homogeneity, from the 26 H$\alpha$-bright star-forming clumps identified in U/LIRGs  we do not consider those in F06206-6315 and F06035-7102 here. These are two ULIRGs which are at a significantly larger distance than the remaining LIRGs. As consequence,  for the clump in F06206-6315 only a limit on its size could be determined. The clump in F06035-6315 is unusually large, and it is likely an aggregate of smaller SF knots (see Table B1). In addition, F06035-6315 has a complex merger-like kinematics (Bellocchi et al. 2013)}. These SF clumps are characterized by a mean observed (i.e., uncorrected by internal extinction) H$\alpha$-derived SFR and half-light radius of 0.30 M$_{\odot}$ yr$^{-1}$, and 0.49 kpc, respectively (see Table B.1 for specific values).  The presence of a secondary blueshifted, H$\alpha$ broad component is identified in the majority (83\%) of the clumps, indicating that outflowing ionized gas appears to be very common in these luminous clumps characterized by a mean (median) observed star formation surface density of 0.43 (0.13) $M_{\odot} yr^{-1} kpc^{-2}$. However, the average properties of these outflows appear to be significantly different and less extreme than for the outflows of the host galaxies, derived from the integrated H$\alpha$ emission. The mean line width (FWHM(B)= 191$\pm$12kms$^{-1}$), velocity offset ($\Delta$V=$-$18$\pm$8kms$^{-1}$), and strength (F(B)/F(N)=0.37$\pm$0.08) for the individual star-forming clumps in these non-AGN LIRGs have more modest values than the corresponding average values of its class, when the entire galaxy is considered: 299$\pm$17kms$^{-1}$ , -34$\pm$6kms$^{-1}$, and 0.80$\pm$0.12, respectively (see Table 2). The same result, i.e., that ionized gas outflows launched directly by the clumps are less strong than those generated in the entire galaxy,  appears when the properties of each clump are compared with those of the host galaxy (see Figure 17). 

This suggests that a non-negligible fraction of the outflowing material in at least this sub-class of LIRGs does not appear to be directly associated with the brightest star-forming clumps, but rather with the star formation occurring in the nucleus of the galaxy, or in the extended, low surface brightness diffuse medium. This scenario is supported by our spatially resolved spectroscopy (Bellocchi et al. 2013), which shows that strong spatially resolved outflows are mostly detected in the nuclear regions of LIRGs\footnote{Note, however, that in a spaxel-by-spaxel analysis the presence of outflows (i.e., broad component in a two component Gaussian fit) is strongly biassed towards the bright inner regions, which have enough S/N for a proper fit.} with typical extensions of about 1-2 kpc$^2$, while in a relevant fraction (22$\%$; generally the most luminous, i.e., ULIRGs), the presence of outflowing material is identified over larger (projected) areas of about 5 to 16 kpc$^2$. A detailed study of the origin of the outflowing (or highly turbulent) material outside the brightest extranuclear star-forming clumps (i.e., diffuse inter-clump medium) for our entire sample is beyond the scope of the present work and will be the subject of a future analysis.

\begin{figure}[h]
 %generated with  r.vs.v.sm
 % los ficheros generados con comp_comp_knots.f
\includegraphics[width=10cm]{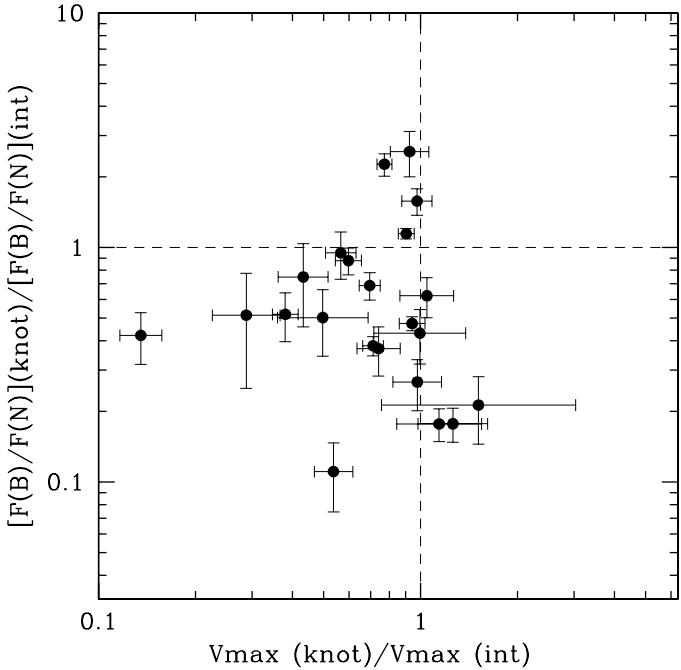}
\caption{V$_{max}$ and relative strength of the broad component with respect to the narrow (i.e., F(B)/F(N)) of the SF clumps in LIRGs. The F(B)/F(N) values for the clumps are normalized by the corresponding value of the host galaxy (integrated) emission.}
\
\label{Fig11}
\end{figure}

\subsubsection{Outflowing ionized gas in low- and high-z star-forming clumps}

Genzel et al. (2011) have recently studied the properties of  outflowing ionized gas coming from giant star-forming clumps in five z$\sim$ 2 galaxies (two large rotating discs with $\sim$ 5 kpc star-forming rings, one large rotating disc with an AGN, one more compact, asymmetric rotating disc, and one dispersion dominated system)  using SINFONI + AO. They have used a similar methodology as the one considered here, with H$\alpha$ as the tracer and fitting two-gaussians per emission line after removing the velocity field smearing, which allow us to make a direct comparison.

As discussed in 4.1.7, the high-z clumps are somewhat larger than the clumps in the LIRGs. While our sample of bright H$\alpha$-clumps have an average half-light radius of 0.49 kpc (see Table B.1), high-z clumps appear to be larger with a radius of 0.8 to 1.5 kpc (Genzel et al. 2011). In terms of their star formation surface density, low and high-z clumps cover to first order a similar range, with the high-z clumps biassed towards the largest star formation densities (i.e., $\sim$ from 0.02 to 2.9 and from 0.14 to 4.9 M$_{\odot}$ yr$^{-1}$ kpc$^{-2}$ for LIRGs and high-z clumps, respectively). Despite the similar star formation densities, the properties of the outflows generated by the clumps in low-z LIRGs appear less extreme than those measured at high redshift. Genzel et al. (2011) find that the broad H$\alpha$ component has line widths (FWHM) in the range 300-1000 km s$^{-1}$, with blueshifts in the -30 to -150 km s$^{-1}$ range, and V$_{max} \sim$ 400-1000 kms$^{-1}$, while carrying between 20\% and 60\% of the total H$\alpha$ flux.  Therefore, while outflows associated with SF clumps in both low- and high-z luminous star-forming galaxies carry out a similar fraction of their H$\alpha$ luminosity, high-z SF clumps appear to be launching outflows with much higher velocities, factors 4-10 higher, than low-z clumps. This would imply that for a given star formation surface density (no internal extinction taken into account), outflows in high-z clumps would be about one to two orders of magnitude more energetic than the outflows launched by clumps in U/LIRGs, i.e., the most luminous starburst galaxies in the low-z universe. Whether these differences are real or related to different systematic factors is not clear yet. The high-z clumps have been identified in galaxies covering the upper range of mass and bolometric luminosity of the z $\sim$ 2 main sequence of star-forming galaxies, and therefore, do not necessary represents the outflow properties of the general population of star-forming clumps at high-z, but only the extreme.  On the other hand, the present sample of low-z clumps are mostly located in low- to intermediate-luminosity LIRGs, therefore not covering clumps in the most luminous low-z systems, i.e., ULIRGs. In addition, the angular resolution (FWHM) of the low-z VIMOS and high-z SINFONI+AO observations correspond to different physical scales (see 4.1.7). It could well be that the larger high-z clumps are really blends of smaller clumps or agglomerates of young stellar clusters that appear unresolved and are more massive than those detected in low-z U/LIRGs. Finally, another possibility could be associated with the location of the clumps. While low-z clumps are at galactocentric distances of about 1-2 kpc, the high-z clumps tend to be at larger galactocentric distances where the gravitational pool and density of the interstellar medium would be smaller than those encountered by the outflows in low-z clumps. These alternatives, among others, should be explored with larger and homogeneous samples of low- and high-z clumps.

\section {Summary and Conclusions}

A detailed study of the integrated kinematic properties of the ionized gas in a large sample (58 systems, 75 individuals) of low-z luminous and ultra-luminous infrared systems is presented. This study is based on high S/N spectra obtained from integral field spectroscopy acquired with the VIMOS/VLT and INTEGRAL/WHT instruments. The properties of a sample of luminous star-forming clumps have also been investigated. This study, which includes the detection and characterization of  global (i.e., host galaxy) and local (i.e., clumps) ionized gas outflows, is the largest so far performed in this class of galaxies. A summary of the main results are given below. 

\begin{enumerate}

\item Global dynamical status of the ambient ionized gas

\begin{itemize}

\item  \textit{The role of star formation regulating $\sigma$}. Isolated non-AGN LIRG discs  have velocity dispersions of the ambient ionized gas (45 $\pm$ 4 kms$^{-1}$) about a factor 2 larger than in less active spirals, likely as a consequence of the increased star formation. However, the heating effects of the star formation tend to saturate in the U/LIRGs luminosity range as indicated by the weak dependency of the velocity dispersion on the IR-derived SFR ($\sigma \propto$ SFR($L_{IR}$)$^{+0.12}$).  The velocity dispersion also has a very slight dependency on the H$\alpha$ star formation surface density ($\sigma \propto \Sigma_{SFR}^{0.06}$), departing significantly from the expected behaviour if  energy release by the starburst were regulating $\sigma$ (i.e., $\sigma \propto \Sigma_{SFR}^{0.5}$).  The relatively small role of the star formation driving the dynamical  status of the ambient ionized gas in U/LIRGs is reinforced by the fact that  the clumps of strong SF do not have an increase in velocity dispersion associated, as we find in the corresponding IFS maps.

\item  \textit{Heating by tidal forces}. The velocity dispersion of the ionized gas follows a sequence of increasing values with the dynamical phase of the system, from isolated discs to interacting pairs and mergers  ($<\sigma>$ of 45, 59, and 73  km s$^{-1}$, respectively). This evolution towards dynamically hotter systems seems to be mainly driven by the gravitational energy release associated with the interacting/merging process, rather than with star formation. 

\item  \textit{The role of AGNs}. The impact of AGNs in the dynamical status of the ionized gas is significant in ULIRGs, with a net  increase in the average velocity dispersion from 74$\pm$6 km s$^{-1}$ for non-AGNs to 109$\pm$14 km s$^{-1}$ for AGNs, i.e., by a factor 1.5. For LIRGs the impact is significantly lower (i.e., a factor of 1.2).  

\item  \textit{Local versus high-z star-forming galaxies}. The low-z U/LIRGs cover a range in velocity dispersion ($\sim$  30 to 100 km s$^{-1}$) and star formation surface brightness ($\sim$ 0.1 to 20 M$_{\odot} yr^{-1} kpc^{-2}$) similar to that of high-z SFGs. Moreover, the observed weak dependency of the velocity dispersion of the ionized gas on the compactness of the star formation for local U/LIRGs ($\sigma \propto \Sigma_{SFR}^{n}$, with $n= 0.06 \pm 0.03$) is in very good agreement with the dependency reported by Genzel et al. (2011) at high-z  ($n= 0.07 \pm 0.025$). These results could indicate that star formation, though non-negligible, is not the main driver regulating the dynamical status of the ambient ionized gas neither at low nor high-z. 

\item  \textit{Star forming clumps properties and scaling laws}. The sizes, luminosities and velocity dispersions of the clumps are in general consistent with the mean scaling  relations derived for (mainly) giant HII regions in spirals, although the smaller clumps (diameter less than 1 kpc) appear more  luminous, at a given size.  When compared with high-z clumps, they are in general smaller, less luminous and have lower velocity dispersions. However, part of these differences could be due to systematic observational effects (e.g., different linear resolution at low- and high-z). 
\end{itemize}

\item  Ionized gas outflows

\begin{itemize}

\item \textit{Detection rate}. The presence of ionized gas outflows in U/LIRGs seems to be universal based on the detection of a secondary broad, usually blueshfited, kinematic component in the H$\alpha$ emission line. 

\item  \textit{The relative role of star formation and AGN}.  In non-AGN U/LIRGs, the maximum velocity of the outflow (V$_{max}$) shows a dependency on the star formation rate of the type V$_{max}(non-AGN) \propto SFR^{+0.24}$. In U/LIRGs with an AGN, the velocity increases by a factor of $\sim$2  with respect to non-AGNs, independent of the global SFR (i.e., V$_{max}(AGN) \propto SFR^{+0.23}$). Moreover, the fraction of  H$\alpha$ flux associated with the outflow increases by a factor of $\sim$ 1.4 in U/LIRGs with an AGN. Thus, AGNs in U/LIRGs are able to generate faster and more massive ionized gas outflows than pure starbursts. 

\item  \textit{Dependency on merger evolution}. The observed differences in  the properties of the outflows for isolated discs, interacting systems, and mergers are consistent with their associated change in SF. Therefore, the release of gravitational potential along the merger evolution does not seem to affect the properties of the outflow, except for its effects on the SF.   

\item  \textit{Outflowing mass and mass loading factors}. The outflowing mass rate covers a wide range from about 0.1 to 100 M$_{\odot}yr^{-1}$  with a close to linear dependency with the IR-derived SFR ($\dot{M_g} \propto SFR^{+1.11}$). Mass loading factors ($\eta$) are in general  below one, with only a few AGNs above this limit. The dependency of  $\eta$ on the dynamical mass of the system agrees well with recent simulations (Hopkins et al. 2012). However, the observed absolute values for $\eta$, as well as its trend with star formation surface density (($\eta \propto \Sigma_{SFR}^{+0.21}$, for all non-AGN U/LIRGs) disagree with these models.

\item  \textit{Escape fractions and IGM metal enrichment}. Only less massive (log(M$_{dyn}/M_{\odot}$) $<$ 10.4) U/LIRGs have outflowing maximum velocities (V$_{max}$) higher than their escape velocities, and therefore part of their ejected metal-rich gas could escape the  gravitational potential of the host galaxy. More massive galaxies (log(M$_{dyn}/M_{\odot}$) $>$ 10.4) would retain all the gas, even if they host an AGN. These results are consistent with chemical evolutionary models.

\item  \textit{Outflows in low- and high-z star-forming galaxies}. The observed average properties (line width, velocity shift, and broad-to-narrow line flux ratio) are similar in low-z U/LIRGs and high-z star-forming galaxies, with hints that the fraction of  H$\alpha$ flux in the broad component could somehow be larger in low-z U/LIRGs. While high-z outflows show mass loading factors above one for galaxies with star formation surface densities above 1 $M_{\odot} yr^{-1} kpc^{-2}$, such a threshold is not observed in low-z  U/LIRGs even after considering the different gas fraction.

\item \textit{Outflows in luminous star-forming clumps}. Ionized gas outflows appear to be very common (detection rate over 80\%) in the bright SF clumps found in LIRGs, which are characterized by a median star formation  surface density of 0.13 $M_{\odot} yr^{-1} kpc^{-2}$ and half-light radius of 0.49 kpc.  Their properties (line width, velocity shift, and fraction of H$\alpha$ flux in the broad component) appear to be less extreme than those associated with the galaxy, and therefore the energy involved in the outflows associated with individual clumps is (in relative terms) more modest than that for  the entire galaxy. 

\item  \textit{Outflows in low- and high-z star-forming clumps}. Ionized gas outflows generated in star-forming clumps at all redshifts  appear to carry a similar fraction of their H$\alpha$ luminosity. However, high-z star-forming clumps launch outflows with much higher velocities, factors 4 to 10 higher, than clumps in low-z U/LIRGs. For a given observed (no internal extinction  correction applied) star formation surface density, outflows in high-z clumps would be about one to two orders of magnitude more  energetic than the outflows launched by clumps in U/LIRGs, i.e. the most extreme starburst galaxies in our low-z universe.

\end{itemize}

\end{enumerate}

 \acknowledgements 
 
We thank Bjorn Emonts for useful discussions and comments after a detailed reading of the manuscript. We also appreciate some information and help provided by Jes{\'u}s Ma{\'{\i}}z-Apell{\'a}niz, Fabi{\'a}n Rosales-Ortega, and Sara Cazzoli. The authors thank an anonymous referee for valuable comments. This paper is based on observations made with the WHT, operated on the island of La Palma by the ING in the Spanish Observatorio del Roque de los Muchachos of the Instituto de Astrof{\'{\i}}sica de Canarias.
This paper is also based on observations carried out at the European Southern observatory, Paranal
(Chile), Programs 076.B-0479(A), 078.B-0072(A) and 081.B-0108(A). This
research made use of the NASA/IPAC Extragalactic Database (NED), which
is operated by the Jet Propulsion Laboratory, California Institute of
Technology, under contract with the National Aeronautics and Space
Administration. This work has been supported by the Spanish Ministry of Science and
Innovation (MICINN) under grants ESP2007-65475-C02-01, AYA2010-21161-C02-01, AYA2012-32295, AYA2012-39408-C02-C01,and by the Marie Curie Initial Training Network ELIXIR of the European Commission under contract PITN-GA-2008-214227.

\clearpage

\appendix
\onecolumn
\section {H$\alpha$-[NII] line fits and kinematic properties from the integrated spectra of galaxies and systems}
%generated with plot3x7_*.sm
% el tercer panel si s—lo los datos de vimos ---> plot3x7_3_solovim.sm
% para los paneles con solo integral plot3x7_#_soloint.sm 
%ficheros generados con @macro2 (@macro1 para los que no tiene ajustes en el NII) en  Astronimia/all_spec/Ajustes_eline. Antes corremos @macro2nd (@macro1nd), para no generar los ficheros (hasta que tenemos el ajuste seguro )
\begin{figure*}[h]
\vspace{0cm}
\includegraphics[width=1\textwidth, height=1\textwidth]{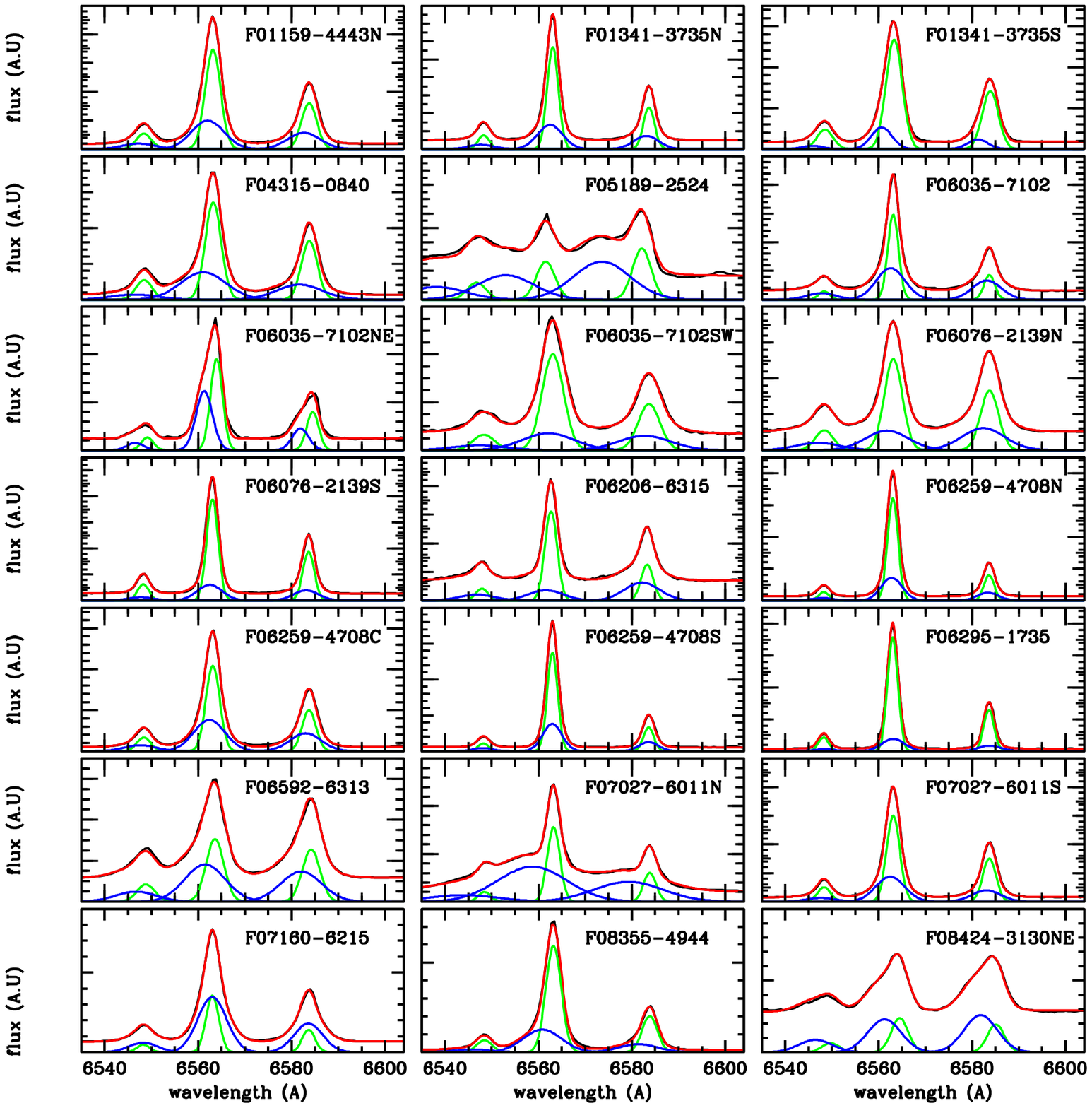}
\vspace{2mm}
\caption{Integrated galaxy spectrum (black) and fit with a two-Gaussian per line model (red) for the VIMOS sub-sample. The narrow (broad) component lines  are represented in green (blue). }
\label{all_panels}
\end{figure*}
\clearpage

\begin{figure*}[h]
\vspace{0cm}
\includegraphics[width=1\textwidth, height=1\textwidth]{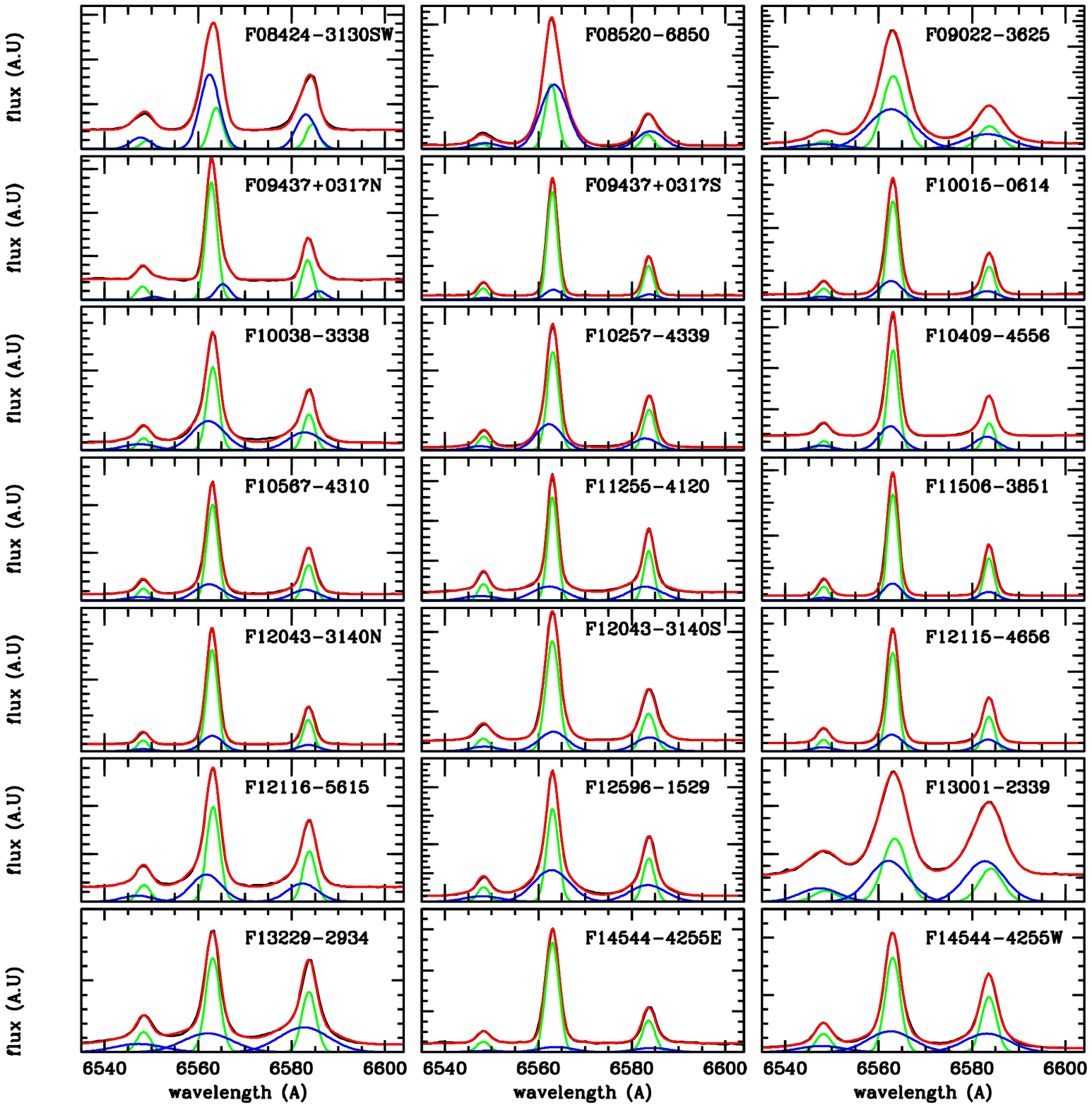}
\vspace{2mm}
%\caption{Integrated galaxy spectrum (black) and fit with a two-Gaussian per line model (red) for the VIMOS subsample. The narrow (broad) component lines  are represented in green (blue). }
%\label{all_panels}
%\centering
{\bf Fig. A.1.} Continued from previous page
\end{figure*}
\clearpage

\begin{figure*}[h]
\vspace{0cm}
\includegraphics[width=1\textwidth, height=1\textwidth]{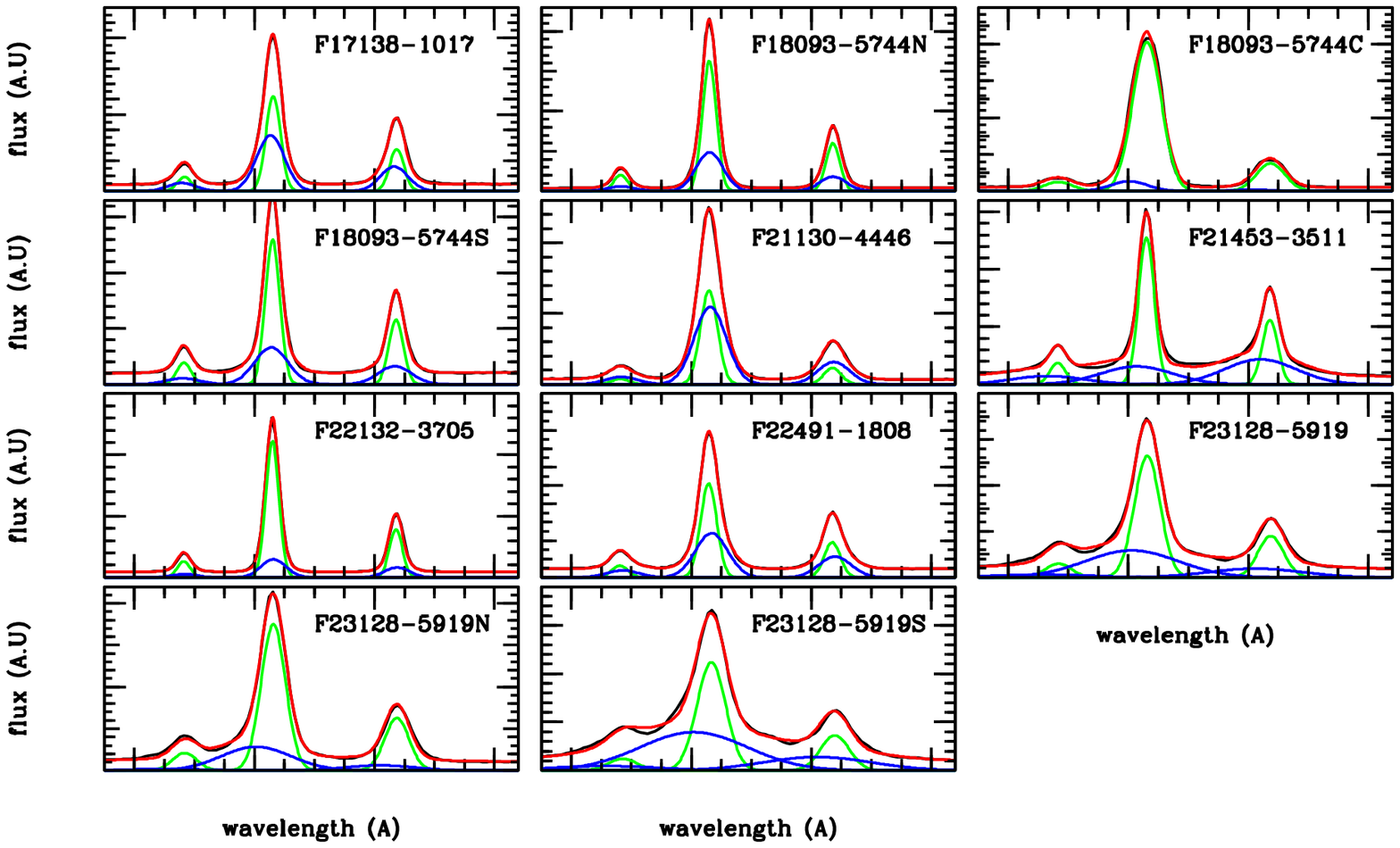}
\vspace{2mm}
%\caption{Integrated galaxy spectrum (black) and fit with a two-Gaussian per line model (red) for the VIMOS subsample. The narrow (broad) component lines  are represented in green (blue).  }
%\label{all_panels}
%\centering
\vskip -6cm
{\bf Fig. A.1.} Continued from previous page
\end{figure*}
\clearpage

\begin{figure*}[h]
\vspace{0cm}
\includegraphics[width=1\textwidth, height=1\textwidth]{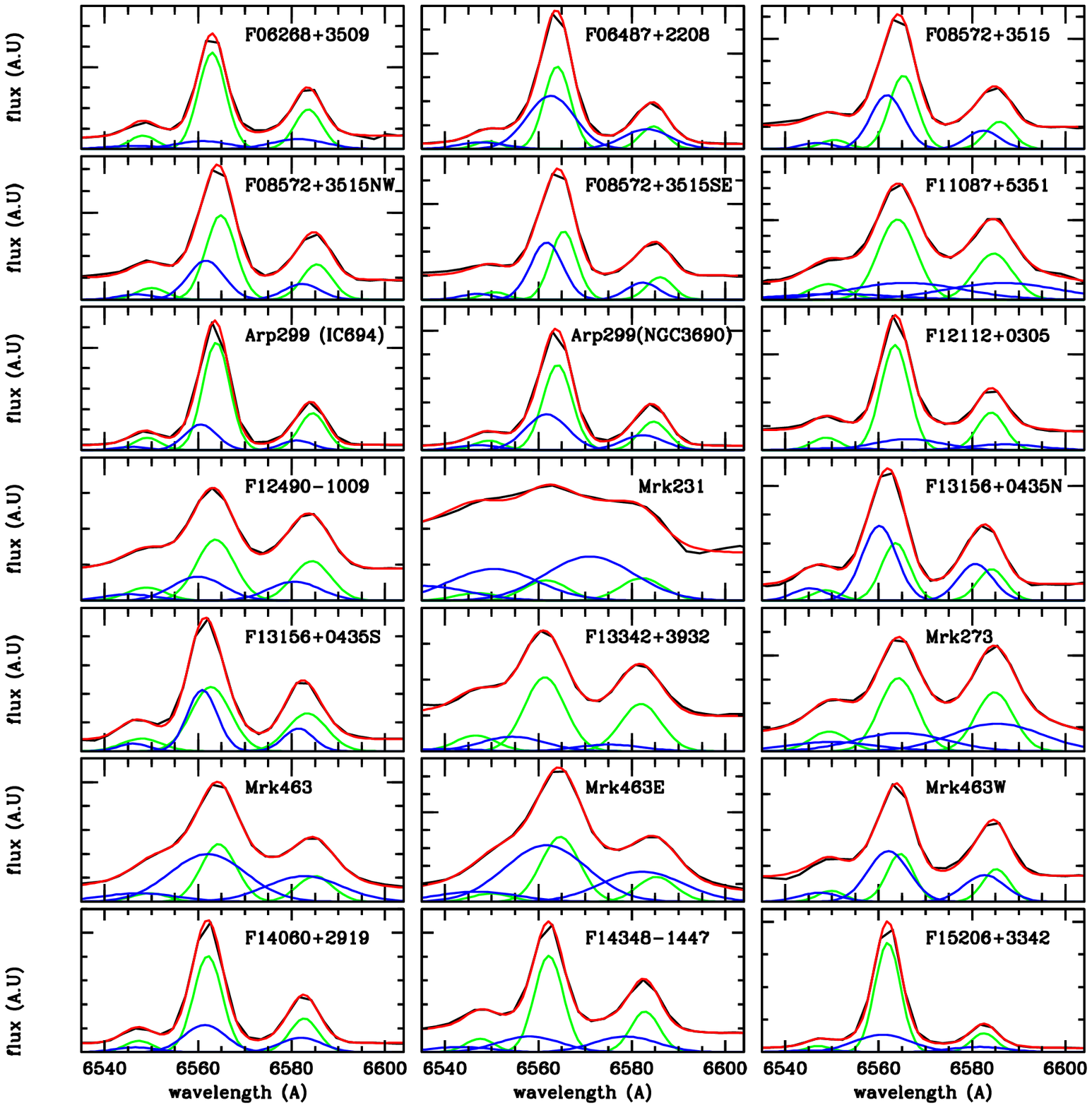}
\vspace{2mm}
\caption{Integrated galaxy spectrum (black) and fit with a two-Gaussian per line model (red) for the INTEGRAL sub-sample. The narrow (broad) component lines  are represented in green (blue).  }
\label{all_panels}
\end{figure*}
%\clearpage

\begin{figure*}[h]
\vspace{0cm}
\includegraphics[width=1\textwidth, height=1\textwidth]{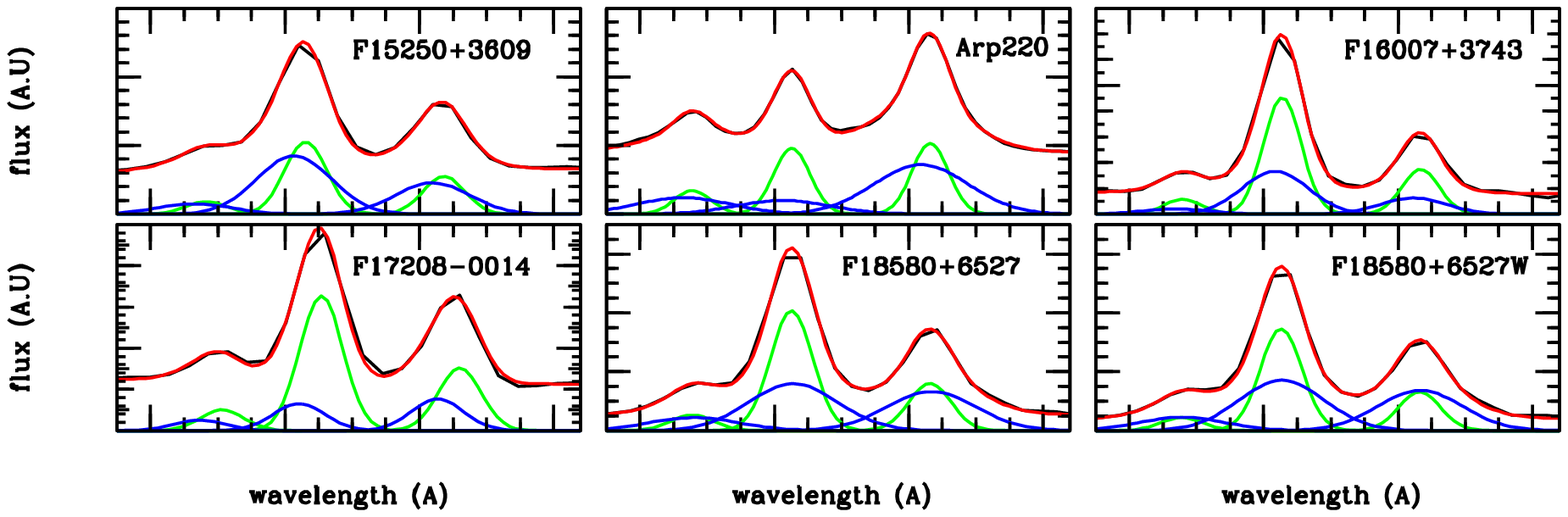}
\vspace{2mm}
\vskip -11cm
{\bf Fig. A.2.} Continued from previous page
\label{all_panels}
\end{figure*}

\clearpage
% This table is generated with comp_comp.f (it needs refinements by hand) 
\begin{table*}[]
\scriptsize
\caption{Kinematic properties from a 2-Gaussian component fit to the integrated H$\alpha$-[NII] spectra}\label{table:tablaA1}
%\centering
%\begin{tiny}
\begin{tabular}{l c c c c c c }
\hline \hline
Sample / ID   & $\sigma$(N) &  FWHM (B) &  $\Delta v$ & V$_{max}$ & F(B) / F(N) & Notes\\               
                & kms$^{-1}$&  kms$^{-1}$&kms$^{-1}$&   kms$^{-1}$&     &        \\               
\hline
\multicolumn{7}{l}{ VIMOS sample}\\
\hline
F01159$-$4443N     &     66 $\pm$ 0.8&  364 $\pm$  8&  -50 $\pm$ 3.3&  232 $\pm$  5& 0.61 $\pm$0.03 & b\\     
F01341$-$3735N     &     42 $\pm$ 0.4&  275 $\pm$  6&  -26 $\pm$ 1.5&  164 $\pm$  3& 0.53 $\pm$0.02 & b\\
F01341$-$3735S     &     71 $\pm$ 0.6&  222 $\pm$ 11& -122 $\pm$12.4&  233 $\pm$ 14& 0.25 $\pm$0.04 & \\
F04315$-$0840      &     70 $\pm$ 0.7&  513 $\pm$  8&  -97 $\pm$ 5.1&  353 $\pm$  7& 0.79 $\pm$0.02 & \\ 
F05189$-$2524      &     85 $\pm$ 2.4&  648 $\pm$ 16& -392 $\pm$ 8.1&  716 $\pm$ 11& 1.95 $\pm$0.11 & b,f\\
F06035$-$7102      &     39 $\pm$ 1.0&  319 $\pm$  9&  -25 $\pm$ 2.2&  185 $\pm$  5& 0.98 $\pm$0.05 & \\
F06035$-$7102N     &     47 $\pm$ 2.0&  170 $\pm$ 14& -120 $\pm$12.7&  204 $\pm$ 14& 0.89 $\pm$0.17 &\\
F06035$-$7102S     &    100 $\pm$ 2.4&  591 $\pm$ 40&  -44 $\pm$16.3&  339 $\pm$ 26& 0.42 $\pm$0.05 & b\\
F06076$-$2139N     &     81 $\pm$ 0.7&  487 $\pm$  8&  -61 $\pm$ 3.7&  304 $\pm$  5& 0.51 $\pm$0.02 &\\
F06076$-$2139S     &     44 $\pm$ 1.0&  280 $\pm$ 24&  -25 $\pm$ 4.7&  165 $\pm$ 13& 0.34 $\pm$0.05 &\\
F06206$-$6315      &     49 $\pm$ 0.7&  396 $\pm$ 11&  -51 $\pm$ 4.5&  249 $\pm$  7& 0.33 $\pm$0.02 & f \\
F06259$-$4708N     &     37 $\pm$ 0.6&  229 $\pm$  9&  -12 $\pm$ 1.8&  126 $\pm$  5& 0.45 $\pm$0.04 &\\
F06259$-$4708C     &     58 $\pm$ 1.1&  348 $\pm$ 11&  -34 $\pm$ 2.9&  208 $\pm$  6& 0.82 $\pm$0.05 &\\
F06259$-$4708S     &     36 $\pm$ 0.7&  204 $\pm$  9&	-1 $\pm$ 1.2&  103 $\pm$  4& 0.51 $\pm$0.05 &\\
F06295$-$1735      &     36 $\pm$ 0.3&  244 $\pm$ 10&	 6 $\pm$ 1.7&  116 $\pm$  5& 0.24 $\pm$0.02 &\\
F06592$-$6313      &     80 $\pm$ 2.0&  460 $\pm$  8&  -98 $\pm$ 6.1&  328 $\pm$  7& 1.36 $\pm$0.08 &\\
F07027$-$6011N     &     51 $\pm$ 0.6&  791 $\pm$  7& -208 $\pm$ 4.4&  604 $\pm$  6& 2.57 $\pm$0.04 & b, f\\
F07027$-$6011S     &     54 $\pm$ 0.6&  322 $\pm$  7&  -29 $\pm$ 2.0&  190 $\pm$  4& 0.63 $\pm$0.03 &\\
F07160$-$6215      &     52 $\pm$ 1.7&  320 $\pm$  6&	-2 $\pm$ 1.3&  162 $\pm$  3& 2.17 $\pm$0.16 &\\
08355$-$4944       &     66 $\pm$ 1.1&  440 $\pm$ 16& -111 $\pm$ 8.9&  332 $\pm$ 12& 0.52 $\pm$0.03 &\\
F08424$-$3130NE    &     90 $\pm$ 2.6&  299 $\pm$ 18& -207 $\pm$15.3&  356 $\pm$ 18& 0.75 $\pm$0.10 &\\
F08424$-$3130SW    &     52 $\pm$ 3.1&  211 $\pm$  4&  -64 $\pm$ 5.3&  170 $\pm$  6& 2.74 $\pm$0.53 &\\
F08520$-$6850      &     48 $\pm$ 1.1&  302 $\pm$  3&	30 $\pm$ 1.2&  121 $\pm$  2& 2.22 $\pm$0.10 & d\\
09022$-$3615       &    101 $\pm$ 1.2&  524 $\pm$  7&  -21 $\pm$ 1.8&  283 $\pm$  4& 1.15 $\pm$0.05 &\\
F09437$+$0317N     &     43 $\pm$ 0.5&  137 $\pm$ 13&  112 $\pm$11.0&	43 $\pm$ 13& 0.16 $\pm$0.03 & d\\
F09437$+$0317S     &     36 $\pm$ 0.4&  171 $\pm$ 15&	12 $\pm$ 2.7&	73 $\pm$  8& 0.15 $\pm$0.03 & d\\
F10015$-$0614      &     44 $\pm$ 0.4&  246 $\pm$  5&  -20 $\pm$ 1.6&  143 $\pm$  3& 0.37 $\pm$0.02 &\\
F10038$-$3338      &     47 $\pm$ 1.0&  368 $\pm$ 11&  -43 $\pm$ 3.2&  227 $\pm$  6& 0.96 $\pm$0.05 &\\
F10257$-$4339      &     44 $\pm$ 0.5&  283 $\pm$  6&  -39 $\pm$ 2.3&  180 $\pm$  4& 0.58 $\pm$0.02 &\\
F10409$-$4556      &     40 $\pm$ 0.4&  244 $\pm$  5&  -26 $\pm$ 1.7&  148 $\pm$  3& 0.49 $\pm$0.02 &\\
F10567$-$4310      &     43 $\pm$ 0.5&  315 $\pm$ 10&  -31 $\pm$ 3.1&  189 $\pm$  6& 0.42 $\pm$0.02 &\\
F11255$-$4120      &     34 $\pm$ 0.3&  377 $\pm$  8&  -34 $\pm$ 2.6&  223 $\pm$  5& 0.46 $\pm$0.01 &\\
F11506$-$3851      &     36 $\pm$ 0.6&  201 $\pm$ 11&	-3 $\pm$ 1.8&  104 $\pm$  6& 0.29 $\pm$0.04 &\\
F12043$-$3140N     &     41 $\pm$ 0.8&  226 $\pm$ 16&	 1 $\pm$ 2.9&  112 $\pm$  9& 0.29 $\pm$0.05 &\\
F12043$-$3140S     &     55 $\pm$ 0.8&  308 $\pm$ 13&	 8 $\pm$ 2.6&  147 $\pm$  7& 0.37 $\pm$0.03 &\\
F12115$-$4656      &     42 $\pm$ 0.3&  241 $\pm$  6&  -10 $\pm$ 1.1&  130 $\pm$  3& 0.33 $\pm$0.02 &\\
12116$-$5615       &     54 $\pm$ 0.8&  339 $\pm$  9&  -67 $\pm$ 4.2&  236 $\pm$  6& 0.66 $\pm$0.03 &\\
F12596$-$1529      &     45 $\pm$ 0.5&  403 $\pm$  6&  -10 $\pm$ 1.6&  212 $\pm$  4& 1.03 $\pm$0.02 &a \\
F13001$-$2339      &    110 $\pm$ 2.5&  440 $\pm$  7&  -60 $\pm$ 4.2&  280 $\pm$  5& 1.06 $\pm$0.10 &\\
F13229$-$2934      &     60 $\pm$ 1.0&  563 $\pm$ 14&  -49 $\pm$ 5.7&  331 $\pm$  9& 0.69 $\pm$0.03 & f \\
F14544$-$4255E     &     54 $\pm$ 0.8&  388 $\pm$ 47&	33 $\pm$17.3&  161 $\pm$ 29& 0.13 $\pm$0.03 &\\
F14544$-$4255W     &     65 $\pm$ 0.6&  510 $\pm$  9&  -21 $\pm$ 2.8&  276 $\pm$  5& 0.66 $\pm$0.02 & f\\
F17138$-$1017      &     42 $\pm$ 0.8&  235 $\pm$  5&  -20 $\pm$ 1.2&  138 $\pm$  3& 1.13 $\pm$0.07 &\\
F18093$-$5744N     &     40 $\pm$ 0.4&  226 $\pm$  4&	 2 $\pm$ 0.7&  111 $\pm$  2& 0.58 $\pm$0.02 &\\
F18093$-$5744C     &     65 $\pm$ 2.4&  205 $\pm$  8& -126 $\pm$ 9.5&  229 $\pm$ 10& 1.91 $\pm$0.37 &\\
F18093$-$5744S     &     42 $\pm$ 0.6&  286 $\pm$  9&	-9 $\pm$ 1.8&  152 $\pm$  5& 0.58 $\pm$0.03 &\\
F21130$-$4446      &     55 $\pm$ 1.3&  274 $\pm$  5&	 9 $\pm$ 1.0&  128 $\pm$  3& 1.53 $\pm$0.11 & d\\
F21453$-$3511      &     48 $\pm$ 0.8&  636 $\pm$ 21&  -74 $\pm$ 9.0&  391 $\pm$ 14& 0.55 $\pm$0.03 & f\\
F22132$-$3705      &     35 $\pm$ 0.4&  205 $\pm$  9&	 8 $\pm$ 1.9&	94 $\pm$  5& 0.25 $\pm$0.03 &\\
F22491$-$1808      &     50 $\pm$ 0.5&  275 $\pm$  3&	22 $\pm$ 0.9&  116 $\pm$  2& 0.94 $\pm$0.03 &d\\
F23128$-$5919      &     91 $\pm$ 1.2&  826 $\pm$ 31& -118 $\pm$11.4&  531 $\pm$ 19& 0.80 $\pm$0.03 &a, f \\
F23128$-$5919N     &     86 $\pm$ 0.9&  618 $\pm$ 22& -130 $\pm$11.6&  439 $\pm$ 16& 0.46 $\pm$0.02 & \\
F23128$-$5919S     &    104 $\pm$ 1.8&  960 $\pm$ 32& -134 $\pm$10.8&  614 $\pm$ 19& 1.32 $\pm$0.05 &a, f \\
 \hline
\multicolumn{7}{l}{ INTEGRAL sample}\\
\hline
06268$+$3509       &     69 $\pm$ 5.0&  560 $\pm$ 90&  -94 $\pm$66.4&  374 $\pm$ 80& 0.16 $\pm$0.10 &\\
06487$+$2208       &     56 $\pm$ 3.4&  475 $\pm$ 13&  -64 $\pm$ 6.9&  302 $\pm$  9& 1.16 $\pm$0.13 &\\
F08572$+$3515      &     68 $\pm$ 7.5&  265 $\pm$ 31& -158 $\pm$51.6&  291 $\pm$ 54& 0.88 $\pm$0.48 &\\
F08572$+$3515N     &     89 $\pm$ 9.2&  310 $\pm$ 45& -145 $\pm$72.2&  300 $\pm$ 76& 0.55 $\pm$0.43 & c\\
F08572$+$3515S     &     53 $\pm$ 8.3&  248 $\pm$ 30& -167 $\pm$48.4&  291 $\pm$ 51& 1.03 $\pm$0.53 &\\
F11087$+$5351      &    147 $\pm$ 5.7& 1274 $\pm$176&	88 $\pm$53.3&  549 $\pm$103& 0.61 $\pm$0.10 & f \\
Arp299E / IC694    &     61 $\pm$ 2.4&  241 $\pm$ 35& -151 $\pm$48.0&  272 $\pm$ 51& 0.28 $\pm$0.12 & \\
Arp299W/ NGC3690   &     82 $\pm$ 2.7&  399 $\pm$ 11& -106 $\pm$13.5&  305 $\pm$ 15& 0.61 $\pm$0.09 & f, g\\
F12112$+$0305      &     75 $\pm$ 2.5&  749 $\pm$103&  128 $\pm$47.6&  247 $\pm$ 70& 0.25 $\pm$0.04 &\\
F12490$-$1009      &    147 $\pm$ 6.4&  529 $\pm$ 36& -170 $\pm$44.1&  434 $\pm$ 48& 0.53 $\pm$0.18 &\\
Mrk 231            &    180 $\pm$40.2&  979 $\pm$140& -509 $\pm$90.7&  999 $\pm$114& 3.14 $\pm$1.44 &f\\
F13156$+$0435N     &     76 $\pm$11.3&  286 $\pm$ 34& -159 $\pm$55.1&  301 $\pm$ 58& 1.58 $\pm$1.14 &\\
F13156$+$0435S     &     88 $\pm$16.4&  364 $\pm$ 19&	85 $\pm$34.2&	97 $\pm$ 35& 1.39 $\pm$1.04 & d\\
F13342$+$3932      &    162 $\pm$11.4&  690 $\pm$279& -312 $\pm$162.&  657 $\pm$214& 0.31 $\pm$0.18 & c,f\\
Mrk273             &    145 $\pm$ 3.2& 1007 $\pm$ 49&	23 $\pm$11.5&  480 $\pm$ 27& 0.60 $\pm$0.05 & d, f\\
Mrk463             &    129 $\pm$ 6.7&  826 $\pm$ 35& -106 $\pm$14.6&  519 $\pm$ 23& 1.76 $\pm$0.19 & f \\
Mrk463E            &    149 $\pm$ 9.5&  899 $\pm$ 48& -139 $\pm$18.3&  589 $\pm$ 30& 1.85 $\pm$0.25 & f \\
Mrk463W            &     62 $\pm$15.8&  382 $\pm$ 30& -116 $\pm$40.1&  307 $\pm$ 43& 1.61 $\pm$0.90 &\\
F14060$+$2919      &     78 $\pm$ 4.5&  391 $\pm$ 38&  -31 $\pm$12.4&  226 $\pm$ 23& 0.41 $\pm$0.18 &\\
F14348$-$1447      &     79 $\pm$ 2.0&  671 $\pm$ 33& -191 $\pm$23.9&  526 $\pm$ 29& 0.36 $\pm$0.03 &\\
F15206$+$3342      &     46 $\pm$ 0.8&  601 $\pm$ 17&  -50 $\pm$ 5.4&  351 $\pm$ 10& 0.36 $\pm$0.02 &\\
F15250$+$3609      &     82 $\pm$14.0&  528 $\pm$ 42&  -79 $\pm$28.0&  343 $\pm$ 35& 1.43 $\pm$0.59 &\\
Arp220             &     44 $\pm$ 2.3&  705 $\pm$ 18&  -61 $\pm$ 5.7&  414 $\pm$ 11& 0.54 $\pm$0.05 &g\\
F16007$+$3743      &     53 $\pm$ 6.5&  453 $\pm$ 54&  -49 $\pm$18.0&  276 $\pm$ 32& 0.65 $\pm$0.21 &\\
F17207$-$0014      &     89 $\pm$ 6.6&  310 $\pm$ 58& -151 $\pm$81.7&  306 $\pm$ 87& 0.24 $\pm$0.19 &c\\
F18580$+$6527      &     84 $\pm$ 4.7&  658 $\pm$ 40&	11 $\pm$ 9.3&  318 $\pm$ 22& 0.82 $\pm$0.12 & f \\
F18580$+$6527W     &     80 $\pm$ 5.3&  622 $\pm$ 34&	 5 $\pm$ 7.9&  306 $\pm$ 19& 1.01 $\pm$0.15 &f\\
\hline
\hline
\end{tabular}
\vskip -0.2cm
\tablefoot{
Columns: (1) Sample and identification according to the IRAS code, or other common name; (2) Intrinsic (i.e., deconvolved from the instrumental profile) sigma (i.e., FWHM/2.35) of the narrower component;  (3) Intrinsic FWHM of the broader component; (4) difference in the central velocity of the broad component with respect to the narrow component; (5) Maximum velocity associated with the broad component defined as V$_{max}$= abs(-$\Delta$V+FWHM/2.) (e.g., Veilleux et al. 2005; Rupke et al., 2005), (6) flux ratio between the broad and the narrow H$\alpha$ lines; (7) Notes:  (a) presence of an extra (very) broad H$\alpha$ line; (b) some evidence for a very broad additional H$\alpha$ line; (c) large uncertainty in the flux ratio; (d) broad component redshifted by more than 2-sigma; (e) bad fit; (f) Seyfert-like nuclear spectral classification (see compilations by Rodriguez-Zaurin et al. 2011 and Garcia-Marin et al. 2009). Quoted errors are only those associated with the fit; (g) spectroscopic evidence for an AGN (Arp299: Garcia-Marin et al. 2006; Arp220: Arribas et al. 2001). 
}
\end{table*}

\onecolumn
\section {H$\alpha$-[NII] line fits and kinematic properties from the integrated spectra of star forming clumps}
%\vskip 2cm

%generated with plot3x7_knots1.sm
%ficheros generados con @macro2 (@macro1 para los que no tiene ajustes en el NII) en VIMOS/INTEGRADOS_KNOTS
\begin{figure*}[h]
\includegraphics[width=1\textwidth, height=1\textwidth]{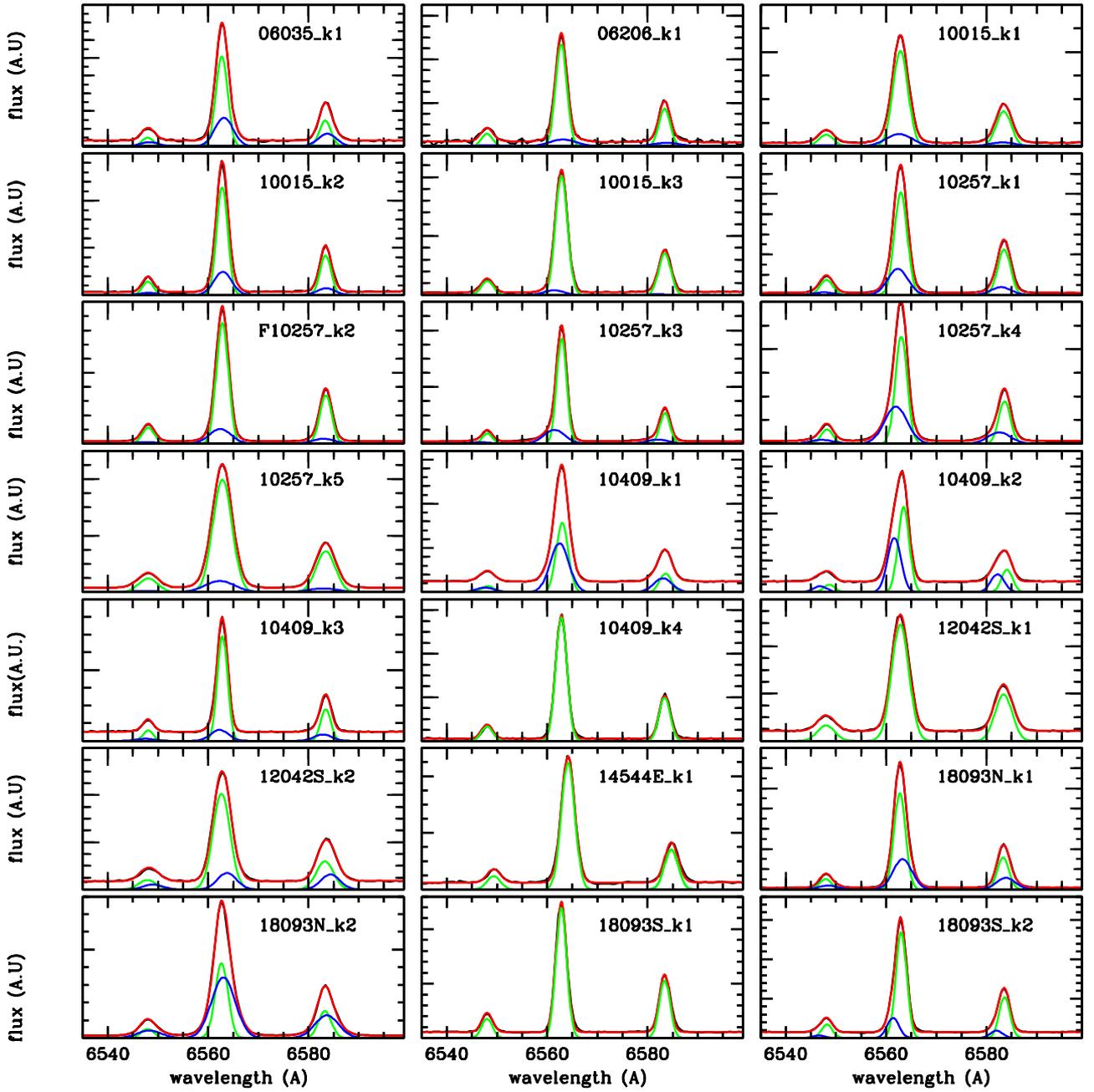}
\caption{Integrated spectrum (black) and fit with a two-Gaussian per line model (red) for the clumps. The narrow (broad) component lines  are represented in green (blue). }
\label{panel1}
\end{figure*}

%generated with plot3x7_knots2.sm
%ficheros generados con @macro2 (@macro1 para los que no tiene ajustes en el NII) en VIMOS/INTEGRADOS_KNOTS

\begin{figure*}[h]
\includegraphics[width=1\textwidth, height=1\textwidth]{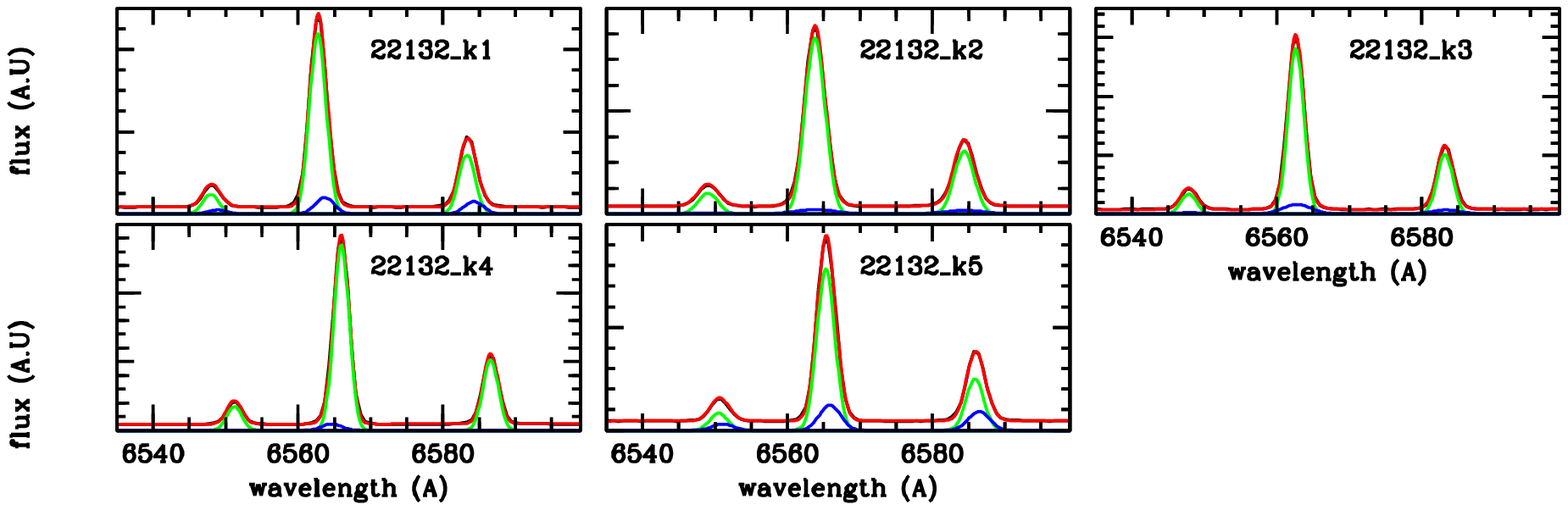}
\vskip -11cm
{\bf Fig. B.1.} Continued from previous page
\end{figure*}

%Esta tabla se ha obtenido del fichero "res_knots.dat" generado con comp_comp_knots.f y posteriormente editada a mano.  
% These data come from t_kinematics.dat generated in /all_spec with comp_comp.f 
\begin{table*}[h]
\scriptsize
\caption{Properties of the selected star forming clumps$^1$}\label{table:tablaB1}
%\centering
%\begin{tiny}
\begin{tabular}{l c c c c c c c c c }
\hline \hline
 ID   & r$_{hl}$& FWHM (N)   &  FWHM (B)  &  $\Delta v$  &  F(B) / F(N) & V$_{max}$ & SFR& $\Sigma_{SFR}$& Notes\\               
        & kpc       & kms$^{-1}$ &  kms$^{-1}$ &  kms$^{-1}$&                    &kms$^{-1}$& $M\odot yr^{-1}$ &  $M\odot yr^{-1}kpc^{-2}$                    \\               
\hline
     6035.1 & 1.69 $\pm$ 0.5 &   90.9 $\pm$  3.2    &  179. $\pm$ 12.0 &   19.3 $\pm$  3.9	&0.51 $\pm$   0.12  &     70.2   $\pm$ 7.2                           & 2.45  $\pm$  0.45 & 0.14 $\pm$ 0.11& c\\
     6206.1  & ( $<$1.84)&  84.6  $\pm$ 2.7    &  280.$\pm$  61.6 &   16.2  $\pm$ 16.2	&0.17 $\pm$   0.05  &    123.7   $\pm$34.8                        &  0.23 $\pm$   0.04 & ($>$ 0.012)  &d \\
    10015.1 & 0.75 $\pm$ 0.22 & 138.$\pm$  2.0     &  276. $\pm$ 21.3 &  -11.2 $\pm$    4.2	&0.23 $\pm$   0.04  &    149.2  $\pm$ 11.4               &  0.39 $\pm$   0.07 & 0.11 $\pm$ 0.09&\\
    10015.2 & 0.38 $\pm$ 0.11 &  82.9$\pm$   2.2   &  173. $\pm$ 11.9 &     5.7  $\pm$   2.7	&0.35 $\pm$   0.08  &     80.7   $\pm$ 6.5                  &  0.16  $\pm$  0.03 & 0.18 $\pm$ 0.14& \\
    10015.3 & 0.46$\pm$ 0.14 &  96.6$\pm$   0.6   & 190. $\pm$ 31.1  &  -67.7 $\pm$   10.0	&0.07 $\pm$   0.01  &    163.0  $\pm$ 18.5               &  0.17 $\pm$   0.03 & 0.13 $\pm$ 0.10 &\\
    10257.1 & 0.31 $\pm$ 0.09 & 108. $\pm$ 1.6     &  197. $\pm$  7.6  & -26.6   $\pm$  3.5	&0.40 $\pm$   0.05  &    125.3  $\pm$  5.2                 & 0.44  $\pm$  0.08  & 0.73 $\pm$ 0.57 & \\
    10257.2 & 0.21 $\pm$ 0.06 &  96.4 $\pm$  0.7    &  217. $\pm$  8.9  & -19.9  $\pm$   2.3	&0.22 $\pm$   0.02 &     128.3  $\pm$  5.0                 & 0.30  $\pm$  0.05  &  1.02 $\pm$ 0.80 &\\
    10257.3 & 0.19 $\pm$ 0.06 &  65.2$\pm$  0.8     & 207. $\pm$  7.2  & -66.0   $\pm$  5.1	&0.27 $\pm$   0.02  &    169.7  $\pm$  6.2                 & 0.24  $\pm$  0.04  & 1.03 $\pm$ 0.81 &\\
    10257.4 & 0.20 $\pm$ 0.06 &  94.6 $\pm$  1.0    &  227. $\pm$  3.2  & -49.0  $\pm$   2.2	&0.66$\pm$    0.03 &     162.6   $\pm$ 2.7                 & 0.73  $\pm$  0.13 &  2.90 $\pm$ 2.27 &\\
    10257.5 & 0.27 $\pm$ 0.08 & 186. $\pm$  1.6     &   307.$\pm$  25.1&   -22.8 $\pm$ 3.6	&0.15$\pm$    0.04  &   176.0   $\pm$13.1                 &  0.22  $\pm$  0.04 & 0.47 $\pm$ 0.37 &\\
    10409.1&  0.98 $\pm$ 0.29 &  96.1 $\pm$  2.3    &  180. $\pm$  3.6 &  -24.5   $\pm$  1.8	&1.11 $\pm$   0.11 &     114.6  $\pm$  2.6                 &  0.34  $\pm$  0.06 & 0.06 $\pm$ 0.04 &\\
    10409.2 & 1.47 $\pm$ 0.44 &  85.5 $\pm$  1.8    &  118. $\pm$  5.0 &  -85.4   $\pm$  5.8	&0.77$\pm$    0.09 &     144.7   $\pm$ 6.3                 &  0.29  $\pm$  0.05 & 0.023 $\pm$ 0.017 &\\
    10409.3 & 0.43 $\pm$ 0.13 &  70.1  $\pm$ 1.4    &  169.$\pm$  14.4&   -25.4  $\pm$   6.5	&0.18 $\pm$   0.04  &    109.8  $\pm$  9.7                 &  0.08  $\pm$  0.02 & 0.07 $\pm$ 0.06 &\\
    10409.4 & 0.35 $\pm$ 0.10 &  83.9  $\pm$  0.6    &                ...           &                  ...                &        ...                       &                    ...                                 &  0.10  $\pm$  0.02 & 0.13 $\pm$ 0.10& a \\
 12042002.1 & 0.50 $\pm$ 0.15 &164.$\pm$  0.8     &                ...           &                  ...                &         ...                      &                    ...                                &  0.18  $\pm$  0.03 & 0.11 $\pm$ 0.09& a\\
 12042002.2 & 0.52 $\pm$ 0.16 &165.$\pm$ 2.0      &   186.$\pm$   8.3&    50.7   $\pm$ 14.7	&0.19 $\pm$   0.10  &     42.2   $\pm$15.3                 &  0.19  $\pm$  0.03& 0.11 $\pm$  0.09& \\
 14544001.1 & 0.76 $\pm$ 0.23 & 117$\pm$ 0.7      &              ...           &                   ...                  &        ...                       &                   ...                                &  0.08 $\pm$ 0.01 & 0.023 $\pm$ 0.016& \\
 18093001.1 & 0.34 $\pm$ 0.10 & 91.6 $\pm$ 1.9   &   183.$\pm$   6.3 &   24.9  $\pm$   3.0	&0.51$\pm$    0.06  &     66.5    $\pm$4.3                   & 0.82   $\pm$ 0.15 & 1.11 $\pm$ 0.87 &\\
 18093001.2 & 0.41 $\pm$ 0.12 &110.$\pm$7.5       &   239.$\pm$  11.8&   16.6  $\pm$   2.4	&1.48$\pm$   0.32  &   103.0   $\pm$ 6.4                    & 1.36   $\pm$ 0.25 & 1.31 $\pm$  1.03 &\\
 18093002.1 & 0.39 $\pm$ 0.12 & 79.4$\pm$   0.6  &               ...               &                ...              &          ...                    &                    ...                                    & 0.19   $\pm$ 0.04 & 0.19 $\pm$  0.15 & a\\
 18093002.2  & 0.30 $\pm$ 0.09 &80.6 $\pm$1.1    &    99.$\pm$   9.0 &    -73.6  $\pm$  11.7&0.21$\pm$    0.06  &    123.0 $\pm$  12.5                     &  0.13  $\pm$  0.02 & 0.23 $\pm$ 0.18&\\
    22132.1 &  0.47 $\pm$ 0.14 & 89.5 $\pm$ 0.5      &   113.$\pm$   4.2&    43.8  $\pm$   5.7	&0.11 $\pm$   0.02  &     12.7   $\pm$ 6.11                  & 0.28   $\pm$ 0.05 & 0.20 $\pm$  0.16&\\
    22132.2 &  0.63 $\pm$ 0.19 &122. $\pm$0.9        &   292.$\pm$  51.5&    3.6  $\pm$   9.4	&0.05$\pm$    0.02  &    142.7  $\pm$ 27.4                 &  0.24  $\pm$  0.04 & 0.10 $\pm$  0.07&\\
    22132.3 &  0.47 $\pm$ 0.14 & 81.4 $\pm$ 1.0      &   201.$\pm$  22.9&	6.7  $\pm$   5.1	&0.11$\pm$    0.03  &     93.8   $\pm$12.5                            & 0.14   $\pm$ 0.03 & 0.10 $\pm$ 0.07&\\
    22132.4 &  0.37 $\pm$ 0.11 & 76.8 $\pm$ 0.4      &   117.$\pm$  10.6 &     -60.4 $\pm$  6.1&0.04$\pm$    0.00 &     118.8   $\pm$ 8.1                     &  0.08  $\pm$  0.02 & 0.10 $\pm$ 0.07& b \\
    22132.5 &  0.54 $\pm$ 0.16 & 101. $\pm$ 1.1      &   132.$\pm$  5.6   &     25.4  $\pm$   6.3&0.19 $\pm$   0.07  &     40.7    $\pm$7.0                      &  0.13   $\pm$ 0.02 & 0.07 $\pm$ 0.06& \\
\hline

   mean (all) & 0.54$\pm$0.07 & 102.$\pm$6              &  195.$\pm$12       & -15$\pm$8           & 0.36$\pm$0.07    &   112. $\pm$ 9               &0.38 $\pm$ 0.10 & 0.43 $\pm$ 0.13& \\
   median (all) & 0.43                    & 93.                            &  188.                       &-16.                          & 0.21                        &  121.                               & 0.22                     & 0.13\\
   range (all) & 0.19 $-$ 1.69    & 65.$-$186.                &  99$-$307              & -85.$-$+51.             & 0.04$-$1.48          &  12.7$-$176                  &0.08$-$2.45        & 0.023 $-$ 2.90\\ 
\hline
   mean (LIRG)     & 0.49$\pm$0.06 & 103.$\pm$6              &  191.$\pm$12       & -18$\pm$8           & 0.37$\pm$0.08    &   113. $\pm$ 10               &0.30 $\pm$ 0.06 & 0.44 $\pm$ 0.13& \\
   median (LIRG)  & 0.42                    & 95.                            &  188.                         &-21.                          & 0.21                        &  121.                               & 0.20                     & 0.13\\
   range (LIRG)     & 0.19 $-$ 1.47    & 65.$-$186.                &  99$-$307             & -85.$-$+51.             & 0.04$-$1.48          &  12.7$-$176                  &0.08$-$1.36        & 0.023 $-$ 2.90\\ 
\hline

\end{tabular}

\tablefoot{
$^1$    The kinematic properties were obtained from a 2-Gaussian component fit to the integrated H$\alpha$-[NII] spectra of the clumps, except when otherwise stated (see notes). Columns: (1) Identification: First digits of the IRAS code for the galaxy and clump identification according to Fig. 1; (2) Half-light radii, r$_{hl}$. They correspond to intrinsic values (i.e., deconvolved by the PSF) calculated from the IFS based H$\alpha$ maps using the CoG method (e.g., Arribas et al. 2012). Their uncertainties are estimated to be 30 percent.  
(3) FWHM of the narrower component;  (4) FWHM of the broad component; (5) difference in the central velocity of the broad component with respect to the narrow component; (6) flux ratio between the broad and the narrow H$\alpha$ lines; (7) Maximum velocity defined as V$_{max}$=abs(-$\Delta$V+FWHM/2) following, e.g., Rupke et al. (2005).
(8) Observed (i.e. no reddening corrected) star forming rate derived from the H$\alpha$ luminosities. (9) Observed (i.e., no reddening corrected) star-forming rate density within half-light radius.
 (10)  Notes:  (a) Only the FWHM of the 1-Gaussian fit is reported as for this clump the 2-Gaussian fit does not give a good fit, which suggests that if a second component exits, it must be very weak. (b) For this clump the fit for the [NII] lines was not good, but the results for H$\alpha$ reported here were okay.  (c)  IRAS06035 was observed under no photometric conditions (see Rodriguez-Zaurin et al. 2011), so the SRF value may be in error. (d) as the clump size is not resolved (i.e., $\sim$ PSF), only an upper limit on the radius (lower limit on $\Sigma_{SFR}$) could be obtained
Quoted errors are only those associated with the fit, except for the SFR for which considered 30 per cent (errors from the fit were too small to be meaningful). SFR are obtained from the observed H$\alpha$ flux  using the Kennicutt (1998) relation.}

\end{table*}

\end{document}